\documentclass[%
11pt,          %Schriftgroesse
parindent,     % Absatz kennzeichnen durch Einrucken
parskip=never,   % Vermeide Abstand zwischen Absaetzen komplett
oneside,         % Einseitige Ausgabe
headinclude=false,              % Kopfbereich zum Textkoerper rechnen
footinclude=false,              % Fussbereich zum Textkoerper rechnen
DIV=11,                         % DIV-Wert festlegen
BCOR=0mm,                       % Bindekorrektur
titlepage=true,                 % Titelseite auf separater Seite
bibliography=totoc,             % Bibliographie ins Inhaltsverzeichnis
a4paper,                        % Papiergroesse
numbers=noenddot,               % Vermeide abschliessenden Punkt bei Kapitelnummer immer
]{scrartcl}

\usepackage[%
english,    %
]{babel}

\usepackage[%
utf8        % Eingabekodierung auswaehlen
]{inputenc}

\usepackage[%
T1          % Mehr Buchstaben laden
]{fontenc}

\usepackage{lmodern}            % Schriftoptimierung fuer Standardschriftart

\usepackage{textcomp}
\usepackage{tikz}               
\usetikzlibrary{%
  calc,         %
  arrows,       %
  decorations,  %
  patterns,     %
  snakes,       %
  trees         %
}

\usepackage{pgf}
\usepackage[%
headsepline,                    % Trennlinie fuer den Kopfbereich
footsepline                     % Trennlinie fuer den Fussbereich
]{scrlayer-scrpage}
\setkomafont{pagehead}{%
  \normalfont\scshape  % Kapitaelchen fuer Kolummnentitel
}

\usepackage[                    %
affil-it                        % Institutsname kursiv
]{authblk}

\usepackage{framed}

\usepackage{enumitem}

\usepackage{color}

\usepackage{eurosym}            % Verwenden mit \officialeuro

\usepackage{calc}

\usepackage{tabularx}

\usepackage[%
final       % Paket aktiv
]{microtype}

\usepackage[%
all         % Paket aktivieren
]{nowidow}

\usepackage[%
format=plain,           % Unterschriften haengen NICHT am Bezeichner
labelsep=colon,               % Bezeichner wird durch Doppelpunkt abgetrennt
justification=justified,      % Blocksatz fuer Unterschrift
singlelinecheck=true,         % Kurze Bildunterschrift automatisch zentrieren
labelfont=bf                  % Bezeichner fett drucken
]{caption}

\usepackage{subcaption}

\usepackage[%
colorinlistoftodos,             % Farbiger Kasten in Todo-Liste
]{todonotes}

\usepackage[%
style=authoryear,               % Bibliographiestil
natbib=true,                    % Zitierstil
safeinputenc=true,              % Korrigiere falsche Eingabekodierung
backend=biber,                  % Wahl des Backends
maxbibnames=20,                 % Maximale Anzahl Namen pro Bibliographieeintrag
minbibnames=10,                 % Minimale Anzahl Namen pro Bibliographieeintrag
maxcitenames=2,                 % Maximale Anzahl Namen pro Zitierung im Text
hyperref=true,                  % Aktiviere Verweisfunktionalitaet
backref=false,                  % Seitenzahl des Zitats in der Bibliographie
isbn=false,                     % ISBN anzeigen
url=false,                      % URL anzeigen
doi=true,                      % DOI anzeigen
eprint=false,                   % E-Print anzeigen
firstinits=true,                % Initialen anzeigen
uniquename=false,               % Eindeutige Namen sicherstellen
uniquelist=false,               % Eindeutige Namen in Bibliographie
sorting=nyt                     % Sortierreihenfolge
]{biblatex}

\bibliography{references.bib}        % Eingabedatei

\DeclareNameAlias{sortname}{last-first} % Reihenfolge der Namen

\AtBeginBibliography{%
}
\renewbibmacro*{bbx:savehash}{} % Technischer Befehl

\usepackage{amsmath}
\usepackage{amsthm}
\usepackage{amsfonts}
\usepackage{amssymb}
\usepackage{amstext}

\usepackage{mathtools}

\usepackage{physics}
\usepackage[version=4]{mhchem}

\renewcommand{\epsilon}{\varepsilon}
\renewcommand{\phi}{\varphi}

\usepackage{siunitx}

\DeclareSIUnit[number-unit-product = {\,}]\cal{cal}

\DeclareSIUnit[number-unit-product = {\,}]\sec{sec} % Neue Einheit: Sekunde in alt

\usepackage[left]{lineno}
\usepackage[%
hidelinks,  % Keine Box und Farben fuer Verweise
]{hyperref}

\begin{document}
%% =============================================================================
%% =============================================================================
%%%%%%%%%%%%%%%%%%%%%%%%%%%%%%%%%%%%%%%%%%%%%%%%%%%%%%%%%%%%%%%%%%%%%%%%%%%%%%%%

% % Vorspann fuer Buchklasse
% \frontmatter

%%%%%%%%%%%%%%%%%%%%%%%%%%%%%%%%%%%%%%%%%%%%%%%%%%%%%%%%%%%%%%%%%%%%%%%%%%%%%%%%
%% Titelseite
\microtypesetup{protrusion=false}
% \subject{WISSENSCHAFTLICHER AUFSATZ}
\title{Ice clouds as nonlinear oscillators}
% \subtitle{UNTERTITEL}
% \titlehead{%
%   \includegraphics[width=0.3\textwidth]{penguin.pdf}\\%
%   Johannes Gutenberg Universität Mainz\\%
%   Institut für Physik der Atmosphäre
% }

\author[1]{Hannah Bergner}
\author[1,*]{Peter Spichtinger}
%\author[1,2]{Juliane Rosemeier}
%\author[2]{Manuel Baumgartner}
%\author[3]{Manuel Baumgartner}

%\affil[1]{Institute for Physics of the Atmosphere, Johannes
%  Gutenberg University, Mainz, Germany} 
\affil[1]{Institute for Atmospheric Physics, Johannes
  Gutenberg University, Mainz, Germany}
\affil[*]{corresponding author}
%\affil[2]{now at University of Exeter, UK} 
%\affil[3]{German Weather Service (DWD), Offenbach, Germany} 
%\affil[2]{German Weather Service (DWD), Offenbach, Germany} 

\date{\today}
%\publishers{to be submitted to Chaos}
\maketitle
\microtypesetup{protrusion=true}
%%%%%%%%%%%%%%%%%%%%%%%%%%%%%%%%%%%%%%%%%%%%%%%%%%%%%%%%%%%%%%%%%%%%%%%%%%%%%%%%

% %%%%%%%%%%%%%%%%%%%%%%%%%%%%%%%%%%%%%%%%%%%%%%%%%%%%%%%%%%%%%%%%%%%%%%%%%%%%%%%%
% % Zusammenfassung
% \begin{abstract}
%   Dies ist die Zusammenfassung
% \end{abstract}
% %%%%%%%%%%%%%%%%%%%%%%%%%%%%%%%%%%%%%%%%%%%%%%%%%%%%%%%%%%%%%%%%%%%%%%%%%%%%%%%%

%%%%%%%%%%%%%%%%%%%%%%%%%%%%%%%%%%%%%%%%%%%%%%%%%%%%%%%%%%%%%%%%%%%%%%%%%%%%%%%%
% Inhaltsverzeichnis
\microtypesetup{protrusion=false}
\tableofcontents
\microtypesetup{protrusion=true}
\clearpage
%%%%%%%%%%%%%%%%%%%%%%%%%%%%%%%%%%%%%%%%%%%%%%%%%%%%%%%%%%%%%%%%%%%%%%%%%%%%%%%%

% % Hauptteil fuer Buchklasse
% \mainmatter

\begin{abstract}
    \textbf{Abstract}: Clouds are important features of the atmosphere, determining the energy budget by interacting with incoming solar radiation and outgoing thermal radiation, respectively. For pure ice clouds, the net effect of radiative effect is still unknown. In this study we develop a simple but physically consistent ice cloud model, and analyze it using methods from the theory of dynamical systems. We find that the model constitutes a nonlinear oscillator with two Hopf bifurcations in the relevant parameter regime. In addition to the characterization of the equilibrium states and the occurring limit cycle, we find scaling behaviors of the bifurcations and the limit cycle, reducing the parameter space crucially. Finally, the model shows very good agreement with real measurements, indicating that the main physics is captured and such simple models are helpful tools for investigating ice clouds. 
\end{abstract}

%%%%%%%%%%%%%%%%%%%%%%%%%%%%%%%%%%%%%%%%%%%%%%%%%%%%%%%%%%%%%%%%%%%%%%%%%%%%%%%%
%%%%%%%%%%%%%%%%%%%%%%%%%%%%%%%%%%%%%%%%%%%%%%%%%%%%%%%%%%%%%%%%%%%%%%%%%%%%%%%%
\section{Introduction}
\label{sec:intro}

Clouds are frequently found to be distributed over the whole globe. A portion of about $60-70\%$ of the Earth is permanently covered by clouds \citep{stubenrauch_etal2017}, which are also distributed over several vertical levels in the troposphere (i.e. up to 12-15 km altitude). Clouds are composed of myriads of water particles of different phases. For low vertical levels in the atmosphere, i.e. at temperatures above the triple point ($T_t=\SI{273.15}{\kelvin}$) clouds contain only liquid water droplets. For the temperature range between the triple point and about $T\sim\SI{235}{\kelvin}$ liquid water can still exist, however the solid phase, i.e. classical hexagonal ice, is the stable phase. At temperatures around $T\sim\SI{235}{\kelvin}$ liquid water freezes spontaneously and below this temperatures, pure liquid water cannot exist anymore \citep{gallo_etal2016}, thus clouds at these temperatures (i.e. at high altitudes close to the tropopause) are completely composed by ice crystals.  Clouds play a major role for the Earth-Atmosphere system. They determine the hydrological cycle by producing precipitation, but also drive atmospheric flows by the huge amount of energy, which is released at phase transitions \citep{gallo_etal2019}. Vigorous systems like thunderstorms are only possible because of this anomaly of water. Finally, clouds also regulate the energy budget of the system itself, controlling the impact of radiation. Clouds interact with incoming solar radiation by scattering and partly reflecting solar radiation; in the presence of clouds less solar energy is transported into the system as compared to the clear sky case, leading to a cooling effect (albedo effect). On the other hand, they partly absorb and re-emit thermal radiation as radiated by the Earth’s surface (as an almost perfect black body), thus keeping more energy in the system as compared to the clear sky case (greenhouse effect). The difference of these two effects leads to the net climate effect of clouds. For low level liquid clouds, the albedo effect is dominant, hence leading to a net cooling of these clouds. For clouds at higher altitudes, i.e. clouds containing ice particles, the net effect is still uncertain; for pure ice clouds we do not even know the sign of the net effect. This results also from the fact that for ice clouds both effects (cooling vs. warming) are of comparable values but with different signs. Thus, small variations in microphysical properties as e.g. size of particles lead to significant changes in the net effect, as demonstrated by \citet{zhang_etal1999}. For high level ice clouds, often a net warming is assumed \citep[e.g.][]{chen_etal2000}, but recent estimations indicate different or even inconclusive outcomes \citep[see Figure 5 in ][]{kraemer_etal2020}. Since for clouds, thermodynamics decides about the mass concentration in first order, the number concentration of ice particles can be directly translated into the size of particles and thus into the radiation calculations; this is crucial, since smaller particles scatter more efficiently than larger ones \citep[e.g.][]{fu_liou1993}. The number concentration of ice particles is determined by the generation mechanisms, i.e. the nucleation pathway. In recent studies \citep{kraemer_etal2016, wernli_etal2016}, a distinction between generally different formation pathways was formulated, termed as  \textit{liquid origin} cirrus and \textit{in situ} cirrus, respectively. The former pathway is characterized by pre-existing liquid cloud droplets at states close to thermodynamic equilibrium (saturation over water), which freeze to ice particles, either triggered by solid aerosols or spontaneously. This pathway is only valid in the temperature range $T>\SI{235}{\kelvin}$, where liquid water can still exist. The latter pathway takes place at low temperature; water vapor is either directly deposited as ice on solid aerosol particles (so-called heterogeneous nucleation), or supercooled aqueous solution droplets freeze spontaneously. The nucleation mechanisms might have a big impact on the resulting number concentration from such a freezing/nucleation event. The resulting number concentration might be limited either by the amount of available ice nucleating particles (particle limited regime) or even by the available water vapor for producing cloud or solution droplets (supersaturation limited regime). The latter one is crucially depending on local dynamics, which leads to a source of supersaturation by adiabatic cooling in upward motions.
Beside the details of nucleation processes, which are barely understood on a molecular level, many other processes in clouds are quite uncertain. Very often, we find that the processes might be understood to some extend on a particle level, but the scaling up to the ensemble is not clear. This points to the fact that in contrast to fluid dynamics where a general and well accepted theory is available (i.e. Navier-Stokes equations or approximations of it), this is not the case for clouds. Actually, there is \textbf{no} generally accepted theory of clouds in terms of a closed system of partial differential equations or something similar. Processes are mostly known on particle basis and reaction rates are formulated often in an ad hoc manner in order to take into account the needs for the respective applications. 

There are only few attempts to formulate a general theory of clouds for some parts of the phase space of clouds, mostly in the field of liquid clouds. For ice clouds, we are aware of some attempts  of deriving a Boltzmann-type evolution equation \citep[][]{khvorostyanov1995}. However, none of them resulted into an overall and generally accepted cloud formulation. For a review of such formulations see \citet{khain_etal2015}.

In this study we focus on pure ice clouds in a low temperature regime ($T<\SI{240}{\kelvin}$), where \textit{in situ} formation takes place at levels in the upper troposphere. For the formation of ice particles at this regime, environmental conditions far away from thermodynamic equilibrium (i.e saturation with respect to ice) are required; even after the formation process, ice clouds remain in a state out of equilibrium, because relaxation towards equilibrium can be quite slow. Thus, ice clouds form a non-equilibrium multiphase system. Ice clouds are also multiscale phenomena, showing structure formation and emergence of pattern, thus not only relevant in terms of microscale processes but also in the propagation of information through all scales. For investigation of such phenomena simple but consistent models are necessary, which contain the main features of the physics, but are simple enough for coupling to fluid dynamics equations. For instance, if we want to investigate issues such as predictability, we have to make sure that the cloud model as coupled to the equations of motion in terms of reaction rates is well-posed. Otherwise, investigations about representing clouds on different scales in numerical models with changing grid resolutions remain a bit arbitrary.

In this study we derive a simple ice cloud model on the basis of a consistent formulation of cloud processes. Here, we include the most important processes, as ice nucleation, growth and evaporation by water vapor diffusion and sedimentation of ice crystals, respectively.  After some simplifications, we end up with a three-dimensional (3D) system of nonlinear autonomous ordinary differential equations, which then can be analyzed by the theory of dynamical systems. There were several former studies, investigating cloud models as dynamical systems, e.g. for liquid and mixed-phase clouds \citep{hauf1993, wacker1992, wacker1995, wacker2006}, but also for aerosol-cloud-systems \citep{koren_feingold2011, feingold_koren2013}. Our investigation is based on the former study by \citet{spreitzer_etal2017}, which evaluated a small part of the phase space of ice clouds, i.e. so-called subvisible cirrus clouds at low vertical updrafts; they used a slightly different model, which could not be analyzed in details. Our approach extends this former study and aims to derive a much more detailed analysis; since our current model is "much simpler", however still realistic, we can determine several properties in an analytical or semi-analytical way, including also the regime treated by \citet{spreitzer_etal2017}. 
We determine equilibrium states and their quality using linearization approaches. In addition, we determine bifurcations and thus periods of limit cycles with numerical evaluations. Finally, the model is compared to real measurements in order to determine its feasibility for the use in extended models.

The study is structured as follows. In Section 2, we present the
physical variables and processes for modeling ice clouds in the low
temperature regime. In Section 3, the model is presented as a 3D
nonlinear system of ordinary differential equations. This system is
characterized in terms of dynamical systems theory in Section 4,
i.e. determining equilibrium states and their quality, and possible
limit cycles. In addition, scaling properties of the system are
investigated. In Section 5, we compare the model results with real
measurements. Finally, we summarize the work, draw some conclusions
and present an outlook for future work.

%%%%%%%%%%%%%%%%%%%%%%%%%%%%%%%%%%%%%%%%%%%%%%%%%%%%%%%%%%%%%%%%%%%%%%%%%%%%%%%%
%%%%%%%%%%%%%%%%%%%%%%%%%%%%%%%%%%%%%%%%%%%%%%%%%%%%%%%%%%%%%%%%%%%%%%%%%%%%%%%%
\clearpage
\section{Physical variables and processes}
\label{sec:physics}

In this section we will define the general settings for the model. We give a general description of clouds as an ensemble of many particles, embedded into atmospheric flows. Nevertheless, later we will simplify the whole system to a 3D dynamical system, describing Lagrangian air parcels transported by prescribed air motion.

%We restrict ourselves to a well-known simplification, i.e. a box or parcel
%model. The general formulation of the model might be extended to
%spatially extended models. However, for a general understanding of the
%relevant processes, we stay with the simple setup of a box or parcel,
%which has a vertical extension of $\Delta z$.

\subsection{Thermodynamic environment}

For modeling ice clouds, we have to set up the thermodynamic
environment. We consider air and water vapor in a control volume $V$
(box or parcel) as ideal gases as described by the ideal gas law:
\begin{equation}
  p_kV=M_kR_kT \Leftrightarrow p_k=\rho_kR_kT,
  ~~\text{with}~\rho_k=\frac{M_k}{V}
\end{equation}
using the partial pressure $p_k$, the specific gas constant
$R_k=\frac{R^*}{M_{\text{mol},k}}$ as derived from the universal gas constant $R^*=\SI{8.314462}{\joule\mol^{-1}\kelvin^{-1}}$, and the temperature $T$,
respectively. The partial masses of the gases are denoted by $M_k$.
The index $a$ indicates dry air, the index $v$ denotes water vapor,
respectively. The total pressure in the volume is determined by
Dalton's law, i.e. $p=p_a+p_v$; however, a good approximation is given
by $p\approx p_a$ (since $p_a\gg p_v$).  The parcel can be seen as one
vertical layer at a certain height within a full atmospheric column.  The
pressure is determined by the hydrostatic balance, i.e. the air mass
in the vertical column
\begin{equation}
  \pdv{p}{z}=-g\rho.
\end{equation}
We consider the low temperature regime in the upper troposphere,
i.e. temperatures in the range
$\SI{190}{\kelvin}\le T\le \SI{240}{\kelvin}$. The corresponding air
pressure values are considered to be within the range
$\SI{200}{\hecto\pascal}\le p\le \SI{300}{\hecto\pascal}$.  In
addition to water vapor, we also consider the mass of the
condensed water phase, i.e. the total mass of ice particles $M_i$. At these
low temperatures, liquid water does not exist in the 
atmosphere~\citep[e.g.][]{gallo_etal2016}, thus we can restrict the
system to gaseous and solid phases of water. For the further
investigations, we use the mass concentration of water vapor
$q_v=\frac{M_v}{M_a}$ and the mass concentration of ice
$q_i=\frac{M_i}{M_a}$, respectively. In fact, the total mass $M=M_a+M_v+M_i$ can be approximated by the mass of air, since the relative contribution of water substances is less than $10^{-3}$, thus $M\approx M_a$.

\subsection{Phase changes and saturation ratio}

We are only interested in the formation of ice crystals
within the respective low temperature region, i.e. the formation of
ice ``directly'' from water vapor without any transition through the
liquid phase. Recently, formation pathways of this type were grouped
together and named as ``\textit{in situ} formation''
\citep[e.g.,][]{kraemer_etal2016,wernli_etal2016}. From these
different pathways we only consider the pathway of homogeneous
freezing of aqueous solution droplets \citep[short ``homogeneous
nucleation'', see, e.g.,][]{koop_etal2000,spichtinger_etal2023}.
At these atmospheric conditions, hexagonal ice ($I_h$) is the stable solid
phase of water \citep{gallo_etal2016}. For assuming water vapor as an ideal gas, the thermodynamic equilibrium of these water phases is
approximately determined by the Clausius-Clapeyron
equation~\citep[e.g.][]{kondepudi_prigogine2014}
\begin{equation}
  \label{eq:clausius-clapeyron}
  \dv{p_\text{si}}{T}=\frac{L\,p_\text{si}}{R_vT^2}
\end{equation}
using the specific latent heat of sublimation $L$, which is slightly
dependent on temperature \citep[see, e.g., ][]{murphy_koop2005};
$p_{\text{si}}$ denotes the saturation vapor pressure over hexagonal
ice (see also appendix~\ref{appA_functions}).  The thermodynamical control variable for determining ice cloud
processes is the saturation ratio over ice
\begin{equation}
  S_i:=\frac{p_v}{p_\text{si}(T)}=\frac{p\,q_v}{\epsilon p_\text{si}(T)}
\end{equation}
using the water vapor mixing ratio $q_v$, and the ratio of molar
masses of water and air
$\epsilon=\frac{M_{\text{mol},v}}{M_{\text{mol},a}}\approx 0.622$.

We will often consider changes in the saturation ratio, which are
governed by changes in temperature, pressure and specific humidity,
respectively. For this purpose, we investigate the total derivative of
the saturation ratio, as given by:
\begin{equation}
  \dv{S_i}{t}=
  \pdv{S_i}{T}\dv{T}{t}+\pdv{S_i}{p}\dv{p}{t}+\pdv{S_i}{q_v}\dv{q_v}{t}
\end{equation}
using the partial derivatives
\begin{equation}
   \pdv{S_i}{T}  = -S_i\frac{L}{R_vT^2},~~
  \pdv{S_i}{p}  =  \frac{S_i}{p},~~
  \pdv{S_i}{q_v}  =  \frac{S_i}{q_v}=\frac{p}{\epsilon p_\text{si}}
\end{equation}
where we have used the Clausius-Clapeyron
equation~\eqref{eq:clausius-clapeyron}. 

We can split the changes in temperature into adiabatic and diabatic
components, i.e. temperature changes driven by adiabatic expansion due
to vertical upward motions and temperature changes induced by latent
heat release during phase changes. These temperature rates can be
represented as follows:
\begin{eqnarray}
  \eval{\dv{T}{t}}_\text{adiabatic} & = &
                                        \dv{T}{z}\dv{z}{t}=-\frac{g}{c_p}w\\ 
  \eval{\dv{T}{t}}_\text{diabatic} & = & -\frac{L}{c_p}\dv{q_v}{t}.
\end{eqnarray}
Note, that we omit other diabatic processes, as e.g. friction or
radiation processes. For our investigation, this restriction is still
meaningful. 
Using these expressions, we can determine the change in $S_i$ due to
temperature changes:
\begin{eqnarray}
  \pdv{S_i}{T}\dv{T}{t}
  &=&-S_i\frac{L}{R_vT^2}
      \left(
      \eval{\dv{T}{t}}_\text{adiabatic}+
      \eval{\dv{T}{t}}_\text{diabatic}
      %\dv{T}{t}\Big\vert_\text{adiabatic}
      %+\dv{T}{t}\Big\vert_\text{diabatic} 
      \right)\\
  &=&\frac{L\,g}{c_pR_vT^2}S_iw
      +\frac{L^2}{c_pR_vT^2}S_i\dv{q_v}{t}
\end{eqnarray}
The adiabatic pressure change, assuming hydrostatic balance, 
\begin{equation}
  \dv{p}{t}=\dv{p}{z}\dv{z}{t}=-g\rho w
\end{equation}
also influences the saturation ratio, thus
\begin{equation}
  \pdv{S_i}{p}\dv{p}{t}=-\frac{S_i}{p}g\rho w=-S_i\frac{g}{R_aT}w
\end{equation}
and changes in the water vapor concentration $q_v$ finally give the
rate 
\begin{equation}
  \pdv{S_i}{q_v}\dv{q_v}{t}=\frac{p}{\epsilon p_\text{si}}\dv{q_v}{t}
\end{equation}
Putting all rates together and sorting leads to the final rate for the
saturation over ice:
\begin{equation}
  \label{eq:si_evolution}
  \dv{S_i}{t}=
  \left[
    \frac{L}{c_pR_vT^2}-\frac{1}{R_aT}
  \right]gwS_i+
  \left[
    \frac{L^2}{c_pR_vT^2}S_i+\frac{p}{\epsilon p_\text{si}}
  \right]\dot{q}_v.
\end{equation}

In our investigations we are often interested in asymptotic states of
the ice cloud model, i.e. the long term evolution. For these
investigations, long time evolution (and integration) is necessary. In
order to approach these states analytically, we make the following
simplification. In Equation~\eqref{eq:si_evolution} the changes of
temperature and pressure are represented by the vertical upward motion
$w$. For the further investigation, we assume that
temperature/pressure rates just appear in the evolution equation of
$S_i$, but we skip evolution equations for $p,T$, i.e. we assume that
both quantities remain constant. As a consequence to be consistent, we
also skip the term in Equation~\eqref{eq:si_evolution} reflecting the
latent heat release due to phase changes, i.e. we use the
approximation:
\begin{eqnarray}
   \dv{S_i}{t}&=&
  \left[
    \frac{L}{c_pR_vT^2}-\frac{1}{R_aT}
  \right]gwS_i+
  \left[
    \frac{L^2}{c_pR_vT^2}S_i+\frac{p}{\epsilon p_\text{si}}
  \right]\dot{q}_v\\
  & \approx & 
              \underbrace{
              \left[
              \frac{L}{c_pR_vT^2}-\frac{1}{R_aT}
              \right]gwS_i   
              }_{\text{Cool}}+
              \underbrace{
              \left[
              \frac{p}{\epsilon p_\text{si}}
              \right]\dot{q}_v    
              }_{\text{Dep}_s}\\
  & = &\underbrace{A_sS_i}_{\text{Cool}}+
        \underbrace{B_s\dot{q}_v}_{\text{Dep}_s} 
\end{eqnarray}
introducing two terms $\text{Cool}$ and $\text{Dep}_s$, respectively,
governing the time evolution of the saturation ratio. These terms will also appear in the complete model. We will later use the notation, setting $s\coloneqq S_i$.
Note here that
the first term constitutes a source for supersaturation, if the
vertical velocity $w>0$, i.e. if there is an upward motion. The
coefficient $A_s$ depends on temperature $T$, and linearly on vertical
velocity $w$, whereas the coefficient $B_s$ depends on temperature and
pressure, respectively. 
This kind of simplification neglecting changes of temperature and
pressure was already used in former investigations of subvisible ice
clouds \citep{spreitzer_etal2017}.

\subsection{General assumptions for ice crystals}
\label{sec:general_assumptions_particle}

Ice crystals usually  have quite irregular shapes with inherent
hexagonal structures due to the stable solid phase (hexagonal ice
$I_h$). The exact and detailed shape and its evolution in time depend
crucially on the environmental conditions \citep[e.g.][]{libbrecht2005}.

% The shape itself influences
% the diffusional growth of the particle (see
% sec.~\ref{sec:processes}); using the electrostatic analogon
% \citep[see, e.g.,][]{spichtinger_gierens2009a, lamb_verlinde2011,
%   baumgartner_spichtinger2018}, a capacity coefficient which
% represents the shape of a single particle can be introduced to the
% growth equation.

Since there is a huge variety of shapes for ice
crystals \citep[see, e.g.,][]{bailey_hallett2009,grulich_etal2021} and
we cannot represent this variability within a simple model, we have to
choose an approximation. There are two different reasons for choosing
a spherical shape for ice crystals in our model. First, we are
interested in the low temperature regime; thus, the available amount
of water vapor is limited, so ice crystals stay quite small and are
almost spherical; from measurements we know that the maximum diameter
should not exceed values of $200-\SI{300}{\micro\metre}$
\citep{kraemer_etal2020}. The complex shapes of ice crystal usually
appear for the warm temperature regime (i.e. $T>\SI{240}{\kelvin}$) together with much larger sizes. Second, from theoretical and
computational investigations
\citep[e.g.][]{baumgartner_spichtinger2017} we know that in a certain
distance to a non-spherical crystal, the resulting water vapor field
is almost the same as for a larger spherical ice crystal. Thus, we use
a spherical shape for ice crystals, with the common relationship
between mass and radius
\begin{equation}
  \label{eq:mass_radius_sphere}
  m=\frac{4}{3}\pi\rho_b r^3
  ~\Leftrightarrow~
  r=\sqrt[3]{\frac{3}{4\pi\rho_b}}m^\frac{1}{3}=c\,m^\frac{1}{3},
  ~\text{with}~\rho_b=\SI{810}{\kilo\gram\,\metre^{-3}} \text{~bulk
    density of ice}. 
\end{equation}
The radius will appear later in the growth equation, when we specify a
spherical shape.

\subsection{Relevant processes for ice cloud modeling}
\label{sec:processes}

\begin{enumerate}
\item Particle formation - ice nucleation

  For ice formation at low temperatures, two nucleation processes
  might be relevant, i.e. (1) homogeneous freezing of aqueous solution
  droplets \citep[see, e.g.,][]{koop_etal2000}, and (2) heterogeneous
  nucleation at solid aerosol particles.  In the first case, we assume
  that supercooled solution droplets (e.g. containing a small amount
  of sulfuric acid or other inorganic substances) freeze spontaneously
  (i.e. stochastically) to small ice crystals. The probability
  of a solution droplet of volume $V$ freezing within a time interval
  $\Delta t$ can be calculated as
  \begin{equation}
    P=1-\exp(-JV\Delta t)
  \end{equation}
  with the nucleation rate $J$. For such solution droplets this rate
  can be formulated in terms of water activity \citep{koop_etal2000},
  which translates into the form
  \begin{equation}
    J(T,S_i)=J_0\exp(A(T)(s-S_c(T)))
  \end{equation}
  with a steepness $A(T)$ and a threshold $S_c(T)$,
  respectively. These two functions can be approximated by a recent
  formulation \citep{spichtinger_etal2023} to
  \begin{equation}
  \label{eq:approximation_nucl}
    A(T)\approx p_1(T)=a_{w0}+a_{w1}T,
    ~~S_c(T)\approx p_2(T)=a_{s0}+a_{s1}T+a_{s2}T^2
  \end{equation}
  using polynomials $p_n(x)$ of degree $n\in\mathbb{N}$.  There is
  still a debate about the relevance or dominance of these different
  nucleation pathways. However, for our purposes we can restrict the
  formulation to homogeneous freezing of solution droplets, since in
  the cold temperature regime of \textit{in situ} formed ice clouds
  \citep[cf.][]{kraemer_etal2016}, homogeneous nucleation produces
  most of the ice particles. Thus, we include a supersaturation limited pathway of ice nucleation.
  
\item Particle growth/evaporation by diffusion of water vapor

  Small water particles (liquid or solid) grow by diffusion of
  surrounding water vapor. The boundary condition at the surface of the particle (i.e. saturation with respect to the phase of the particle)
  determines the direction of the water vapor flux, leading either to
  evaporation or growth of the particle. The mathematical details of
  the problem and a proof of uniqueness of the solution can be found
  in \citet{baumgartner_spichtinger2018}. The growth equation of a
  single water particle with mass $m$ can be written as
  \begin{equation}
    \label{eq:growth_particle}
    \dv{m}{t}=4\pi C D_vf_{k} G_v(s-1)f_v.
  \end{equation}  
  The shape of the particle influences the diffusional growth; using
  the electrostatic analog \citep[see,
  e.g.,][]{spichtinger_gierens2009a, lamb_verlinde2011,
    baumgartner_spichtinger2018}, a capacity coefficient $C$
  representing the shape of a single particle can be introduced to the
  growth equation. $D_v$ denotes the diffusion coefficient of water
  vapor in air,
  \begin{equation}
    D_v=D_{v0}\qty(\frac{T}{T_0})^2\qty(\frac{p_0}{p}),~
    D_{v0}=\SI{2.1422e-5}{\metre^2\per\second},~
    T_0=\SI{273.15}{\kelvin},~
    p_0=\SI{101325}{\pascal}
  \end{equation}
  with a correction $f_{k}$ of the flux in the kinetic regime; this correction would reduce the uptake of water vapor, since $f_k\le 1$. The fit
  of $D_v$ is based on the description by
  \citet{hall_pruppacher1976}. The effect of latent heat release is
  captured by the Howell factor
  \begin{equation}
    \label{eq:howell}
    G_v=\left[
      \left(\frac{L}{R_vT}-1\right)\frac{L}{T}\frac{D_v}{K_T} +
      \frac{R_vT}{p_{si}}
    \right]^{-1},
  \end{equation}
  with the latent heat of sublimation $L$, the ideal gas constant of
  water vapor $R_v$, and the heat conduction of air $K_T$
  \citep[cf.][]{dixon2007}, respectively. Note, that in this
  formulation we already omit possible corrections for small particles for $D_v,K_T$, respectively
  \citep[see, e.g.,][]{lamb_verlinde2011}. For large particles, a
  correction $f_v$ for the ventilation effect (wakes behind falling
  particles enhance water vapor diffusion) is applied; this correction would enhance the uptake of water vapor, since $f_v\ge 1$.

\item Sedimentation - a particle falls out due to gravity

  Water particles in general, and also ice crystals are accelerated by
  gravity; since the friction of air leads to a drag and thus a balance
  of forces, ice crystals very quickly reach a so-called terminal
  velocity without any further acceleration. For low Reynolds numbers
  using the Stokes law \citep[see][pp. 381]{lamb_verlinde2011} and
  assuming a spherical shape, the terminal velocity is proportional to
  the squared radius.  For irregularly shaped particles as ice
  crystals, the terminal velocity can be approximated by power laws
  $v_t(m)=\alpha_k m^{\beta_k}$ with piecewise constant coefficients
  $\alpha_k,\beta_k$ \citep[see][]{spichtinger_gierens2009a}. We have to apply a correction factor~\citep[see,
  e.g.,][]{spichtinger_gierens2009a, book_pruppacher_klett2010},
  taking into account the friction of the air at different densities,
  expressed as a function of temperature and pressure:
  \begin{equation}
    c(p,T)=\left(\frac{p}{p_c}\right)^{a_c}
    \left(\frac{T}{T_c}\right)^{b_c}, ~~
    p_c=\SI{30000}{\pascal}, ~~ T_c=\SI{233}{\kelvin}, ~~
    a_c=-0.178, ~~ b_c=-0.394
  \end{equation}
  
\item Collision of particles

  Collision of particles is usually the major process for growing
  particles to large sizes. For liquid particles, collision is very efficient, but for ice particles in the cold temperature regime
  aggregation is not a relevant process \citep{kajikawa_heymsfield1989, kienast_etal2013}. 
\end{enumerate}

\subsection{General assumptions for cloud ensembles}
\label{sec:general_assumptions_emsemble}

% \begin{itemize}
% \item Boltzmann approach
% \item general moments -> equations for moments
% \item closure for ice clouds
% \end{itemize}

Clouds are usually composed by myriads of water particles, thus we
have to consider a cloud as a statistical ensemble. As described in
details by \citet{spreitzer_etal2017}, a generic way for describing
the cloud ensemble would be to use a probability distribution
$f(m,x,t)$ for the masses of cloud particles and investigate the
evolution of the distribution in time and space. This would lead to a
Boltzmann type evolution equation for the distribution, i.e.
\begin{equation}
  \pdv{(\rho f)}{t}+\nabla_x\vdot\qty(\rho f u)
  +\pdv{(\rho f g)}{m}=\rho h
\end{equation}
with transport with velocity $u$, growth with a rate function $g$, and
other sources and sinks, described by $h$.
From this equation we can derive
(infinitely many) equations for the general moments
\begin{equation}
  \mu_k[m]\coloneqq\int_0^\infty m^k f(m,x,t)\,\dd m,~~k\in\mathbb{N}_0,
\end{equation}
 of the distribution if the integral converges.
The general formulation of moment equations as derived from the
Boltzmann approach is presented in Appendix~\ref{appA}.\\
%Solving the equations for all $k\in\mathbb{N}_0$ is equivalent to
%solving the Boltzmann equation for the mass distribution. Actually,
Since we can neither solve infinitely many
equations, nor do we have a general and consistent description of the
sources and sink term $h$ on the right hand side, we make
the following assumptions for simplification:
\begin{itemize}
\item We only use two equations, i.e. for the moments $k=0,1$ and use a
  closure condition. Usually, the distribution is normalized by the
  cloud particle number concentration~$n$, thus we obtain equations
  for the number concentration $\mu_0=n$ and the mass concentration
  $\mu_1=q$. For ice crystals in the low temperature regime
  ($T<\SI{240}{\kelvin}$) we assume a fixed type of distribution,
  i.e. a lognormal distribution
  \begin{equation}
    \label{eq:lognormal_mass}
    f(m)=\frac{n}{\sqrt{2\pi}\log(\sigma_m)m}
    \exp(-\frac{1}{2}\qty(\frac{\log(\frac{m}{m_m})}{\log(\sigma_m)})^2)
  \end{equation}
  with the modal mass $m_m$ and the geometric standard deviation
  $\sigma_m$.  Thus, the general moments can be calculated
  analytically as
  \begin{equation}
    \mu_k[m]=%\int_0^\infty f(m)m^k \dd m =
    nm_m^k\exp(\frac{1}{2}\qty(k\log\sigma_m)^2).
  \end{equation}
  The closure is determined by a fixed ratio of moments $r_0$, i.e. 
  \begin{equation}
    r_0=\frac{\mu_2\mu_0}{\mu_1^2}=\exp(\qty(\log(\sigma_m))^2)
  \end{equation}
  which can be used for calculating the moments
  \begin{equation}
    \label{eq:general_moments_lognormal}
    \mu_k[m]= n m_m^kr_0^{\frac{k^2}{2}}=
    n \bar{m}^k
    r_0^{\frac{k(k-1)}{2}}
  \end{equation}
  with the mean mass $\bar{m}\coloneqq\frac{q}{n}$.
  We set $r_0=$ const. and thus the geometric standard deviation
  $\sigma_m$ is constant. Motivated by former investigations
  \citep[][]{spichtinger_gierens2009a} we set $r_0=3$. 
\item For the right hand side, i.e. the formation and annihilation
  processes of ice particles, we will use meaningful
  parameterizations, which are partly derived from the moment approach (see also Appendix~\ref{appA}).
\end{itemize}
Thus, we will formulate our model using the mean variables number
concentration of ice particles $n=\mu_0$ and mass concentration of
ice particles $q=\mu_1$, respectively.
% In the next section, we
% will investigate the different relevant processes and their
% mathematical description.

% \subsection{Derivation of ensemble process rates}
% \label{sec:ensemble_processes}

\subsection{Basic structure of the model system}
\label{sec:ODE_System_structure}

We now determine the general structure of the resulting model system,
which then turns into a system of ordinary differential equations (ODEs). The process of nucleation affects
number and mass concentrations as a source of the system, thus we
formally introduce the terms $\text{Nuc}_n,\text{Nuc}_q$ for the two
variables respectively. Growth and evaporation act in different ways
on the variables $n$ and $q$. For a supersaturated regime ($s>1$) the ice
particles can grow by diffusion, thus we need a deposition term $\text{Dep}_q$; the number concentration is not affected by
growth. For evaporation, both mass and number concentrations are
depleted; however, the effect is treated differently. From the general
derivation of the growth term for ensembles (see
Appendix~\ref{appA_growth}), we see that evaporation of mass is
symmetric to growth, thus the term $\text{Dep}_q$ already treats the
loss of mass correctly. For the effect of evaporation on the number concentration,
we introduce the term $\text{Evap}_n$. In fact,
it is quite tricky to introduce a meaningful ansatz for this term,
since the type of the underlying distribution is changing by
evaporation. However, in the following investigations, we will
concentrate on the supersaturated regime, thus we do not need to
investigate the evaporation term in details for this study.
Finally, we have to consider
sedimentation of particles and introduce the sedimentation terms $\text{Sed}_n$ and $\text{Sed}_q$ for the change rates in 
number and mass concentration due to sedimentation. We will use an
approximation in order to avoid a vertical transport equation (i.e. a system of
partial differential equations). This approximation will be introduced
later based on a vertical
discretization. \\
The \textbf{basic structure of the ice cloud model} 
%for the ice cloud model 
is then as follows
\begin{align}
  \label{eq: structure_ni} 
  \dv{n}{t} &\ =\  \text{Nuc}_n\  +\ \text{Evap}_n\  +\ \text{Sed}_n \\
  \label{eq: structure_qi}
  \dv{q}{t} &\ =\  \text{Nuc}_q\  +\ \text{Dep}_q\  ~~ +\ \text{Sed}_q \\
  \label{eq: structure_Si}
  \dv{s}{t} &\ =\  \text{Cool}\   +\ \text{Dep}_s,  
\end{align}
with total derivatives $\dv{\psi}{t}$.
We will see in the following section that the system (after some further
simplifications) becomes an autonomous nonlinear ODE system.

%\textbf{---- old stuff -- end}

%%%%%%%%%%%%%%%%%%%%%%%%%%%%%%%%%%%%%%%%%%%%%%%%%%%%%%%%%%%%%%%%%%%%%%%%%%%%%%%%
%%%%%%%%%%%%%%%%%%%%%%%%%%%%%%%%%%%%%%%%%%%%%%%%%%%%%%%%%%%%%%%%%%%%%%%%%%%%%%%%
\clearpage
\section{Simple model}
\label{sec:simple_model}

In this section, we finally derive the model, which will be used for
analytical and numerical considerations.

\subsection{Simplifications}
\label{sec:simplifications}

% \begin{itemize}
% \item nucleation -> no change in aerosol distribution
% \item growth -> no corrections
% \item sedimentation -> terminal velocity as simple power law
% \end{itemize}

For nucleation of particles, we prescribe a size distribution of
solution droplets. Instead of tracking aerosol properties with water
uptake controlled by Köhler theory \citep{koehler1936}, we use a lognormal distribution
with fixed modal radius and geometric standard deviation, normalized
by the number concentration of aerosols. Details are given in
Appendix~\ref{appA_nucleation}. This leads to a nucleation term $\text{Nuc}_n$ for
the ice crystal number concentration as given below. In addition, we assume a constant nucleation mass $m_{\text{nuc}}=\SI{e-16}{\kilo\gram}$ for each newly nucleated ice crystal, which
gives the nucleation term $\text{Nuc}_q$ for the mass concentration. In summary, we
obtain
\begin{eqnarray}
  \text{Nuc}_n  =  n_a\, V_{sol}\, J_0 \exp(p_{1e}(s-p_2)) 
               & = & A_n \exp(p_{1e}(s-p_2))\\
  \text{Nuc}_q =  m_{\text{nuc}}\,n_a\, V_{sol}\, J_0 \exp(p_{1e}(s-p_2)) 
               & = & A_q \exp(p_{1e}(s-p_2))
\end{eqnarray}
using two polynomials $p_{1e},p_2$ depending on temperature solely
(see Section~\ref{sec:processes} and Appendix~\ref{appA_functions}). 
Note, that the mass of the newly
nucleated ice particles is obtained from the surrounding water vapor
(since the solution droplets deplete some water vapor), thus we
introduce a sink term for the saturation ratio as
\begin{equation}
  - B_{s1}\exp(p_{1e}(s-p_2)),~~\text{with}
  ~~B_{s1}=A_q\frac{p}{\epsilon p_{\text{si}}}.
\end{equation}
We use a spherical shape of the ice particles, which in turn affects
the description of the growth and sedimentation, respectively. For the
representation of the growth term, we apply additional
simplifications:
\begin{itemize}
\item We neglect the kinetic correction, which would apply for very
  small ice particles, and especially affect nucleation events in
  clear air at very
  high velocities and/or low temperature regimes \citep[see discussion
  in][]{spichtinger_gierens2009a}. However, since we are interested in
  the qualitative behavior of the model, and especially in the long
  term behavior (where cloud particles pre-exist), this correction is
  not very relevant; thus, we set $f_k=1$
\item We neglect the ventilation correction, which is quite important
  for large particles. However, in our low temperature regime with
  \textit{in situ} formation process of homogeneous nucleation, ice
  crystals do not grow to such large sizes. Thus, we set $f_v=1$.
\item For a spherical particle, we obtain the radius as capacity
  factor $C=r$, which can be expressed by the cubic root of the mean
  mass. 
\end{itemize}
Thus, we can express the growth term for the mass concentration (see
also Appendix~\ref{appA_growth}) as
\begin{equation}
  \text{Dep}_q=B_qn^{\frac{2}{3}}q^\frac{1}{3} (s-1).
\end{equation}
The growth term models the transfer of water vapor to ice
particles, thus we have to introduce a sink term for the saturation
ratio, i.e.
\begin{equation}
  -B_{s2}n^{\frac{2}{3}}q^{\frac{1}{3}}(s-1),~~\text{with}
  ~~B_{s2}=B_q\frac{p}{\epsilon p_{\text{si}}};
\end{equation}
in combination with the nucleation term above, we can set
\begin{equation}
  \text{Dep}_s=
  - B_{s1}\exp(p_{1e}(s-p_2))
  -B_{s2}n^{\frac{2}{3}}q^{\frac{1}{3}}(s-1).
\end{equation}
For the evaporation term of the number concentration, we make the ad
hoc ansatz
\begin{equation}
  \text{Evap}_n= \frac{n}{q}\text{Dep}_qH(1-s)
\end{equation}
with the assumption that the number concentration of ice crystals reduces proportionally to the
loss of mass concentration. This only happens for
subsaturation, thus a Heaviside function $H$ is used for switching off
the process at $s>1$.

The terminal velocity of the ice particles as formulated in
\citet{spichtinger_gierens2009a} by piecewise power laws can be
approximated by a simple approach
\begin{equation}
  v_t(m)\approx
  v_{\text{final}}\,a_s\,m^{b_s}\frac{1}{v_{\text{final}}+a_s\,m^{b_s}},
\end{equation}
mimicking the major increase of the velocity. However, even this
approach is too sophisticated, since in our regime, we do not find
such large particles approaching the final value 
\hbox{$v_{\text{final}}\approx \SI{2.4641}{\metre\per\second}$},
thus a simple power law
\begin{equation}
  v_t(m)\approx\,a_s\,m^{b_s}
\end{equation}
works well. Since we consider a spherical shape of ice crystals, we
find a meaningful fit
\begin{equation}
  \label{eq:vt_approx}
  v_t(m)=a_sm^\frac{2}{3},
  a_s=\SI{6.e5}{\metre\,\second^{-1}\kilo\gram^{-\frac{2}{3}}},
\end{equation}
which is consistent with the expressions for falling spheres \citep[velocity
proportional to the squared radius as in][]{lamb_verlinde2011}.
The sedimentation is a vertical transport in a column. However, in our
simple model we concentrate on a single level and restrict ourselves to a box of a
certain vertical extension $\Delta z$. Then we can approximate the
change of the flux $j_\psi=\rho v_\psi \psi$ (for a quantity
$\psi=n,q$) by the finite difference
\begin{equation}
  \pdv{j_\psi}{z}\approx
  \frac{j_{\psi}^{\text{top}}-j_{\psi}^{\text{bottom}}}{\Delta z}
\end{equation}
with the flux $j_{\psi}^{\text{top}}$ through the top boundary into
the box and the flux $j_{\psi}^{\text{bottom}}$ through the bottom out
of the box, respectively. Generally, we can only determine the
outgoing flux, but we might add constraints for determining the
incoming flux; otherwise, the incoming flux can be seen as a variable
parameter. Thus, the sedimentation terms can be formulated as
\begin{equation}
  \text{Sed}_n=F_n-\frac{1}{\Delta z}v_nn,~~
  \text{Sed}_q=F_q-\frac{1}{\Delta z}v_qq
\end{equation}
with fluxes from above $F_n,F_q$. 
Using the approximation of the terminal velocity
\eqref{eq:vt_approx}, and the formulation of general moments of the
lognormal distribution \eqref{eq:general_moments_lognormal} we can
derive the averaged terminal velocities
\begin{eqnarray}
  v_n=\frac{1}{n}\int_0^\infty f(m)v_t(m)\dd m
  &=&
      \frac{1}{n}\qty(a_sr_0^{-\frac{1}{9}}n^\frac{1}{3}q^\frac{2}{3})\\
  v_q=\frac{1}{q}\int_0^\infty f(m)mv_t(m)\dd m
  &=&
      \frac{1}{q}\qty(a_sr_0^{\frac{5}{9}}n^{-\frac{2}{3}}q^\frac{5}{3})
\end{eqnarray}
and thus the sedimentation terms
\begin{eqnarray}
  \text{Sed}_n=F_n-\frac{1}{\Delta z}v_nn
  &=& F_n - C_nn^\frac{1}{3}q^\frac{2}{3},
      ~~C_n=\frac{a_sr_0^{-\frac{1}{9}}}{\Delta z}\\
  \text{Sed}_q=F_q-\frac{1}{\Delta z}v_qq
  &=& F_q - C_qn^{-\frac{2}{3}}q^\frac{5}{3},
      ~~C_q=\frac{a_sr_0^{\frac{5}{9}}}{\Delta z}  
\end{eqnarray}
The fluxes from above, $F_n$ and $F_q$, appear as additional variables (or
parameters). Since we would like to avoid too many parameters for our
investigations, we adopt a strategy by \citet[][]{spichtinger_cziczo2010}:
We assume that the sedimentation flux into the box/layer from above through the top boundary is a fixed fraction $f_{\text{sed}}$ of the sedimentation flux through the bottom  out of
the box/layer, i.e. we
can set \begin{equation}
  j_{\psi}^{\text{top}}= f_{\text{sed}}\cdot j_{\psi}^{\text{bottom}}
\end{equation}
with a constant $0\le f_{\text{sed}}\le 1$. The extreme values
indicate either no flux into the box from above ($f_{\text{sed}}=0$) or no net flux
out of the box ($f_{\text{sed}}=1$), i.e. the flux into the box from above and the flux out of the box are identical in magnitude. Thus, we find
\begin{equation}
  F_n=f_{\text{sed}}C_nn^\frac{1}{3}q^\frac{2}{3},
  F_q=f_{\text{sed}}C_qn^{-\frac{2}{3}}q^\frac{5}{3}
\end{equation}
and set $F\coloneqq 1-f_{\text{sed}}$ for computational convenience later on with $0<F\le 1$. We end with the final formulation of
the sedimentation terms
\begin{eqnarray}
  \text{Sed}_n&=&-FC_nn^\frac{1}{3}q^\frac{2}{3}\\
  \text{Sed}_q&=&-FC_qn^{-\frac{2}{3}}q^\frac{5}{3}.
\end{eqnarray}
In our investigations, we set the vertical extension to $\Delta
z=\SI{100}{\metre}$, which is a typical extension of ice cloud layers 
\citep[see discussion in][]{spichtinger_cziczo2010, kay_etal2006}.

\subsection{Final model for investigations}
After applying all the simplifications formulated above in
%sections~\ref{sec:general_assumptions_particle} 
Sections~\ref{sec:physics}
and
\ref{sec:simplifications}, we obtain the final model for our
investigations as a 3D system of ODEs with ice crystal number concentration $n$, ice crystal mass
concentration $q$, and saturation ratio over ice $s$ as variables:
\begin{eqnarray}
  \label{eq:3D_system_n}
  \dv{n}{t} &=& A_n\exp(p_{1e}(s-p_2))
                  + B_q
                  n^{\frac{5}{3}}q^{-\frac{2}{3}}H(1-s)(s-1)
                  -FC_nn^{\frac{1}{3}}q^{\frac{2}{3}}\\\
  \label{eq:3D_system_q}
  \dv{q}{t} &=& A_q\exp(p_{1e}(s-p_2))
                  + B_qn^{\frac{2}{3}}q^{\frac{1}{3}}(s-1)
                  -FC_qn^{-\frac{2}{3}}q^{\frac{5}{3}}\\
  \label{eq:3D_system_s}
  \dv{s}{t} &=& A_ss
                  - B_{s1}\exp(p_{1e}(s-p_2))
                  -B_{s2}n^{\frac{2}{3}}q^{\frac{1}{3}}(s-1)
\end{eqnarray}
This is a nonlinear autonomous ODE system, depending on the
environmental parameters pressure $p$, temperature $T$, vertical
velocity $w$, and sedimentation parameter $F$, respectively. The
ranges for the parameters are given in Table~\ref{tab:parameters}, which are representative for the environmental conditions in the upper troposphere
\begin{table}[tb!]
  \centering
  \begin{tabular}{c||c|c}
    Parameter & Dimension & Range\\\hline\hline
    Pressure $p$ & hPa & $200\le p\le 300$\\\hline
    Temperature $T$ & K & $190\le T\le 240$\\\hline
    Vertical velocity $w$ & m/s & $0.0005\le w\le 2$\\\hline
    Sedimentation parameter $F$ &1& $0.01 \le F\le 1$
  \end{tabular}
  \caption{Ranges of relevant parameters for the 3D ice cloud model
    \eqref{eq:3D_system_n}-\eqref{eq:3D_system_s}.}
  \label{tab:parameters}
\end{table}
Using the notation $x=\qty(n,q,s)^T$ we could write the system with a
suitable 3D vector field $V$ as an autonomous ODE system
$\dot{x}=V(x)$. Note, that this system is not well-defined for initial
conditions without a cloud, i.e. for $x_0=(0,0,s_0)$; the terms on the
right hand side include singularities of the form $n^{-\alpha}$ or
$q^{-\beta}$ (with $\alpha,\beta>0$). For the supersaturation regime
($s>1$) a regularization of the sedimentation term
\begin{equation}
  -FC_qn^{-\frac{2}{3}}q^{\frac{5}{3}}\approx
  -FC_q\qty(n+n_{\text{small}})^{-\frac{2}{3}}q^{\frac{5}{3}}
\end{equation}
might be applied. Nevertheless, even if the singularities are removed,
the right hand side is not Lipschitz continuous for initial values
$n=0$ or $q=0$ because of terms $n^\alpha q^\beta$ with $0\le\alpha<1, 0\le\beta<1$. Thus, for initial value problems starting at such
states the standard theory \citep[general theorem for existence and
uniqueness, see, e.g.,][p.68]{book_walter_1998} cannot be applied. For
other starting points, existence and uniqueness of solutions is
guaranteed, since the right hand side is smooth; this is also true for the system without a regularization.
We want to note here, that even for the initial values $n=0$ or $q=0$,
uniqueness might be obtained, following the arguments by
\citet{hanke_porz2020} to produce a mathematically unique solution with additional physical conditions. However, for our investigations these details
might be not relevant, since we are mainly interested in the long time
behavior, i.e. the qualitative properties of the system; then we can assume that $n,q$ are positive.

%%%%%%%%%%%%%%%%%%%%%%%%%%%%%%%%%%%%%%%%%%%%%%%%%%%%%%%%%%%%%%%%%%%%%%%%%%%%%%%%
%%%%%%%%%%%%%%%%%%%%%%%%%%%%%%%%%%%%%%%%%%%%%%%%%%%%%%%%%%%%%%%%%%%%%%%%%%%%%%%%
\clearpage
\section{Qualitative properties of the model}
\label{sec:quality}

In this section, we investigate the quality of the system in
terms of dynamical systems theory.

\subsection{Equilibrium states}
\label{sec:equilibrium_states}
We start with the determination of the equilibrium states of the
system. Equilibrium points (stationary points/fixed points) of an ODE
system are states for which the right hand side of the ODE system
vanishes, i.e. $\dot{n}=\dot{q}=\dot{s}=0$, 
so we have to determine the roots of the 3D vector field $V(x)$.
In order to (approximately) calculate the equilibrium points, we make
two more simplifying assumptions. First, since we are interested in
the supersaturated regime (i.e. $s>1$), we can omit the evaporation
term
\begin{equation}
  \text{Evap}_n = B_q n^{\frac{5}{3}}q^{-\frac{2}{3}}H(1-s)(s-1)
\end{equation}
from Equation~\eqref{eq:3D_system_n}. Since we assume $w>0$, this is
not an approximation; the supersaturation source $A_ss$ will
always drive the system into states with $s>1$, and thus
$H(1-s)=0$. Second, we estimate the contribution of the nucleation
rate to the mass concentration $q$. This term is usually very small,
thus we can also neglect the term
\begin{equation}
  \text{Nuc}_q = A_q\exp(p_{1e}(s-p_2))\approx 0
\end{equation}
in Equation~\eqref{eq:3D_system_q}; the same argument holds for the term
\begin{equation}
    B_{s1}\exp(p_{1e}(s-p_2))\approx 0.
\end{equation}
Thus we now have to find roots for
the reduced system
\begin{eqnarray}
  \label{eq:3D_system_n_reduced}
  \dv{n}{t} &=& A_n\exp(p_{1e}(s-p_2))
                  -FC_nn^{\frac{1}{3}}q^{\frac{2}{3}}\\\
  \label{eq:3D_system_q_reduced}
  \dv{q}{t} &=& B_qn^{\frac{2}{3}}q^{\frac{1}{3}}(s-1)
                  -FC_qn^{-\frac{2}{3}}q^{\frac{5}{3}}\\
  \label{eq:3D_system_s_reduced}
  \dv{s}{t} &=& A_ss
                  %- B_{s1}\exp(p_{1e}(s-p_2))
                  -B_{s2}n^{\frac{2}{3}}q^{\frac{1}{3}}(s-1).
\end{eqnarray}
This slight simplification has the convenient property that finding
its roots can be reduced to finding the roots of a scalar function of
the saturation ratio $s$. From this root $s_0$ the values of the
equilibrium state (i.e. $n_0,q_0$) can be calculated recursively.  The
respective function for the saturation ratio is as follows (see Appendix~\ref{appB}):
\begin{equation}
  \label{eq:scalar_s}
   f(x) =  ax - b(x-1)^{\frac 3 4}\exp\left( c(x-x_0)\right),
 \end{equation}
 where
\begin{equation}
  a = A_s,~~
  b = \frac{ B_{s2} A_n}{C_n} \left(\frac{C_q}{B_q}\right)^{\frac 1 4}
  \left(\frac 1 F \right)^{\frac 3 4},~~
  c = p_{1e}, ~~
  x_0 = p_2.
\end{equation}
For a given value of the saturation ratio $s$ we can calculate the
respective values for $n$ and $q$ (see Appendix~\ref{appB}) as 
\begin{equation}
  \label{eq:equilibrium_n}
  n = \frac{A_n}{F C_n}\exp(p_{1e}(s - p_2))
  \left(\frac{B_q}{FC_q}(s -1 )\right)^{-\frac 1 2}
\end{equation}
and
\begin{equation}
  \label{eq:equilibrium_q}
  q= \frac{A_n}{F C_n}\exp(p_{1e}(s - p_2))
  \left(\frac{B_q}{F C_q}(s -1 )\right)^{\frac 1 4}.
\end{equation}

Now we can show, that the 3D system has exactly one equilibrium
point. For the existence of at least one root $s>1$, we make the
following argument: We investigate the function $f$ from
Equation~\eqref{eq:scalar_s} on the interval $[1,\infty)$. For $s=1$
we obtain $f(s)=a=A_s>0$. Since the coefficients $a,b,c>0$, the
exponential term is the fastest growing one, such that $f(s)<0$ for
sufficiently large values of $s$. Then the intermediate value theorem
provides the existence of a root $s_0>1$.
In order to see that $f$ has exactly one root, we consider the
function
\begin{equation}
  g(x)=\frac{f(x)}{x}=a - bx^{-1}(x-1)^{\frac 3 4}\exp(c(x-x_0))
\end{equation}
and note that any $x=s>1$ is a root of $f$ if and only if $x$ is a root
of $g$. We calculate the derivative of $g$ as
\begin{equation}
  g'(x)=- bx^{-1}(x-1)^{\frac 3 4}\exp(c(x-x_0))\cdot
  \left( c - \frac{1}{x} \right) 
  - \frac{3}{4}bx^{-1}(x-1)^{- \frac 1 4}\exp(c(x-x_0)).
\end{equation}
Since $c=p_{1e}(T)>1$ in the considered temperature range, we can
conclude that $g'(x)<0$ for all $x>1$, i.e. $g$ is strictly decreasing
for $x>1$ and thus there is at most one root. Altogether, we can
conclude that there is exactly one root $s_0$ of $f$ and thus exactly
one equilibrium point of
system~\eqref{eq:3D_system_n_reduced}-\eqref{eq:3D_system_s_reduced}.

For a meaningful approximation of the equilibrium states, we fix
temperature and pressure, and let $s_0$ denote the saturation ratio at the fixed point of the system for a given
vertical velocity $w_0$.
Remember that among the coefficients of the function $f$, only the coefficient $a = A_s$ depends on the vertical velocity $w$ and and moreover this dependence is linear. 
If we slightly vary the saturation ratio, the
main changes in the values of the function $f$ stem from the exponential term
$\exp(c(x-x_0))$, while changes in the other terms can be considered
to be minor. If we slightly vary the vertical velocity $w_0$ to a new
value $w_1$, we may assume that the corresponding value $s_1$ of the saturation ratio at the
equilibrium point satisfies
\begin{equation}
  a(w_1)s_1=b(s_1-1)^{\frac 3 4}\exp\left( c(s_1-x_0)\right)
\approx b(s_0-1)^{\frac 3 4}\exp\left( c(s_1-x_0)\right)
\end{equation}
and also
\begin{equation}
  a(w_1) \cdot s_1 = \frac{a(w_0)}{w_0}\cdot w_1 \cdot s_1 \approx
  \frac{a(w_0)}{w_0}\cdot w_1 \cdot s_0,
\end{equation}
where we also used the fact that $a$ depends linearly on $w$.
Combined this leads to the approximation
\begin{equation}
  \exp\left( c(s_1-x_0)\right) 
  = \frac{a(w_0)}{w_0}\cdot w_1 \cdot s_0 \cdot b^{-1} \cdot (s_0-1)^{-\frac 3 4}
\end{equation}
and thus an expression for the saturation ratio $s_1$ as follows
\begin{equation}
  s_1=p_2
  + \frac{1}{p_{1e}}\log(
  \frac{A_s(w_0)}{w_0}\cdot w_1 \cdot s_0 (s_0-1)^{-\frac 3 4} 
  \cdot \frac{C_n}{ B_{s2} A_n}
  \left(\frac{B_q}{C_q}\right)^{\frac 1  4} F^{\frac 3 4}
  ).
\end{equation}

\medskip

We introduce two further functions:
\begin{eqnarray}
  \sigma(T,p)
  & \coloneqq& p_2+ \frac{1}{p_{1e}}\log(
        \left(\frac{L}{c_p R_v T^2} - \frac{1}{R_a T} \right) \cdot g
        \cdot \frac{C_n}{ B_{s2} A_n} \left(\frac{B_q}{C_q}\right)^{\frac 1 4}
        )\\   
  \label{eq:h_contraction}      
  h(x)\coloneqq h(x,F,w,T,p)
  & = & \sigma+\frac{1}{p_{1e}}\log(w)
        + \frac{3}{4p_{1e}}\log(F)
        + \frac{1}{p_{1e}}\log(\frac{x}{(x-1)^{\frac 3 4}})
\end{eqnarray}
such that $s_1=h(s_0,F,w_1,T,p)$. A direct calculation shows that
$x^*$ is a fixed point of $h$ (i.e. $x^*=h\qty(x^*)$) if and only if
$x^*$ is a root of $f$ (i.e. $f(x^*)=0$). Therefore, fixed points $x^*$
of $h$ correspond precisely to the values of the saturation ratio at
equilibrium points of the ice cloud model.

We now show that function $h$ is a contraction on a suitable compact interval
(which in turn narrows the range for the equilibrium value $s_0$) and
use this property for a suitable approximation.
Estimating the ranges of functions $\sigma$ and $p_{1e}$ (see Appendix~\ref{appC}) we find that

\begin{equation}
  h(x)=\kappa_0+\frac{1}{p_{1e}}\log(\frac{x}{(x-1)^{\frac{3}{4}}})
\end{equation}
for some (variable) $\kappa_0\in[ 1.375170 , 1.585966]$.
Let us now specify the following interval
\begin{equation}
  [s_\text{min},s_\text{max}]
  ~\text{with}~s_\text{min}=1.37,~  s_\text{max}=1.59.
\end{equation}
We can show (see Appendix~\ref{appC}) that 
\begin{eqnarray}
  h\qty([s_\text{min},s_\text{max}]) &\subset &
  [s_\text{min},s_\text{max}].
\end{eqnarray}
In addition, we calculate the derivative of $h$ as
\begin{equation}
  h'(x)=\frac{1}{4p_{1e}}\frac{x-4}{x(x-1)}
\end{equation}
and thus
\begin{equation}
  \abs{h'(x)}\le
  \frac{\pi_2}{4}
  \frac{4-s_{\text{min}}}{s_{\text{min}}(s_{\text{min}}-1)}\approx
  0.004415\ll 1.
\end{equation}
Therefore, $h$ is a contraction on the compact interval
$[s_\text{min},s_\text{max}]\subset\mathbb{R}$.
Consequently, the Banach Fixed Point Theorem yields the existence of a unique fixed point, and the sequence~$x_n$ defined by
\begin{equation}
  x_n=h\qty(x_{n-1})=\sigma+\frac{1}{p_{1e}}\log(w)
        + \frac{3}{4p_{1e}}\log(F)
        + \frac{1}{p_{1e}}
        \log(\frac{x_{n-1}}{(x_{n-1}-1)^{\frac 3 4}})
\end{equation}
converges to this fixed point $s_0$. 
This implies in particular that the saturation ratio $s_0$ at an equilibrium
point (which has the value of the unique fixed point of $h(x)$) is always contained in the interval
$[s_\text{min},s_\text{max}]$ for any given environmental conditions in the relevant ranges as previously specified and summarized in Table~\ref{tab:parameters}.

In practice, we can determine an
approximation for the value of $s_0$ while approximating the function
$l(x)=\log(\frac{x}{(x-1)^{\frac{3}{4}}})$ on the interval
$[s_\text{min},s_\text{max}]$ by the midpoint of possible values
$\ell_m=\frac{1}{2}\qty(l(s_\text{min})+l(s_\text{max}))\approx
0.959979$, such that
\begin{equation}
  \label{eq:equilibrium_s_approx}
  s_0\approx \sigma(T,p)
  +\frac{1}{p_{1e}(T)}\qty(\ell_m+\log(w)+\frac{3}{4}\log(F))
\end{equation}
with an absolute error of approximation smaller than $4\cdot
10^{-4}$. Hence, the proof of uniqueness using the fixed point theorem
also leads to an approximate description of the equilibrium states as 
a by-product in our setting here. Note here, that the contraction property assures that the approximation converges towards the equilibrium point, i.e. iteration of the function leads to an improved approximation. 

\begin{figure}[bt!]
  \centering
  \includegraphics[width=0.32\linewidth, bb = 50 60 540 455, clip]
  {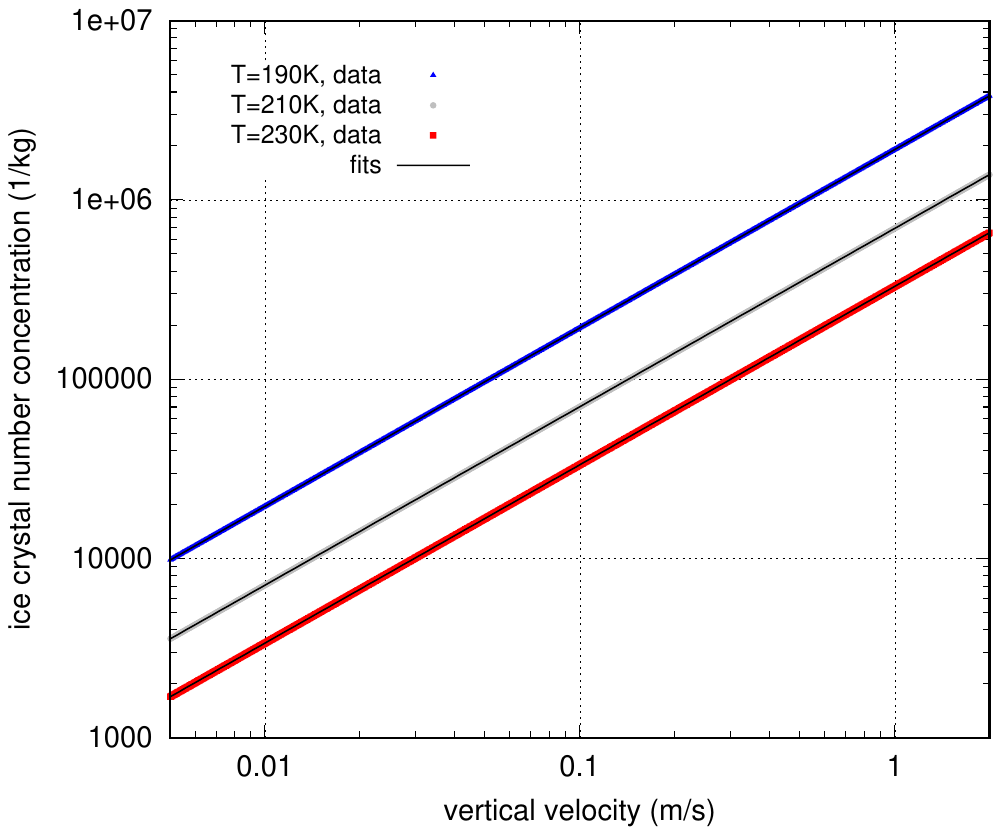}
  \includegraphics[width=0.32\linewidth, bb = 50 60 540 455, clip]
  {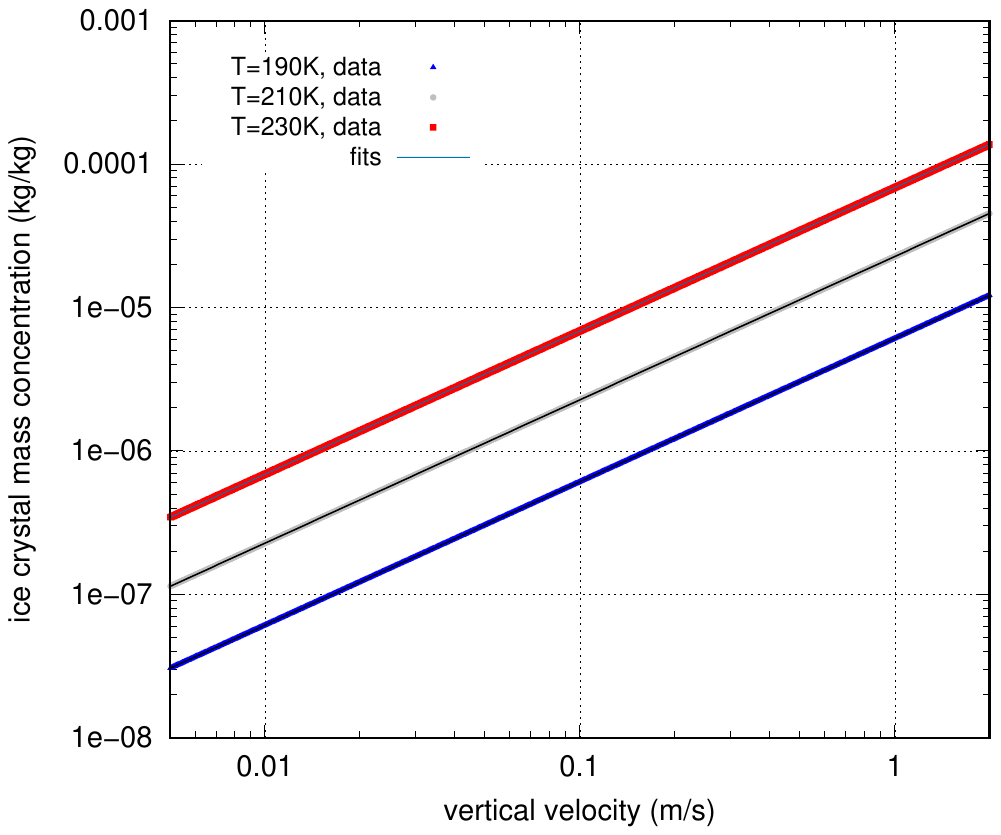}
  \includegraphics[width=0.32\linewidth, bb = 50 60 540 455, clip]
  {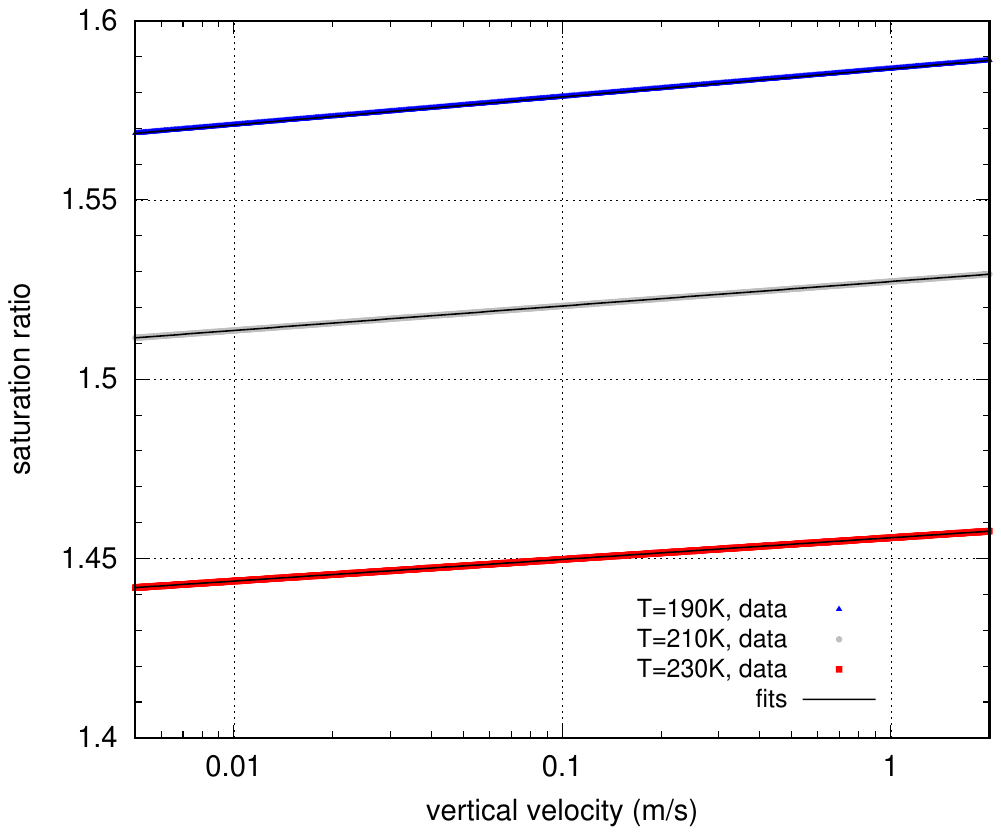}
  \caption{Equilibrium states for the ice cloud model at
    $p=\SI{300}{\hecto\pascal}$. Colors indicate different
    temperatures, i.e. red: $\SI{230}{\kelvin}$, grey:
    $\SI{210}{\kelvin}$, blue: $\SI{190}{\kelvin}$. Black lines
    indicate approximations as given in Equations~\eqref{eq:equilibrium_s_approx}-\eqref{eq:equilibrium_q_approx}. Left: Number concentration (in
    1/kg), middle: mass concentration (in kg/kg), right: saturation
    ratio.}
  \label{fig:eq_states}
\end{figure}

Using the Equations~\eqref{eq:equilibrium_n} and
\eqref{eq:equilibrium_q} for the calculation of the other values of
the equilibrium state $n_0,q_0$ from the value $s_0$, we finally get
the equations
\begin{eqnarray}
  \label{eq:equilibrium_n_approx}
  n_0
  & = & \frac{g}{B_{s2}}\left(\frac{C_q}{B_q}\right)^{\frac 1 4}
        \left(\frac{L}{c_p R_v T^2} - \frac{1}{R_a T} \right)
     \cdot w \cdot F^{\frac 1 4} \cdot s_{0}(s_{0}-1)^{-\frac 5 4}\\
  \label{eq:equilibrium_q_approx}
  q_0
  & = & \frac{g}{B_{s2}}\left(\frac{B_q}{C_q}\right)^{\frac 1 2}
        \left(\frac{L}{c_p R_v T^2} - \frac{1}{R_a T} \right)
        \cdot w \cdot F^{-\frac 1 2 }\cdot s_{0}(s_{0}-1)^{-\frac 1 2}.
\end{eqnarray}
Since the changes in the saturation ratio $s_{0}$ at the equilibrium
point for varying $w$ and $F$ within the specified ranges are only
very minor compared to the changes in $w$ and $F^{\frac 1 4}$ or
$F^{- \frac 1 2}$ in Equations~\eqref{eq:equilibrium_n_approx}
and~\eqref{eq:equilibrium_q_approx}, good approximations for the ice
crystal number and mass concentrations $n_{0}$ and $q_{0}$ at the
equilibrium point for varying $w$ and $F$ are given by calculating
$s_{0}$ for one choice of vertical velocity $w$ and sedimentation
parameter $F$ (e.g. for $w = \SI{0.1}{\metre\per\second}$ and
$F = 1.0$) and then approximating $n_{0}$ and $q_{0}$ by
Equations~\eqref{eq:equilibrium_n_approx}
and~\eqref{eq:equilibrium_q_approx} using the fixed $s_{0}$.
In Figure~\ref{fig:eq_states}, the equilibrium states for different
temperatures are represented. The dependency of all quantities $n,q,s$
on $\log(w)$ is clearly visible, since we use a logarithmic axis for
the vertical updrafts. The quasi-linear fits are almost perfect.

\subsection{Quality of equilibrium states}
\label{sec:quality_equilibrium_states}

The quality of the equilibrium point of the model depends on the
choice of the parameters, and can be deduced from the eigenvalues of
the Jacobi matrix of the system at the equilibrium point; this is true
at least in case of non-zero eigenvalues, which is the case here, as
we show later. The Jacobi matrix of the reduced
system~\eqref{eq:3D_system_n_reduced}-\eqref{eq:3D_system_s_reduced}
at an arbitrary point $(n,q,s)$ is given by
\begin{equation}
  \label{eq:jacobi_3d}
 J = \left(\begin{matrix}
     - \frac 1 3 F C_n \bar m^{\frac 2 3} & 
     - \frac 2 3 F C_n \bar m^{-\frac 1 3} &
     p_{1e} A_n \exp (p_{1e}(s - p_2))\\
     \frac 2 3 \left( B_q \bar m^{\frac 1 3}(s - 1)
       + F C_q \bar m^{\frac 5 3 } \right) &
     \frac 1 3 B_q \bar m^{- \frac 2 3}(s -1 )
     - \frac 5 3 F C_q \bar m^{\frac 2 3} &
     B_q n^{\frac 2 3} q^{\frac 1 3}\\
     - \frac 2 3 B_{s2} \bar m^{\frac 1 3}(s -1 )&
     - \frac 1 3 B_{s2} \bar m^{- \frac 2 3}(s -1 ) &
     A_s - B_{s2} n^{\frac 2 3}q^{\frac 1 3}
      \end{matrix} \right).
\end{equation}
Here, $\bar{m}=\frac{q}{n}$ denotes the mean mass of ice
crystals. Since the expressions become too complicated there is no
chance to calculate the eigenvalues analytically at the equilibrium
points. However, we can partly determine the general quality of the
solution, e.g. properties such as the sign of the eigenvalues. For this
purpose, we reformulate the Jacobi matrix.\\
Using the conditions for an equilibrium point, i.e.
$(\dot{n},\dot{q},\dot{s})=(0,0,0)$, we obtain the following
equations (see Appendix~\ref{appB})
\begin{eqnarray}
  A_n \exp (p_{1e} (s - p_2 )) &=& F C_n n^{\frac 1 3} q^{\frac 2 3}\\
  \bar m  &=&\left( \frac{B_q}{F C_q} (s -1) \right)^{\frac 3 4}\\
  B_{s2} (s - 1) &=& A_s s \bar m^{- \frac 1 3 } n^{-1}
\end{eqnarray}
These identities can be used for rewriting the Jacobi matrix as
\begin{equation}
  J=\left(
    \begin{matrix}
      - \frac 1 3 F C_n \bar m^{\frac 2 3} & 
      - \frac 2 3 F C_n \bar m^{-\frac 1 3} &
      p_{1e} F C_n \bar m^{\frac 2 3} n\\
      \frac 4 3 F C_q \bar m^{\frac 5 3} &
      - \frac 4 3 F C_q \bar m^{\frac 2 3}  &
      \frac{1}{(s -1 )} F C_q \bar m^{\frac 5 3}  n\\
      - \frac 2 3 A_s s n^{-1}&
      - \frac 1 3 A_s s q^{-1}&
      - A_s \frac{1}{(s -1 )} 
    \end{matrix}\right).
\end{equation}
It turns out that this matrix $J$ is similar to a slightly simpler
matrix $J_1$, which has the same eigenvalues,
i.e. $J=D^{-1}J_1D$, and finally $J=a\cdot D^{-1}J_2D$, with
\begin{equation}
  \label{eq:matrix_J2}
  a=\frac{1}{3}FC_n\bar m^\frac{2}{3},~~
  D=\mqty(1 & 0 & 0 \\ 0 & \bar m^{-1} & 0 \\ 0 & 0 & n),~~
  J_2=\left(\begin{matrix}
      - 1 & 
      - 2 &
      3 p_{1e} \\
      4 r_0^{\frac 2 3} &
      - 4 r_0^{\frac 2 3} &
      3 r_0^{\frac 2 3} \frac{1}{(s -1 )} \\
      - 2 \gamma \frac{s}{\sqrt{s-1}} &
      - \gamma \frac{s}{\sqrt{s-1}} &
      - 3 \gamma \frac{1}{(s -1)^{\frac{3}{2}}}
    \end{matrix}\right)
\end{equation}
using the relations
\begin{equation}
  \label{eq:constants_cq_gamma}
  C_q=r_0^{ \frac 2 3}C_n,~~
  \gamma=A_s F^{- \frac 1 2} %(s -1 )^{- \frac 1 2}
  B_q^{ - \frac 1 2} r_0^{ \frac 1 3} C_n ^{ -\frac 1 2}
\end{equation}
For the calculation of the determinant we use the formulation
$J=a\cdot D^{-1}J_2D$, thus we find
\begin{equation}
  \det J = \frac{1}{3} F^2 C_n^2 A_s \bar m^{\frac 4 3} r_0^{\frac 2 3}\frac{1}{s -1}
  \cdot \qty(-4+s-4s(s -1) p_{1e})
\end{equation}
Since $F,C_n,A_s,\bar m$ are all positive and $s>1$ and $p_{1e}>1$ in
the considered range, we find
\begin{equation}
  -4+s-4s(s -1) p_{1e}=-3+(s-1)(-4sp_{1e}+1)<0 ~\Rightarrow \det J<0
\end{equation}
Let $\lambda_1,\lambda_2,\lambda_3$ denote the eigenvalues of $J$ for
a given equilibrium state. After rearranging/renumbering we can assume
that either all eigenvalues are real or one ($\lambda_3$) is real and two are
complex conjugates ($\lambda_1=\lambda_2^*$). Since $\det
J=\lambda_1\lambda_2\lambda_3<0$, we conclude that there has to be
at least one negative real eigenvalue $\lambda_i<0$. In addition, all
eigenvalues are non-zero ($\lambda_i\neq 0$ for all $i=1,2,3$).

The eigenvalues are then calculated numerically using routines of the
package ``linalg'' in ``numpy'' \citep{harris2020}. The real negative
eigenvalue is denoted by $\lambda_3$. For most parameters,
we find two complex conjugate eigenvalues $\lambda_1,\lambda_2$; in a
small regime (higher temperature and/or high vertical velocities) we
obtain two real negative eigenvalues $\lambda_1,\lambda_2<0$. We find
two supercritical Hopf bifurcations
%\citep[following the nomenclature by][]{kuznetsov1998}
of the system, i.e. the real part of the complex conjugate eigenvalues
changes its sign, such that there is a transition from a stable
equilibrium point to the case of an unstable equilibrium point and an
attracting limit cycle~\citep[see, e.g.,][]{kuznetsov1998}. For the
local existence of the periodic solutions (i.e. limit cycles), we can
apply a theorem based on Lyapunov-Schmidt reduction \citep[Theorem 2.1
in][page 341]{golubitsky_schaeffer1985}, since at the bifurcation
point we have exactly two conjugate imaginary eigenvalues and no other
eigenvalues on the imaginary axis. However, we will investigate the 
the limit cycles later in Section~\ref{sec:limit_cycle}, using
numerical investigations.

Using the description of the trace of the Jacobi matrix, we can derive
additional properties of the eigenvalues of the linearized
system. The trace can be calculated as
\begin{equation}
  \label{eq:trace_jacobian}
  \tr(J)=\tr(aD^{-1}J_2D)=a\tr(J_2)=
  -a\qty(1+4r_0^{\frac{2}{3}}+3\gamma\frac{1}{(s -1)^{\frac{3}{2}}})%\frac{1}{s-1})
  =\lambda_1+\lambda_2+\lambda_3
\end{equation}
using the description of the matrix $J_2$ in
Equation~\eqref{eq:matrix_J2}. Since the constants $a,r_0,\gamma$ are all
positive
%, i.e.
%\begin{equation}
%  a>0,~ r_0^{\frac{2}{3}}>0,~\gamma>0
%\end{equation}
and in the considered range we have $s>1$, we find a negative trace of
the Jacobi matrix
\begin{equation}
  \tr(J)=a\tr(J_2)=\lambda_1+\lambda_2+\lambda_3<0.
\end{equation}
This condition can be reformulated as
$  \lambda_1+\lambda_2<-\lambda_3$
with the distinct real and negative eigenvalue
$\lambda_3\in\mathbb{R}_{<0}$. We can distinguish different cases:
\begin{enumerate}
\item For a pair of complex conjugate eigenvalues
  $\lambda_1=\lambda_2^\ast=\alpha+i\beta\in\mathbb{C}$
  ($\alpha,\beta\in\mathbb{R}$) we find
  \begin{equation}
    \label{eq:trace_realpart}
    \lambda_1+\lambda_2=2\alpha<-\lambda_3\Leftrightarrow
    \alpha=\Re(\lambda_{1,2})<-\frac{\lambda_3}{2}\in\mathbb{R}_{>0}
  \end{equation}
  For negative real parts $\alpha$ this is not a restriction, since
  then we find $\alpha<0<-\frac{\lambda_3}{2}$. However, for positive
  real parts (i.e. in case of an unstable fixed point), we determine the
  condition $0<\alpha<-\frac{\lambda_3}{2}$. Thus, the absolute value
  of the growth rate $\Re(\lambda_{1,2})=\alpha$ of the instability
  close to the equilibrium point is always smaller than the value of
  the contraction rate $\lambda_3$, i.e. close to the equilibrium
  state the flow contraction in $\lambda_3-$direction is always faster than
  the blowup in $\lambda_{1,2}-$ directions. This points to a quasi 2D
  system for longer time scales.
\item For real eigenvalues $\lambda_1,\lambda_2\in\mathbb{R}$, we still
  have the property $\lambda_1+\lambda_2<-\lambda_3$. In the numerical
  determination we only find the case of two negative real
  eigenvalues  $\lambda_1,\lambda_2<0$; thus, there is again no
  restriction for the eigenvalues, since the eigenvalues always fulfill
  $\lambda_1+\lambda_2<0<-\lambda_3$.
\end{enumerate}

\begin{figure}[h]
  \centering
  \includegraphics[width=0.5\linewidth, bb = 50 60 540 455, clip]
  {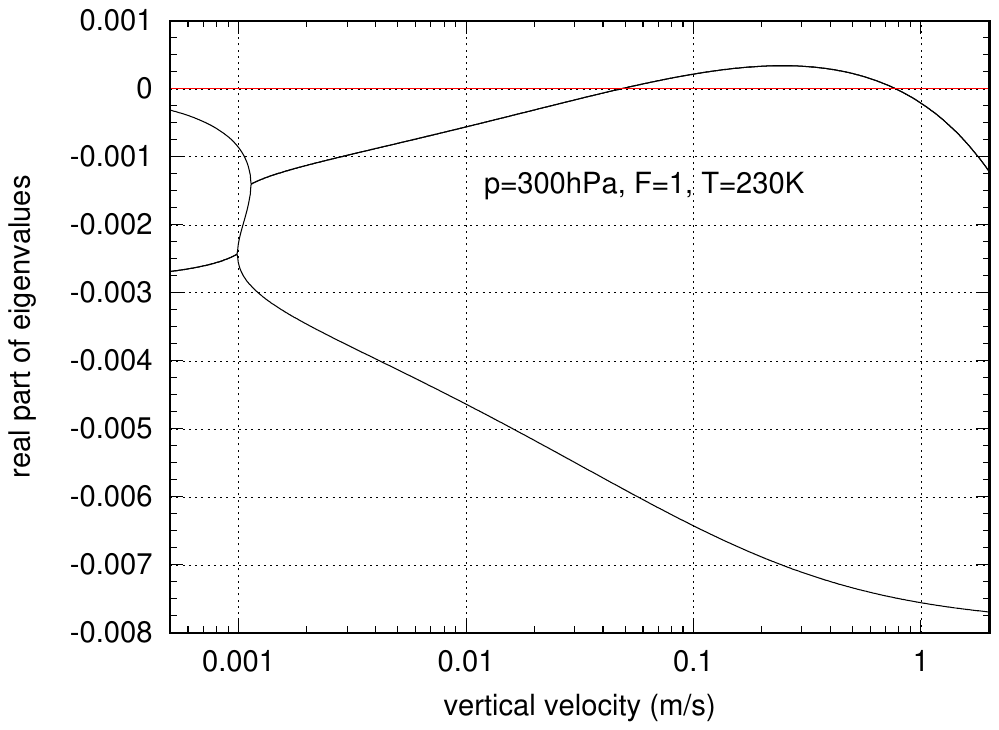}%
  \includegraphics[width=0.5\linewidth, bb = 50 60 540 455, clip]
  {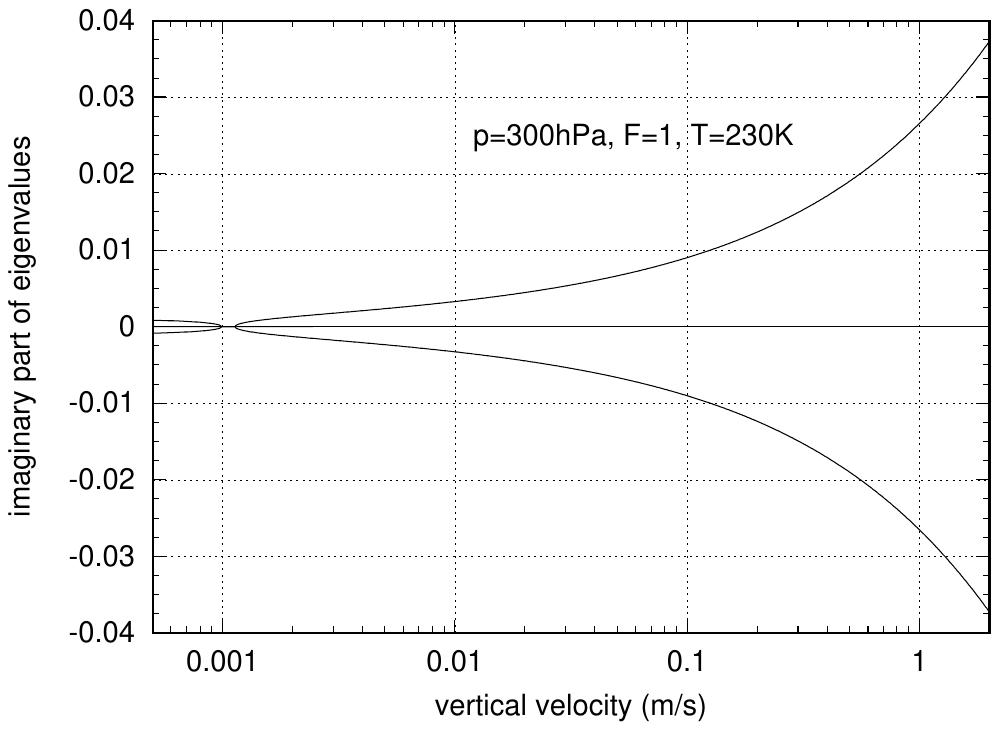}
  \caption{Example of real (left) and imaginary (right) parts of numerically
    determined eigenvalues at the equilibrium state of the system for
    a fixed set of parameters
    $p=\SI{300}{\hecto\pascal}, T=\SI{230}{\kelvin}, F=1$}
  \label{fig:example_eigenvalues}
\end{figure}

In Figure~\ref{fig:example_eigenvalues} we show an example of real and
imaginary parts of the numerically determined eigenvalues for a fixed
parameter set of $p=\SI{300}{\hecto\pascal}, T=\SI{230}{\kelvin}, F=1$
and values of vertical updrafts in the interval
$\SI{0.0005}{\metre\per\second}\le w\le
  \SI{2}{\metre\per\second}$. Note, that we show a regime with all
possible occurring variants of eigenvalues, i.e. conjugate-complex pair
and three real eigenvalues, respectively. For the unstable regime (for
positive real part, i.e. in the range between
$w\sim\SI{0.048}{\metre\per\second}$ and
$w\sim\SI{0.77}{\metre\per\second}$) we clearly see, that the
relation~\eqref{eq:trace_realpart} is fulfilled, i.e. the positive real part of $\lambda_{1,2}$ is much smaller than the absolute value of $\lambda_3$.\\
\textbf{Remark:} If we could assure that the real part of the complex-conjugate eigenvalues changes with non-zero speed at the bifurcation points, a theorem by Hopf \citep[1942, see][Theorem 3.4.2, p. 151]{guckenheimer_holmes2002} would guarantee the existence of a two dimensional centre manifold, on which the limit cycle lies.

\begin{figure}[bt!]
  \centering
  \includegraphics[width=0.55\linewidth, bb = 50 60 540 455, clip]
  {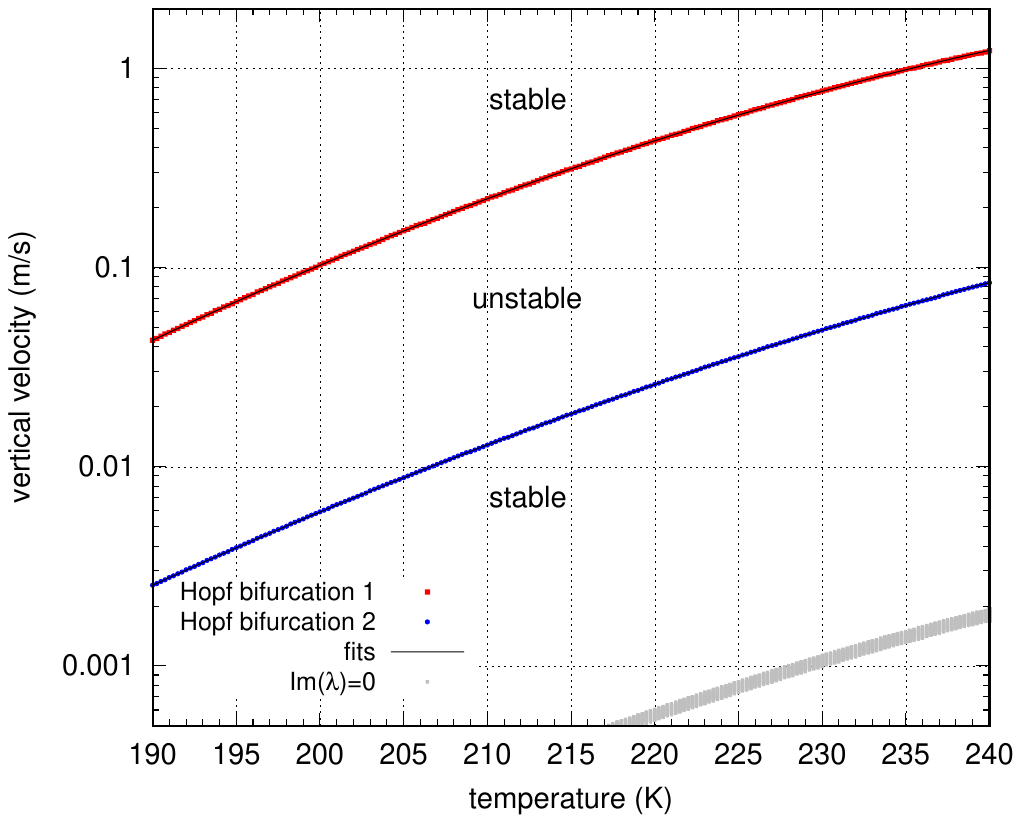}  
  \caption{Bifurcation diagram in the parameter space $T-w$ for
    $p=\SI{300}{\hecto\pascal}$ and $F=1$. The red and blue dots
    indicate the numerically calculated transition in the real parts
    of the complex-conjugate eigenvalues $\lambda_{1,2}$, i.e. the two
    Hopf bifurcations. The black lines indicate fits to the
    bifurcations using a quadratic polynomial in $T$. Above the red
    points and below the blue points, the unique equilibrium state is
    stable. Between the two lines, the equilibrium state is unstable
    and a limit cycle occurs. Grey dots indicate the region with
    purely real eigenvalues (i.e. three negative real eigenvalues). }
  \label{fig:bifurcation_diagram}
\end{figure}

We calculate the bifurcation points numerically. For a fixed
temperature $T$ we scan through the $w$-interval, calculate the
eigenvalues of the respective Jacobi matrix $J$, and determine the
position where the sign of the real part of $\lambda_{1,2}$ changes.
The
bifurcation diagram for values $p=\SI{300}{\hecto\pascal}$ and $F=1$
is represented in Figure~\ref{fig:bifurcation_diagram}. Blue and red
dots mark the numerically determined Hopf bifurcations, grey dots mark
the region of three real and negative eigenvalues.

We find that the two bifurcations for $F=1$ can be fitted using exponential
functions with quadratic polynomials in the argument, i.e. functions
of the form
\begin{equation}
  \label{eq:fits_bifurcations}
  w_a(T)=\bar{w}\exp(a_2T^2+a_1T+a_0),~~w_b(T)=\bar{w}\exp(b_2T^2+b_1T+b_0)
\end{equation}
with coefficients $a_i,b_i$ as represented in Table~\ref{tab:coeff_w},
and the ``constant'' $\bar{w}=\SI{1}{\metre\second^{-1}}$ determining
the correct unit of the functions.
\begin{table}[tb]
  \centering
  $
  \begin{array}{l||c|c|c}
    & k=0 & k=1 & k=2\\\hline
    a_k & \SI{-38.30947}{}
          &  \SI{0.278555}{\kelvin^{-1}}
                & \SI{-0.00049191}{\kelvin^{-2}}\\\hline
    b_k & \SI{-36.15046}{}
          & \SI{0.229111}{\kelvin^{-1}}
            & \SI{-0.00036997}{\kelvin^{-2}}
  \end{array}
  $
  \caption{Coefficients for the quadratic fit functions $w_a(T),w_b(T)$ of the
    two Hopf bifurcations }
  \label{tab:coeff_w}
\end{table}
% \begin{equation}
%   a_2=\SI{-0.00049191}{\metre\second^{-1}\kelvin^{-2}},~
%   a_1=\SI{0.278555}{\metre\second^{-1}\kelvin^{-1}},~
%   a_0=\SI{-38.30947}{\metre\second^{-1}};
%   ~~b_2=-0.00036997,b_1=0.229111,b_0=-36.15046.
% \end{equation}
The fits are shown in Figure~\ref{fig:bifurcation_diagram} as black
lines; however, blue dots ``belong'' to fit $w_a(T)$, whereas red dots
``belong'' to fit $w_b(T)$. The bifurcation diagram for other values of $F$ will be
discussed later in the Section~\ref{sec:scaling}, where we investigate
the scaling properties of the system. For different values of pressure
(i.e. $p=\SI{200}{\hecto\pascal}$) we find qualitatively similar
bifurcation diagrams, i.e. the supercritical Hopf bifurcations are
still there, however slightly shifted in their position. In fact, the quality
of the system stays the same. 

\noindent\textbf{Remark:}
For the Jacobi matrix of the equilibrium state at the bifurcation we can use basic relations from linear algebra
\begin{equation}
    \lambda_{1,2}=\pm i\beta, ~\Rightarrow \tr(J)=\lambda_3,~\Rightarrow
    \det(J-\lambda_3I)=0
\end{equation}
to derive a quasi-quadratic equation from the determinant for the bifurcation value $w$, i.e. 
\begin{equation}
\label{eq:w_bifurcation}
    a_2w^2+a_1w+a_0=0
\end{equation}
with coefficients $a_2,a_1,a_0$ only slightly depending on $w$. This gives also an analytical hint, that we should find two bifurcations for the system. When calculating the values of $w$ from Equation~\eqref{eq:w_bifurcation}, we find values very close to the bifurcation points as numerically calculated from the complete system.

\begin{figure}[h]
  \centering
  \includegraphics[width=0.32\linewidth]
  {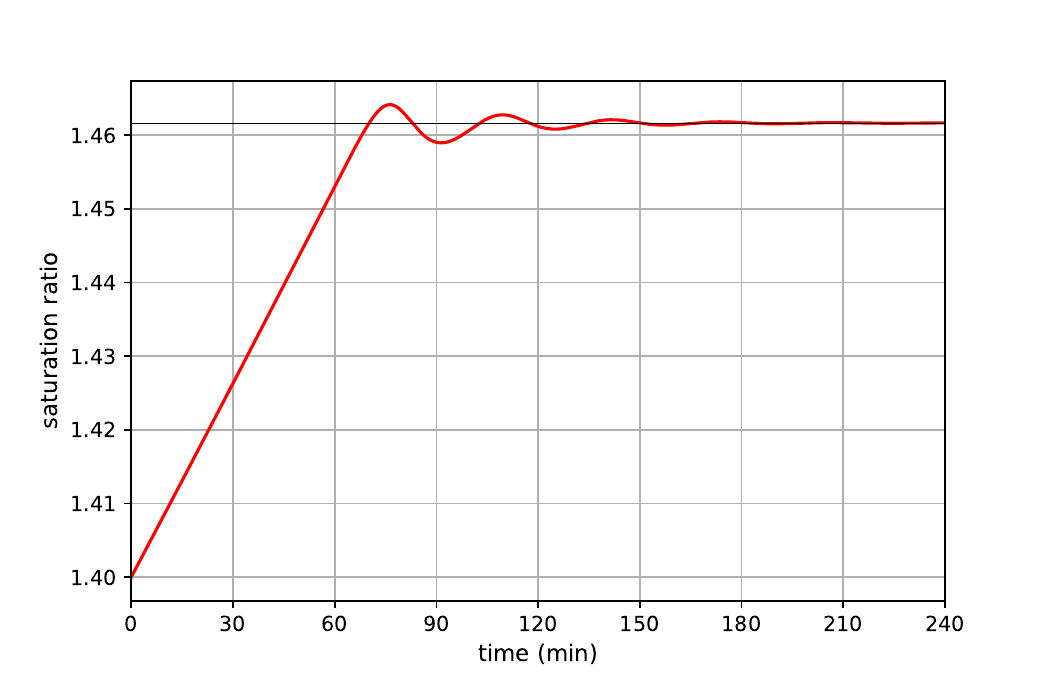}        
  \includegraphics[width=0.32\linewidth]
  {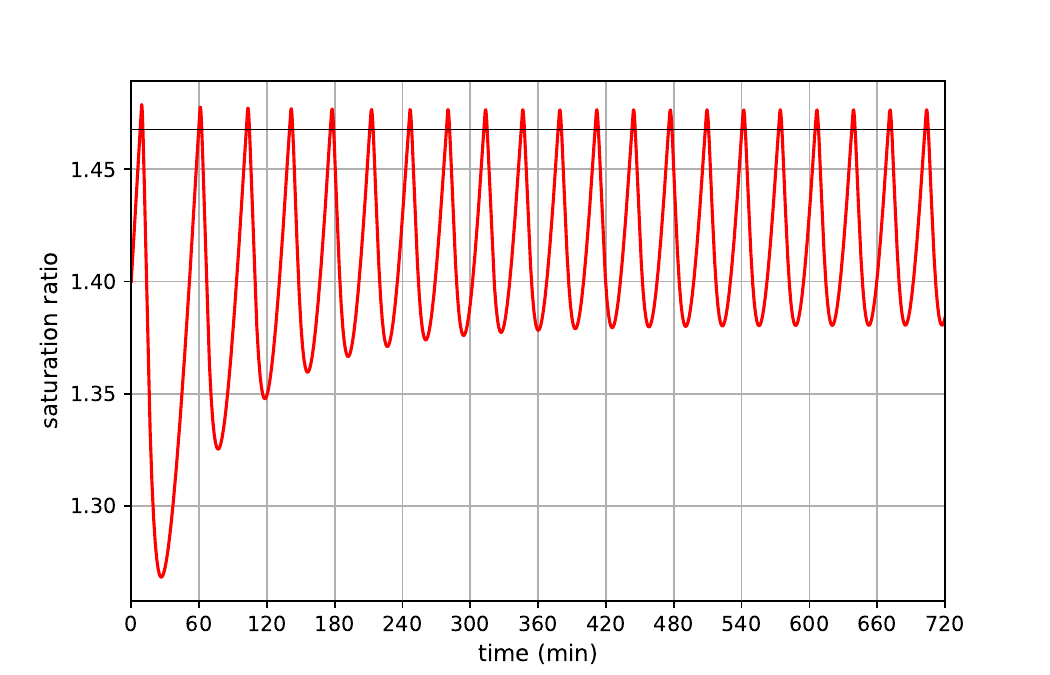}    
  \includegraphics[width=0.32\linewidth]
  {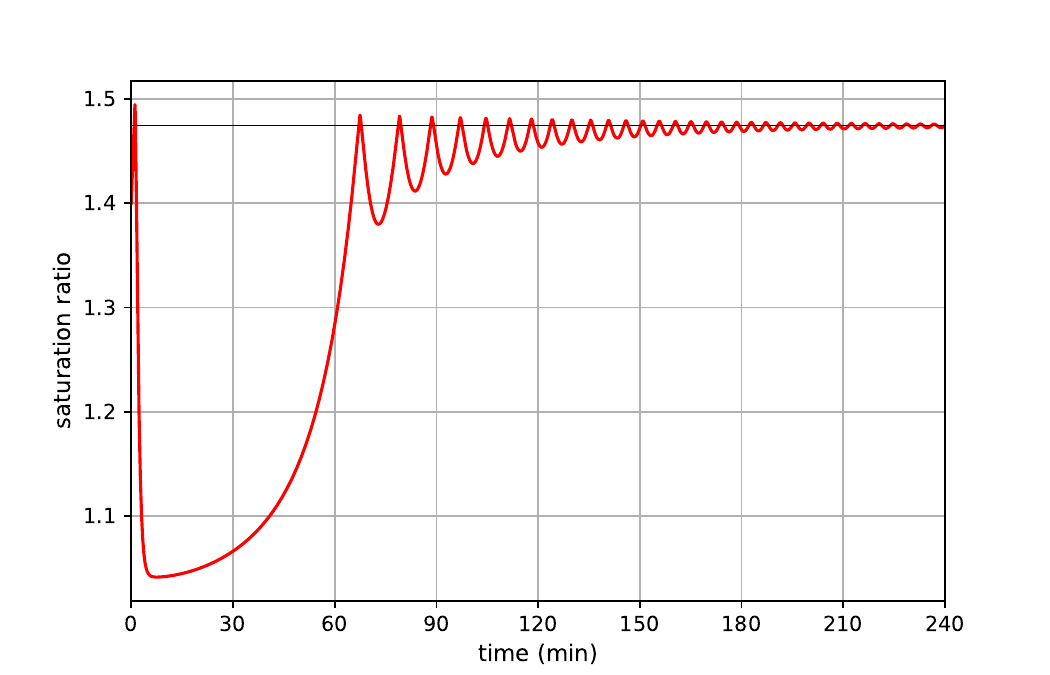}  
  \caption{Examples of solutions for different regimes. Left: damped
    oscillator ($w=\SI{0.01}{\metre\per\second})$, middle: undamped
    oscillator ($w=\SI{0.1}{\metre\per\second}$), right: damped
    oscillator ($w=\SI{1}{\metre\per\second}$)}
  \label{fig:examples_solutions}
\end{figure}

In Figure~\ref{fig:examples_solutions}, some examples for the three
different regimes are represented, i.e. the time evolution of the
saturation ratio $s$ is shown for values
$w=\SI{0.01}{\metre\per\second}$ (stable equilibrium, lower regime),
$w=\SI{0.1}{\metre\per\second}$ (unstable equilibrium, limit cycle), and
$w=\SI{1}{\metre\per\second}$ (stable equilibrium, higher regime). 
In all simulations we see the same general evolution of a first and undisturbed nucleation event followed by a (more or less) strong reduction of the supersaturation, and afterwards subsequent nucleation events, which are less vigorous since pre-existing ice slows down the explosion term. This behavior was observed in former studies \citep[e.g.][]{spichtinger_cziczo2010}. All variables of the simulations are shown in Appendix~\ref{appD}, i.e. 
Figures~\ref{fig:examples_solutions_full001}, \ref{fig:examples_solutions_full010}, and \ref{fig:examples_solutions_full100}.

Qualitatively we can describe the behavior with the three different processes. Nucleation, as driven by the cooling/forcing term of $s$, acts as a source for ice crystals, whereas sedimentation is the sink for ice crystals. The growth process acts as a mediator in between, shifting mass from the saturation ratio to the mass concentration, which (1) shuts down the nucleation event because of less supersaturation as discussed in \citet{spichtinger_kraemer2013}, and (2) enhances the sedimentation sink. These feedback can also be seen calculating the divergence of the vector field, i.e. the trace of the general Jacobi matrix.
\begin{equation}
    \div{V}=\tr(J)=
    A_s+\frac{1}{3}B_qm^{-\frac{2}{3}}(s-1)-B_{s2}nm^{\frac{1}{3}}
    -\frac{1}{3}FC_n\qty(1+5r_0^{\frac{2}{3}})m^{\frac{2}{3}}
\end{equation}
The first term, i.e. the cooling/forcing of the system is clearly positive, the last term, i.e. sedimentation, is negative, and the growth terms in the middle could change signs. For small particles, the first term of growth dominates (positive), whereas for large (or many) particles the second term of growth dominates (negative). However, the divergence can change sign, thus a clear asymptotic behavior cannot be seen from this analysis. 

As
described in the next section, in the unstable case we observe an
undamped oscillation, which indicates an attracting limit cycle.

\subsection{Limit cycles}
\label{sec:limit_cycle}

There is no method available for calculating the limit cycles, or even
their periods analytically for this 3D nonlinear system, at least to
our best knowledge. Therefore we use a numerical method to determine
the period of the limit cycles. %, based on a Poincare map approach. 
For
this purpose we integrate the system again for quite a long time, but
with initial conditions close to the unstable equilibrium state. To be
precise, for a (numerically determined) equilibrium point
$x_0=(n_0,q_0,s_0)^T$ we determine the initial condition as
$x_{\text{init}}=x_0\cdot (1-\epsilon)$ with a small deviation
$\epsilon=0.01$. We then run the simulation for a time $T_\text{final}$, which is
approximately determined as $T_\text{final}=300\cdot \tau_\lambda$ with the period
at the equilibrium point $\tau_\lambda=\frac{2\pi}{\Im(\lambda_1)}$,
using the imaginary part of the two complex-conjugate eigenvalues. The
simulations are carried out using a Gaussian grid with a timestep,
adaptively but constant determined by the empirical relation
$\Delta t=\frac{\SI{0.01}{\metre}}{w}$, using the prescribed vertical
velocity $w$.

\begin{figure}[bt!]
  \centering
  \includegraphics[width=0.54\linewidth]
  {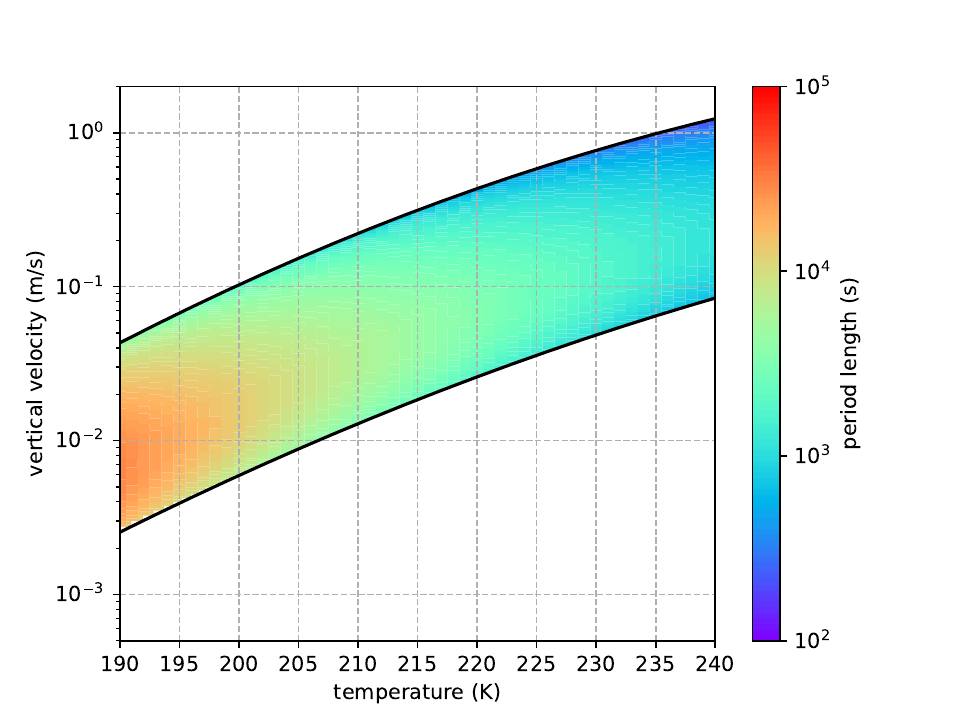}
  \includegraphics[width=0.45\linewidth,  bb = 50 60 540 455, clip]
  {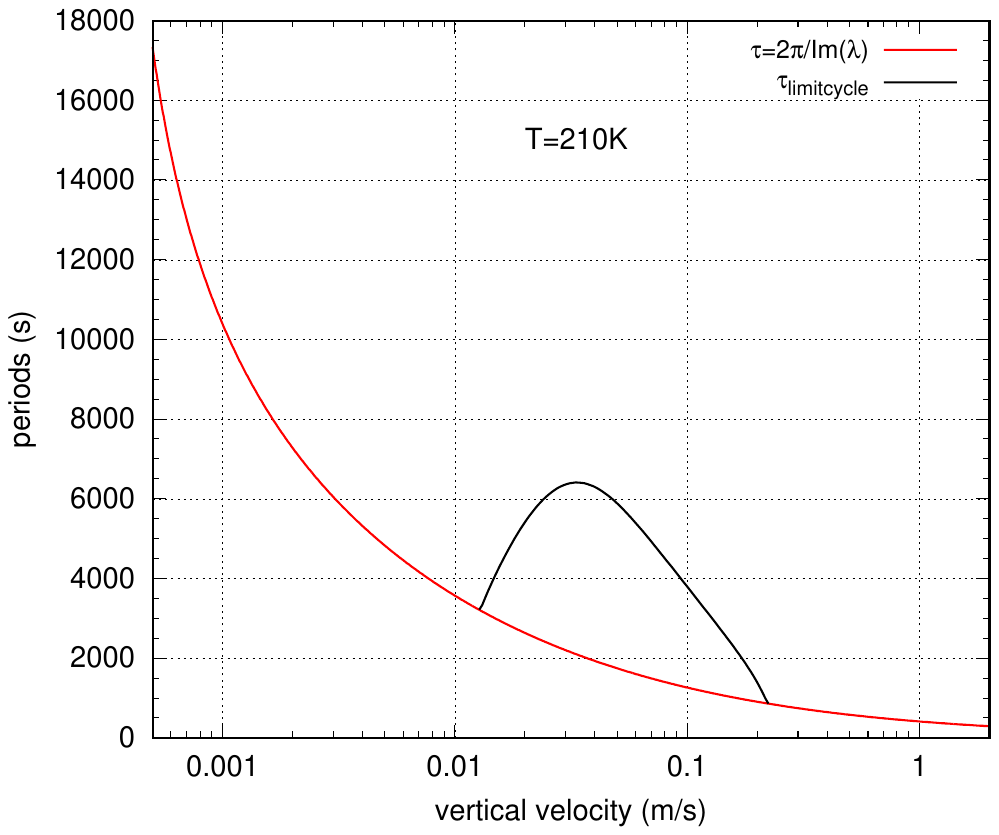}
  \caption{Left: Periods of limit cycles in the bifurcation
    diagram. Right: Example of section through the parameter space at
    fixed temperature $T=\SI{210}{\kelvin}$, representing the
    oscillation period as calculated from the imaginary part of the
    complex eigenvalues (red line), and the period of the respective
    limit cycle in the unstable regime (black line).}
  \label{fig:periods_bifurcation_diagram}
\end{figure}

Since the variable $s$ is the most ``linear'' one, we then determine
minima and maxima of this variable over time. From the difference
between consecutive minima or maxima, respectively, at the end of the
simulation we determine two periods
$\tau_{\text{min}},\tau_{\text{max}}$. The mean value
$\tau=\frac{1}{2}\qty(\tau_{\text{min}}+\tau_{\text{max}})$ is then
used as the numerically determined period of the limit cycle, for a
given set of parameters $(p,T,w,F)$, respectively. In
Figure~\ref{fig:periods_bifurcation_diagram} (left panel) the periods
of the limit cycles are represented. For the domain between the two
Hopf bifurcations, the color code indicates the periods of the limit
cycles. Note that there is a huge range of periods, ranging from very
short periods at high temperatures and high vertical velocities (in
the order of few hundreds of seconds) up to large periods at low
temperatures and low vertical updrafts (in the order of $10^4$ to
$10^5$ seconds). The periods depending on $w$ show a similar shape for a distinct
temperature, as will be investigated later in terms of scaling
properties in Section~\ref{sec:scaling}. In the right panel of
Figure~\ref{fig:periods_bifurcation_diagram} an example for periods at
a fixed temperature of $T=\SI{210}{\kelvin}$ but varying vertical
updraft is displayed. The red curve shows the ``oscillation period''
as derived from the imaginary part of the respective complex-conjugate
eigenvalues ($\tau=\frac{2\pi}{\Im(\lambda_k)}$) at the equilibrium
point, whereas the black curve shows the period of the limit cycle in
the region of the unstable equilibrium point. Both curves touch each
other at the bifurcation, as expected. The periods of the limit cycle
always have a distinct maximum in the interval between the two
bifurcation lines.  Note that for the small regime of three real
eigenvalues, as indicated in Figure~\ref{fig:bifurcation_diagram} by
the grey shading, $\tau$ shows a singularity when approaching an
imaginary part $\Im(\lambda_k)=0$.

\subsection{Scaling properties}
\label{sec:scaling}

Now we investigate some scaling properties of the system.

\subsubsection{Scaling property of bifurcations}
\label{sec:scaling_bifurcations}

Until now we
investigated the bifurcations for variations in $p,T,w$, but left the
sedimentation parameter unchanged, i.e. $F=1$ (no flux from above). Now we investigate
possible changes in the bifurcations due to variation of $F$ in the
interval $0.01<F<1$. This is a realistic range; smaller values do not
make sense because then we would effectively shut down the
sedimentation sink, which is the crucial process for producing an
oscillating behavior. Beside the numerical calculations of the
equilibrium states, the linearization and eigenvalues leading to the
bifurcations as determined by the numerical procedure, we evaluate
the variation in a semi-analytical approximate way.

\begin{figure}[h]
  \centering
  \includegraphics[width=0.45\linewidth,  bb = 50 60 540 455, clip]
  {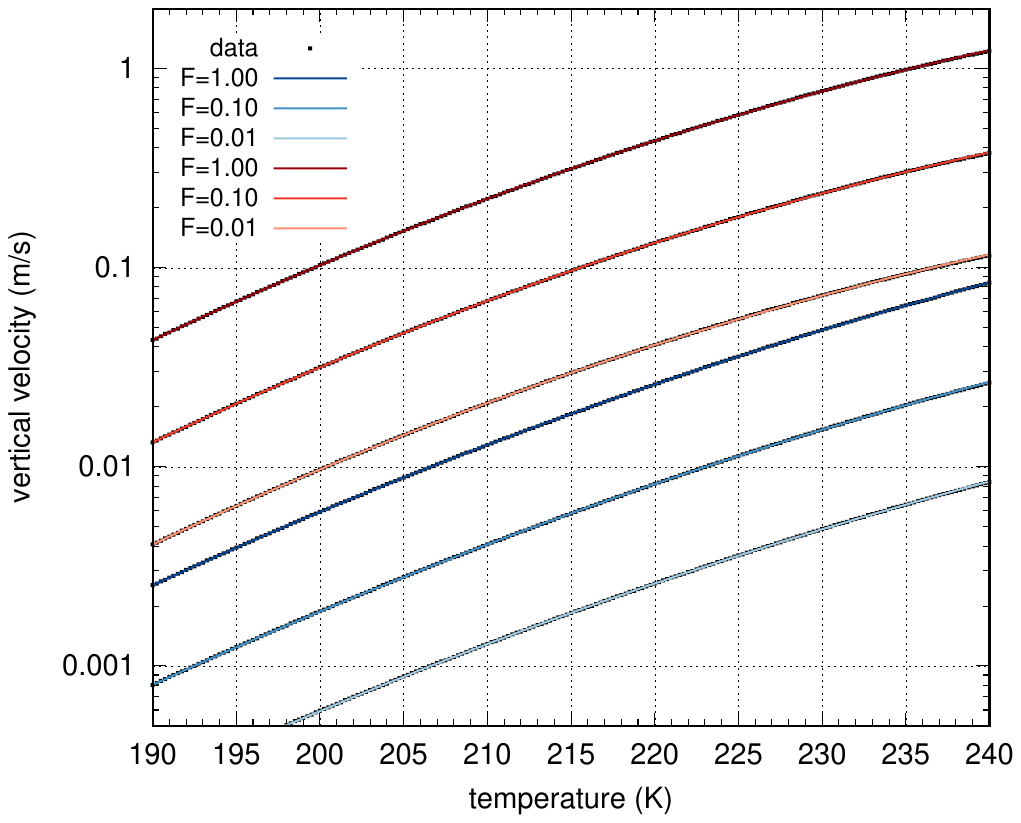}
  \caption{Scaling of bifurcations for different values of the
    sedimentation parameter $F$. Black dots show the numerically
    derived values, whereas colored lines represent the scaling of the
    fit curves as derived for the parameter $F=1$. Reddish colors show
    the change in the upper bifurcation, whereas bluish colors show
    the change in the lower bifurcations; in both cases, sedimentation
    parameters of $F=1,0.1,0.01$ are shown. }
  \label{fig:scaling_bifurcation}
\end{figure}

We investigate pairs of bifurcation coordinates $(w,F)$ for fixed
values of the environment~$p,T$. First, suppose that we have already determined
a bifurcation point $(w_0,F_0)$, i.e. for the Jacobi matrix at the
equilibrium state we find eigenvalues
$\lambda_{1}=\lambda_2^\ast=i\beta$, and for small perturbations their
real parts change the sign. Let us now show that we can
approximately determine another bifurcation point $(w_1,F_1)$ just
from the original pair $(w_0,F_0)$ via the Jacobi matrix. Here, we
observe that beside the actual values of the equilibrium state and the sedimentation parameter $F$ only the
constants $A_s,B_q,C_n,p_{1e}$ impact the value of the Jacobi
matrix. Out of this set of parameters only $A_s$ depends
linearly on $w$, while all others are completely independent of $w$ and~$F$.\\
We try to find another pair $(w_1,F_1)$ close to the original one,
i.e. we apply a small perturbation to $w$ and $F$. From the
approximation Equation~\eqref{eq:equilibrium_s_approx} we see that the
value of $s$ does not change substantially for moderate/small
variations in $w$ and/or $F$. Thus, from the special form of the
Jacobi matrix $J_2$ we might assume that in a first approximation
$J_2$ changes only in $\gamma$ via changes in $A_sF^{-\frac{1}{2}}$
(see Equation~\eqref{eq:constants_cq_gamma}), when varying updraft and
sedimentation parameter. For convenience we write
$A_s=\tilde{A}_sw$. Hence, passing to a different sedimentation
parameter $F_1$ and a vertical velocity $w_1$ we observe that $\gamma$
and consequently $J_2$ are approximately equal for $(w_0,F_0)$ and
$(w_1,F_1)$ if and only if
\begin{equation}
  \label{eq:scaling_F1}
  \tilde{A}_sw_0F_0^{-\frac{1}{2}}=A_s(w_0)F_0^{-\frac{1}{2}}=
  A_s(w_1)F_1^{-\frac{1}{2}}=\tilde{A}_sw_1F_1^{-\frac{1}{2}}
  \Longleftrightarrow
  \frac{w_0}{w_1}=\qty(\frac{F_0}{F_1})^{\frac{1}{2}}
\end{equation}
Now we consider the ``special case'' of setting $F_0=1$. Thus, we
obtain
\begin{equation}
  \label{eq:scaling_F2}
  \frac{w_0}{w_1}=\qty(\frac{1}{F_1})^{\frac{1}{2}}
  \Longleftrightarrow
  w_1=F_1^{\frac{1}{2}}w_0=\sqrt{F_1}w_0
\end{equation}
and for a bifurcation point $(w_0,1)$ also the pair
$\qty(\sqrt{F_1}w_0,F_1)$ is (approximately) a bifurcation point, so
we get a scaling of the updrafts by the square root of the
sedimentation parameter.  However, this relation relies on small
perturbations and approximations, so one would like to evaluate the
``real'' values of bifurcations as determined numerically with the
procedure explained above. These bifurcation values are displayed as
black dots in Figure~\ref{fig:scaling_bifurcation}. In
Section~\ref{sec:quality_equilibrium_states}, we found fits for the
bifurcation points ($F=1$) of the form
\begin{equation}
  w(T)=\bar{w}\exp(c_0+c_1T+c_2T^2).
\end{equation}
On the other hand, from the (semi-)analytical description above we
know that approximately $w_1=\sqrt{F}w_0$. For evaluating this
approximation we make the general ansatz of a power law for the
scaling, i.e. $w_1=F^\alpha w_0$ with a positive real number
$\alpha\in\mathbb{R}_{>0}$. Then we can derive:
\begin{eqnarray}
  w_1(T) = F^\alpha w_0(T)
  & = &  F^\alpha \bar{w}\exp(c_0+c_1T+c_2T^2)\\
  & = & \bar{w}\exp(\bar{c}_0+c_1T+c_2T^2)
\end{eqnarray}
with the expression
\begin{equation}
  F^\alpha \exp(c_0)=\exp(\bar{c}_0) \Leftrightarrow
  \log(F^\alpha \exp(c_0))=\bar{c}_0\Leftrightarrow
  \alpha\log(F)+c_0=\bar{c}_0\Leftrightarrow
  \alpha=\frac{\bar{c}_0-c_0}{\log(F)}.
\end{equation}
We determine fits for sedimentation factors
$F\in\qty{0.01,0.02,0.05,0.1,0.2,0.3,0.4,0.5,0.6}$ for both
bifurcations and determine the exponent $\alpha$ (also in terms of a
mean value). We find for the upper bifurcation
$\alpha\approx\frac{1}{2}$ as from theory, whereas for the lower
bifurcation the values changes slightly to
$\alpha\approx 0.512\approx\frac{1}{2}$.  Thus, the scaling property
of the bifurcation due to $F^\alpha$ is a robust feature. In
Figure~\ref{fig:scaling_bifurcation} the scaled fits using the respective
ansatz $w(T)=F^\alpha w_\ast(T)$ are shown for $F=1,0.1,0.01$; reddish
colors show the upper bifurcations, whereas bluish colors display the
lower bifurcation. The fits match the data points quite well, corroborating this type of scaling. Finally, we can state that the change of the
sedimentation parameter~$F$ does not change the quality of the
bifurcation diagram. Actually, the quality of the phase diagram and
thus the bifurcations remain the same, only the transition
lines/bifurcation curves change their position, but even not their
shape. This is a remarkable feature of this nonlinear oscillator.

\subsubsection{Scaling property of limit cycle periods}
\label{sec:scaling_periods}

\begin{figure}[hb]
  \centering
  \includegraphics[width=0.32\linewidth,  bb = 50 60 540 455, clip]
  {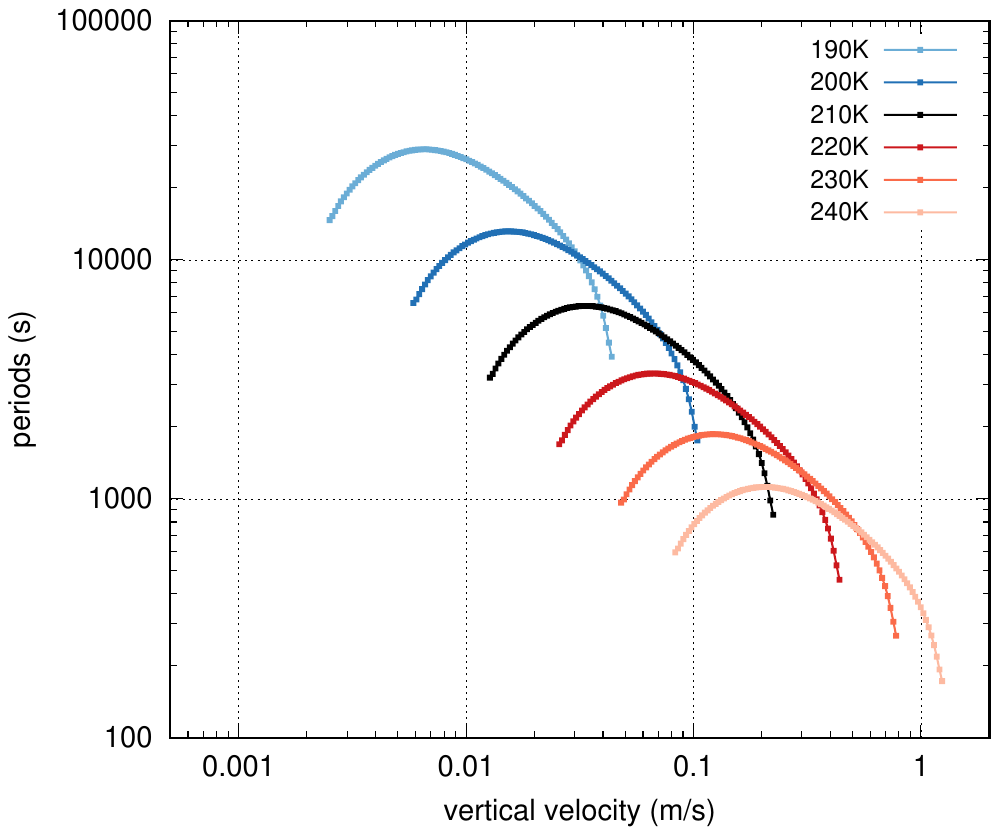}
  \includegraphics[width=0.32\linewidth,  bb = 50 60 540 455, clip]
  {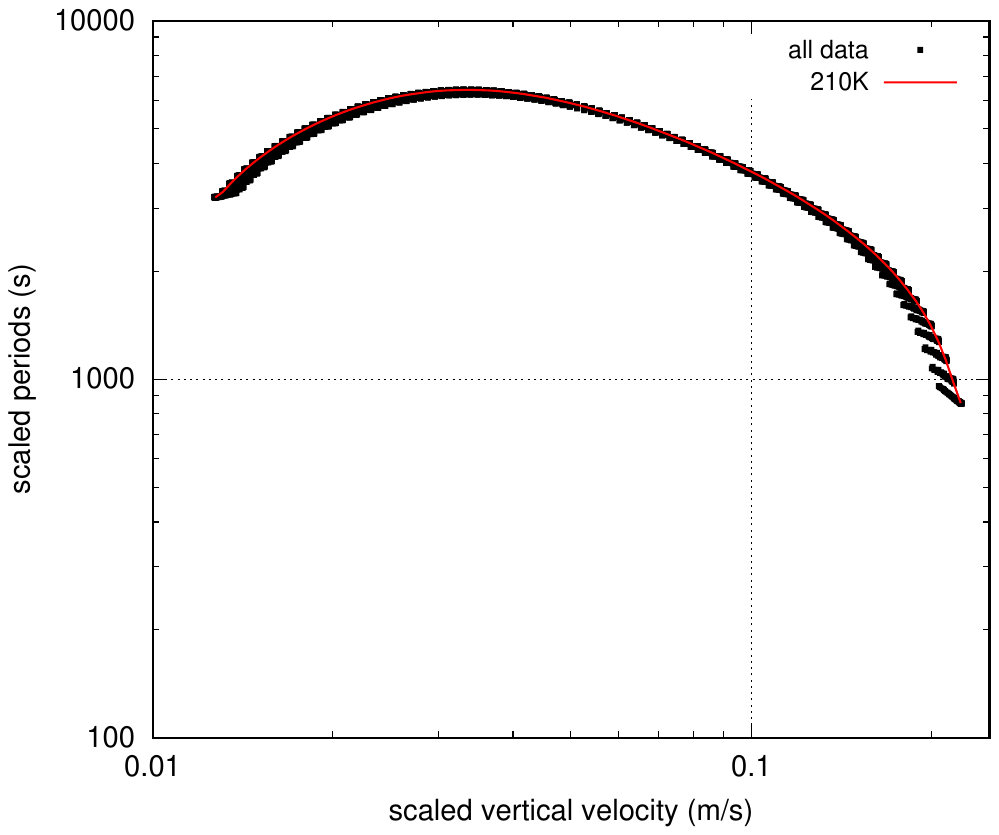}
  \includegraphics[width=0.32\linewidth,  bb = 50 60 540 455, clip]
  {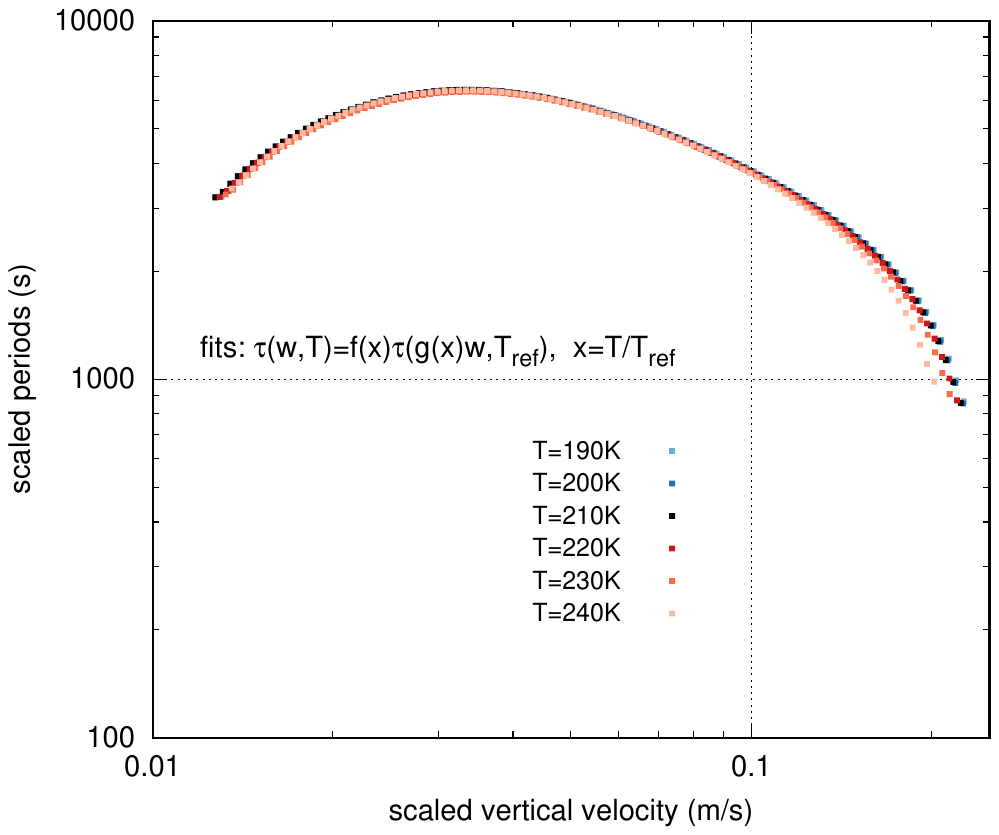}
  \caption{Scalings of limit cycle's periods. Left: periods of limit cycles for different temperatures; middle: numerical scaling of the data together with the reference curve for $T=\SI{210}{\kelvin}$; right: scaled data using the scaling functions from Equations~\eqref{eq:scaling_f_complex} and \eqref{eq:scaling_g_complex}.}
  \label{fig:scaling_periods}
\end{figure}

\begin{figure}[h]
  \centering
  \includegraphics[width=0.5\linewidth,  bb = 50 60 540 455, clip]
  {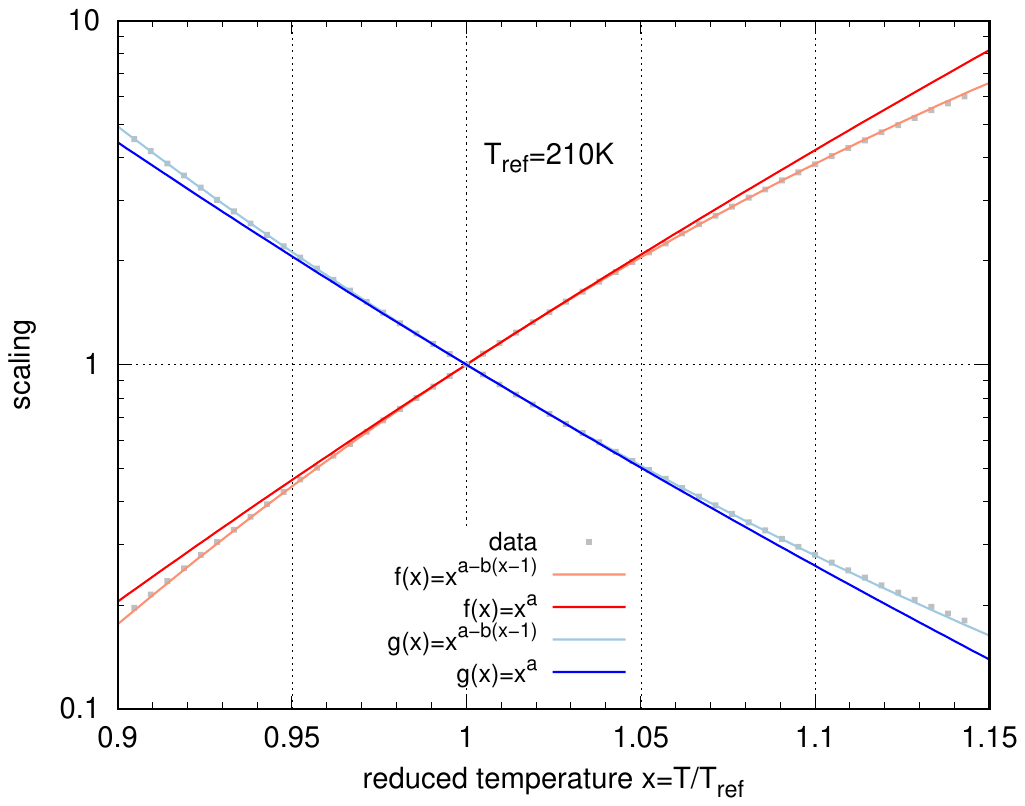}
  \caption{Fit of exponents of scaling functions. Grey dots: empirical scaling as derived from the numerical calculations. Dark red/blue curves: exponents of the empirical fits using scalings of the functional form $f(x)\sim x^a$, exact values are given in Equation~\eqref{eq:scaling_fg_simple}. Light red/blue curves: exponents of the empirical fits using scalings of the functional form $f(x)\sim x^{a-b(x-1)}$, exact values are given in Equations~\eqref{eq:scaling_f_complex} and \eqref{eq:scaling_g_complex}.}
  \label{fig:scaling_exponents}
\end{figure}

In Section~\ref{sec:limit_cycle}, we determined the periods of the limit cycles for the
regime of unstable equilibrium states. We already noted that the shape
of the curves at fixed but different temperatures $T$ is very
similar. In the left panel of Figure~\ref{fig:scaling_periods}, the
curves for the periods of the limit cycles at different temperatures
are displayed.
In the double-logarithmic representation the similarity of the curves
is very evident. Thus, we make an empirical scaling ansatz, using the
temperature value $T_{\text{ref}}=\SI{210}{\kelvin}$ as reference
temperature. We hypothesize that the curve scale with a
non-dimensional temperature. Using the reduced temperature
$x\coloneqq\frac{T}{T_{\text{ref}}}$, we might find scaling functions
$f(x),g(x)$ such that the periods $\tau(w,T)$ can be expressed as
\begin{equation}
  \label{eq:scaling_periods}
  \tau\qty(w,T)=f(x)\tau\qty(g(x)w,T_{\text{ref}}).
\end{equation}
The scaling function $g(x)$ shifts the $w-$position of the periods,
whereas the function $f(x)$ changes the amplitude of the periods,
respectively. This kind of scalings are often found in phase
transitions and self-organized criticality \citep[see,
e.g.,][]{pruessner2012}, and thus we would expect a function of the
form $f(x)\sim x^a$ with a so-called critical exponent
$a\in\mathbb{R}$. A similar form is expected for the function
$g(x)$. Comparing the numerically derived periods and calculating
shifts for matching the curves, we can map the curves on
each other; this is shown in the middle panel of
Figure~\ref{fig:scaling_periods}, with the calculated and shifted data
as black points and the ``reference'' period at $T_{\text{ref}}$ as a
red curve. We can calculate numerically the apparent shifts in position
and amplitude, respectively, for the different reduced temperatures
$x$ in the temperature interval
$\SI{190}{\kelvin}\le T\le \SI{240}{\kelvin}$, i.e. in the interval
$0.9\le x\le 1.15$. These values are plotted in
Figure~\ref{fig:scaling_exponents} as grey dots. Using the expected
form for functions $f(x),g(x)\sim x^a$ we find the exponents
\begin{equation}
\label{eq:scaling_fg_simple}
  f(x)=x^{15.0617},~~g(x)=x^{-14.1228};
\end{equation}
these fits are shown in Figure~\ref{fig:scaling_exponents} as curves
in light red or blue. For values $0.95<x<1.05$ this approach fits
quite well.

However, the scaling functions show deviations
to the numerically optimized shift in the data for values $x$ at the
boundaries of the interval. Empirically, we find a (much) better fit
using the functional form
\begin{equation}
  f(x)=x^{a-b(x-1)},~~a,b\in\mathbb{R}~\text{constant},
\end{equation}
as indicated in Figure~\ref{fig:scaling_exponents} with curves in dark
red or blue. There is small deviation from the fits with a
constant scaling exponent, since the functions' coefficients have the
following values for $f$
\begin{equation}
\label{eq:scaling_f_complex}
  f(x)=x^{a-b(x-1)},~~~ a=15.2984,~~~ b=12.013
\end{equation}
and for $g$
\begin{equation}
\label{eq:scaling_g_complex}
  g(x)=x^{a-b(x-1)},~~~   a=-14.2715,~~~  b=-8.65253
\end{equation}
respectively.  Finally, we apply the adjusted scaling functions
$f(x),g(x)$ for matching the periods of different temperatures on each
other. As can be seen in the right panel of
Figure~\ref{fig:scaling_periods}, this scaling works very well; only
at the ``boundaries'', i.e. for high or low values of $x$, there are
some small deviations, which are probably due to the numerical
derivation of the bifurcations.  However, we have to state here that
the fits are purely empirically, and there is no obvious reason why
the periods should have such a scaling behavior. Nevertheless, this is a
very interesting property, which in turn also reduces the parameter
space for the investigations dramatically.
The complete range of bifurcations in the system then boils
down to the range of a period determined by just a vertical velocity
as shown in the right panel of Figure~\ref{fig:scaling_periods} for
the reference temperature $T_{\text{ref}}$ or even the reduced
temperature $x=1$. All other periods in the unstable regime can be
derived by the scaling law Equation~\eqref{eq:scaling_periods}.

%\begin{itemize}
%\item Scaling of bifurcations
%\item scaling of time scales/periods of limit cycles
%\end{itemize}

%%%%%%%%%%%%%%%%%%%%%%%%%%%%%%%%%%%%%%%%%%%%%%%%%%%%%%%%%%%%%%%%%%%%%%%%%%%%%%%%
%%%%%%%%%%%%%%%%%%%%%%%%%%%%%%%%%%%%%%%%%%%%%%%%%%%%%%%%%%%%%%%%%%%%%%%%%%%%%%%%
\clearpage
\section{Comparison to measurements}
\label{sec:comparison}

In the sections above, we investigated the qualitative properties of
this simplified model. However, the question arises how relevant such
a model is in terms of representing ice clouds in the upper troposphere sufficiently well. Thus, we would like to know if the model produces
realistic values of the relevant quantities of
ice crystal number and mass concentrations. We now compare simulations
of our model with real measurements as obtained from research
aircraft, i.e. \textit{in situ} measurements in the tropopause region; these
measurements were described in former studies
\citep{kraemer_etal2016,kraemer_etal2020}, and are freely available
for scientific evaluations. We use data of ice crystal number
concentrations, as obtained from various instruments, which are
compiled to a consistent data set \citep[see discussion
in][]{kraemer_etal2016}. Since the data are available in units of
numbers per volume, we have to recalculate our model values,
which are in the conserved variable version of numbers per mass dry
air; the conversion is as follows
\begin{equation}
  N\vert_{\text{per volume}} = \rho\cdot n\vert_{\text{per mass}}
\end{equation}
For this purpose we have to decide about a given pressure, since
the conversion relies on the ideal gas law, thus providing density for
given temperature and pressure, respectively. Since most measurements
are obtained in the pressure range
$\SI{200}{\hecto\pascal}\le p \le \SI{300}{\hecto\pascal}$, and the model
results only weakly depend on the given pressure parameter, we show
our results for the two limiting pressure values separately.  We
compile data for the parameter range of vertical updrafts
$\SI{0.01}{\metre\per\second}\le w \le \SI{2}{\metre\per\second}$,
the temperature range $\SI{190}{\kelvin}\le T\le\SI{230}{\kelvin}$,
two pressure regimes $p=200,\SI{300}{\hecto\pascal}$, and two
sedimentation regimes $F=1,0.1$, respectively. For comparison,
we do not indicate full model simulations, but rather the values at the
equilibrium points and the ranges of the limit cycles (depending on whether we have a stable equilibrium point or an unstable equilibrium point with limit cycle) as key features of the long term behavior. Moreover, we also include the values of an undisturbed nucleation event as calculated with the model by \citet{spichtinger_etal2023}; the temperatures are chosen for comparison with simulations by \citet{kaercher_lohmann2002} and \citet{spichtinger_gierens2009a}.
%\citep[similar to the
%nucleation simulations in][]{kaercher_lohmann2002,
%  spichtinger_gierens2009a}.

\begin{figure}[h]
  \centering
  \includegraphics[width=0.49\linewidth]
  {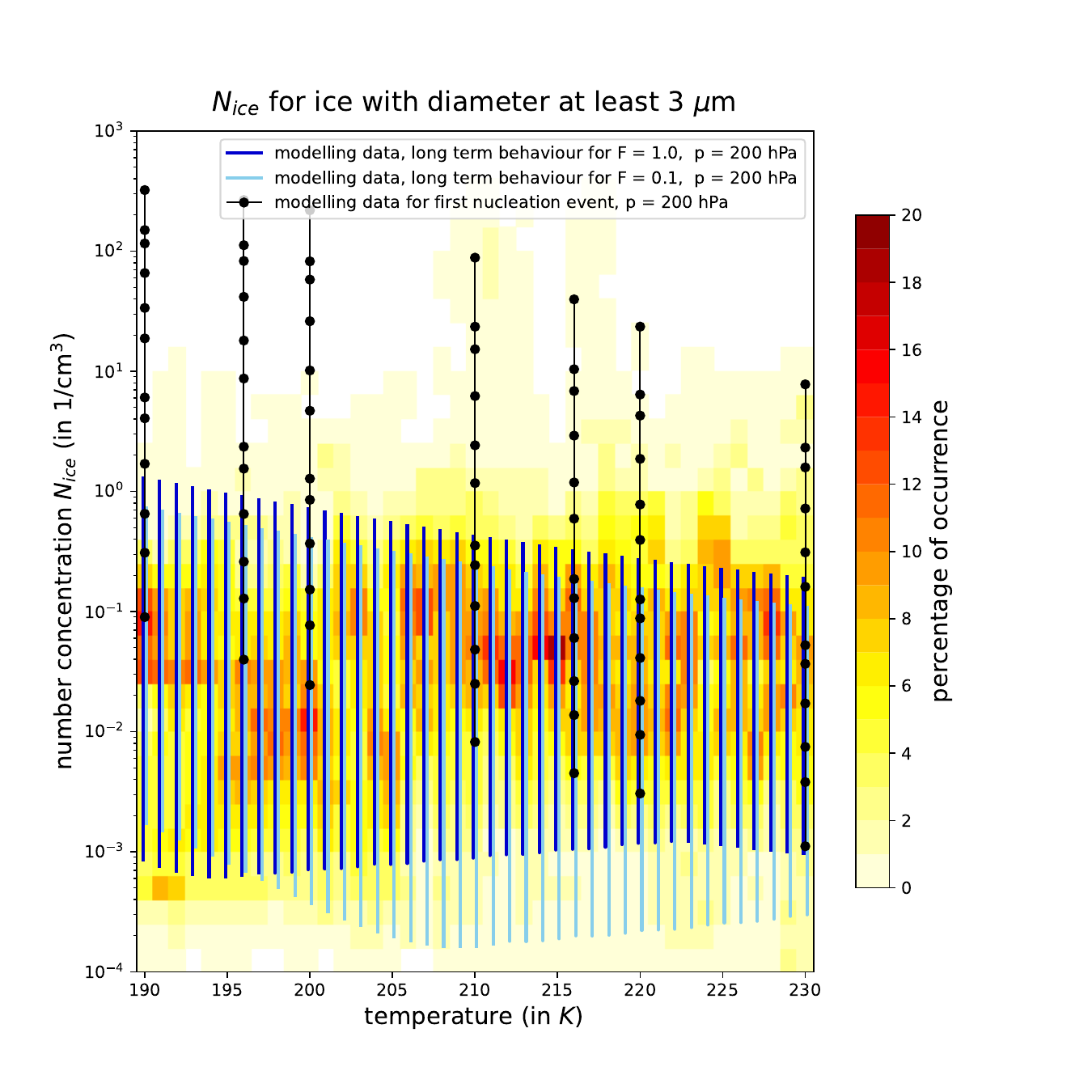}
  \includegraphics[width=0.49\linewidth]
  {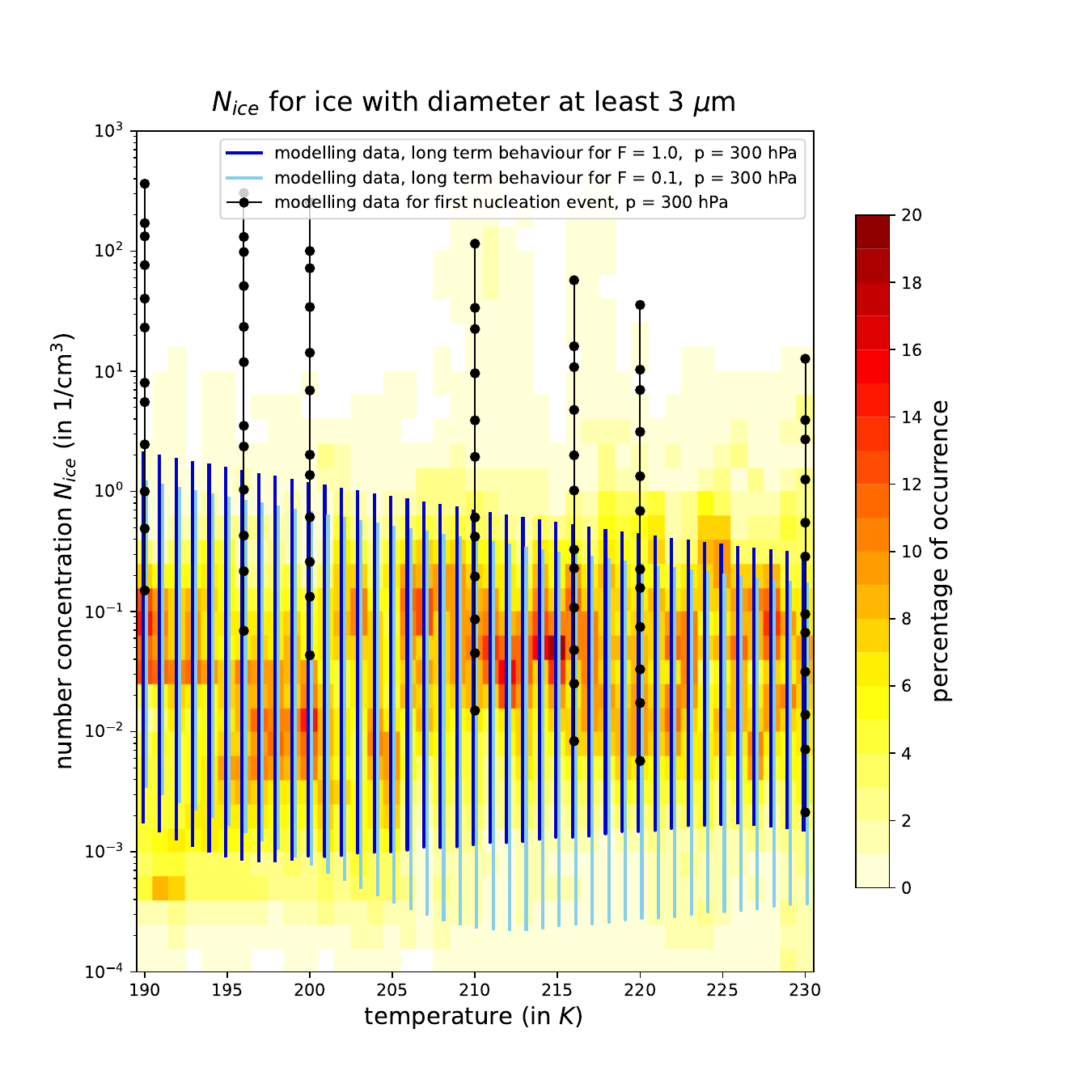}
  \caption{Comparison of the model results with measurements. Colors indicate the frequency of occurrence of ice crystal number concentrations in measurements \citep[][]{kraemer_etal2016, kraemer_etal2020}. Bluish colors show the values of the model, i.e. equilibrium states and range of the limit cycle calculations. Dark blue lines indicate conditions for $F=1$, light blue indicate conditions for $F=0.1$, respectively. Black points show undisturbed nucleation events as calculated by \citet{spichtinger_etal2023}. }
  \label{fig:comp_meas}
\end{figure}
In Figure~\ref{fig:comp_meas}, the comparison of measurements (colors,
indicating frequency of occurrence) and model results (blue lines and
black linepoints) is displayed. The measurements are sorted in
temperature bins of $\SI{1}{\kelvin}$, the frequency of occurrence is represented by colors. Our model results for the sedimentation
factor $F=1$ are shown in dark blue, whereas light blue indicates
results for $F=0.1$.

We find a good agreement of our model results with the
observations. In fact, the qualitative behavior of the model,
i.e. equilibrium points and limit cycles, fits very well with the
observations. Especially, we want to emphasize that our model
simulations are mostly located in the regimes of frequently observed
measurements. We see higher values of $n$ for the regime of strong
sedimentation $F=1$, and slightly reduced values for weak
sedimentation $F=0.1$, respectively. Both simulation regimes seem to
be reasonable and realistic. The variation due to different pressure
regimes is quite marginal, we just see small changes in the picture
for different pressure values. In contrast, the values of undisturbed
nucleation events in the model are not found in the real
observations. The high values at the temperature regime
$\SI{210}{\kelvin}\le T\le \SI{220}{\kelvin}$ can be attributed to
contrails in the measurements \citep{kraemer_etal2016}. For the higher values of $n$ at higher temperatures, we can assume a different formation mechanism, i.e. liquid origin ice clouds, which are formed by freezing of pre-existing water cloud droplets \citep[see discussions in][]{kraemer_etal2016}. Generally, it
might be that such high values occur in fresh and undisturbed
nucleation events. Nevertheless, it is very hard to observe such
events, because they are kind of invisible and now-casting of such events
for the aircraft measurements is almost impossible. 
Overall, the model produces realistic results, as compared to the real measurements. Even the strong simplifications do not crucially affect the
results. This makes us confident to use this kind of reduced model
for further investigations. Since the model is relatively simple, 
the analysis is quite easy but still the model simulations are realistic enough 
for qualitative or even quantitative behavior of ice clouds in the cold temperature regime.

%%%%%%%%%%%%%%%%%%%%%%%%%%%%%%%%%%%%%%%%%%%%%%%%%%%%%%%%%%%%%%%%%%%%%%%%%%%%%%%%
%%%%%%%%%%%%%%%%%%%%%%%%%%%%%%%%%%%%%%%%%%%%%%%%%%%%%%%%%%%%%%%%%%%%%%%%%%%%%%%%
\clearpage
\section{Summary}
\label{sec:summary}

We develop a simple ice cloud model on the basis of general moments, using number and mass concentrations as cloud variables; in addition, we use the saturation ratio over ice as thermodynamic control variable for driving relevant ice cloud processes such as nucleation and growth of ice particles. The final model consists of a three dimensional system of nonlinear autonomous ordinary differential equations. The model is then evaluated using the theory of dynamical systems. We show that in the relevant parameter space, there is exactly one equilibrium state. The main feature is here that the problem of finding a root of the vector field can be reformulated to finding a root of a scalar function for the saturation ratio $s$. The existence and uniqueness of the root are then derived by simple analysis. Furthermore, we can reformulate the scalar function into a sequence generating function and this sequence converges towards the unique fixed point by the Banach Fixed Point Theorem, 
giving us approximate formulas for the equilibrium states.
We also determine the quality of the equilibrium state, using linearization around the equilibrium states and then calculating the respective eigenvalues of the Jacobi matrix numerically.
In addition, we prove the existence of a real negative eigenvalue, whereas the other two eigenvalues are either conjugate-complex or both real and negative in the relevant parameter range.
The quality of the equilibrium for a variable but fixed temperature $T$ changes with the parameter $w$ (vertical velocity), i.e. switching from a damped oscillation (stable equilibrium) to an undamped oscillation (unstable equilibrium with attracting limit cycle) and back to the damped regime. 
Thus, we find two supercritical Hopf bifurcations. The bifurcation points in the phase space can be approximated by functions $w(T)$. Using a numerical approach, we derive the periods of the limit cycles in the relevant regime. The range of periods is quite large, producing values between some hundreds of seconds up to several ten thousands of seconds, depending on the environmental conditions.
We find a scaling behavior of the bifurcations in terms of the sedimentation parameter $F$. Using some approximations and properties of the Jacobi matrix, there is a scaling with the square root of $F$, which can be corroborated by the numerical investigations. 
Finally, we find an unexpected scaling property for the periods of the limit cycles. Using empirical scaling functions, the periods of limit cycles can be mapped onto a single function, using a reduced temperature scaled by a reference temperature $T_\text{ref}=\SI{210}{\kelvin}$. The origin of this scaling property is still open.

We compare the qualitative results of the model (i.e. equilibrium states and limit cycle values) with real measurements from aircraft.  The model results agree very well with the measurements, placed in the correct region of the parameter space. Hence, we conclude that our very simplified model seems to be able to represent the main features of ice clouds.  

%\begin{itemize}
%\item development of a simple model for ice cloud research
%\item characterisation of the model
%\item two supercritical Hopf bifurcations
%\item scaling properties
%\item good agreement with real measurements
%\end{itemize}

\clearpage
\section{Conclusion and outlook}
\label{sec:conclusion}

We have seen that even a quite simplified ice cloud model as a 3D dynamical system can produce realistic cloud states. However, the model is simple enough for theoretical analysis of the general behavior and even for determining important processes and their impact on the time evolution of the system. For instance, we clearly see the importance of the sedimentation as a sink process for the whole system, leading to oscillating behavior. The model might be used for several applications and estimations. For instance, we can estimate the expected values of cloud variables for a certain regime of environmental conditions; this might guide the interpretation of measurements or even help for planning measurements in the upper troposphere, in terms of what cloud regimes we would have to expect. The model can also help to constrain the radiative properties of ice clouds. From values of equilibrium states and potential limit cycles, we could derive estimations of possible optical properties or even robust estimates for the expected radiative impact. The simplest approach would be the calculation of optical depths from the occurring values \citep[as, e.g., carried out in][]{joos_etal2009, joos_etal2014}.

Also further reduction seems to be an option. Since the analysis showed that the main dynamics takes place on a quasi 2D manifold (the contraction in one direction is quite strong), the resulting behavior might be emulated by a suitable nonlinear oscillator, e.g., a Stuart-Landau oscillator as a normal form of Hopf bifurcations \citep[][]{kuznetsov1998} with parameters as derived from the analysis, i.e. prescribing periods from the limit cycle calculations; here, the scaling properties would be very useful. This might result into a reduced order model of ice clouds for coarse resolution models, avoiding e.g. the parameterization of nucleation processes, which would require quite small time steps.

Beyond the "simple" box model approach of this ice cloud model, we could think of some conceptual extensions. The ice model might be coupled with transport equations or even with mixing terms \citep[similarly to the work by][]{rosemeier_spichtinger2020} to a spatial extension, leading to reaction-diffusion equations, which can be analyzed in terms of possible structure formation such as e.g. Turing instabilities. Since the model is well characterized as an ODE system, linearization or even the use of weakly nonlinear theory might lead to conceptual understanding of potential emergence of structures.

\clearpage
\noindent
\textbf{Acknowledgements:}\\
We acknowledge support by the Deutsche Forschungsgemeinschaft (DFG)
within the Transregional Collaborative Research Centre TRR301 TPChange
(Project-ID 428312742), project B7. We thank Martina Krämer for some
discussions about ice crystal measurements.  We thank Philipp Reutter
for help with plotting. We thank Friederike Schmid for the helpful
comment about deriving scaling laws from logarithmic plots. PS thanks
Eberhard Bodenschatz for his hospitality at MPI-DS, where parts of
this article were written. 

\medskip

\noindent
\textbf{Author contributions:}\\
PS developed the simple model, HB and PS carried out the mathematical
analysis, HB carried out the comparison with measurements. Both
authors wrote together the manuscript.

%TC:ignore
% % Anhang fuer Buchklasse
\clearpage
\appendix

\section{Derivation of the model equations}
\label{appA}

\subsection{Moment equations}
\label{appA_moments}

We start with the general Boltzmann equation for the mass distribution
$f(m,t,x)$; in the following we will suppress the dependency on
$(t,x)$ in the notation. The equation
\begin{equation}
  \pdv{(\rho f)}{t}+\nabla\vdot\qty(\rho f u)
  +\pdv{(\rho f g)}{m}=\rho h  
\end{equation}
is equipped with a 3D fluid velocity $v$ and a terminal velocity of particles $v_t$
relative to the center of gravity of the flow, i.e. $u=v-e_zv_t$, thus
we find
\begin{equation}
  \pdv{(\rho f)}{t}+\nabla\vdot\qty(\rho f v)
  -\pdv{~}{z}\qty(\rho f v_t)
  +\pdv{~}{m}\qty(\rho f g)=\rho h.
\end{equation}
Using the continuity equation $\rho_t+\nabla\vdot\qty(\rho v)=0$, we
can reformulate the Boltzmann equation in a more convenient form
\begin{equation}
  \pdv{f}{t}+v\vdot\nabla f+\pdv{~}{m}\qty(\rho f g) =
  \frac{1}{\rho}\pdv{\qty(\rho f v_t)}{z}+h,
\end{equation}
where we set $\dot{m}=g$ in a more suggestive notation. Now, we
multiply the whole equation by powers of the mass coordinate $m^k$,
thus 
\begin{equation}
  \pdv{~}{t}\qty(m^kf)+v\vdot\nabla(m^kf)+m^k\pdv{~}{m}\qty(\dot{m}f) =
  \frac{1}{\rho}\pdv{~}{z}\qty(\rho m^kf v_t)+m^kh.
\end{equation}
Now, all terms are integrated over the whole mass range
$(0,\infty)$. For the first two terms we obtain directly the general
moments, i.e.
\begin{equation}
  \int_0^\infty\qty(
%  \pdv{\qty(m^kf)}{t}+v\vdot\nabla(m^kf)
\pdv{~}{t}\qty(m^kf)+v\vdot\nabla(m^kf)
  )\dd m=
  \pdv{\mu_k}{t}+v\vdot\nabla(\mu_k)
\end{equation}
in case of convergence of the integrals.
For the growth term we have to treat different cases. First we
investigate the case $k\in\mathbb{N}$. Then we can use partial
integration in order to obtain
\begin{equation}
  \int_0^\infty m^k\pdv{~}{m}\qty(\dot{m}f)\dd m =
  m^k\dot{m}f(m)\eval_0^\infty-k\int_0^\infty m^{k-1}\dot{m}f\dd m.
\end{equation}
For the convergence of the integrals we have to assume boundary
conditions
\begin{equation}
  \lim_{m\rightarrow \infty}m^\alpha f(m)=0
  ~\forall \alpha\in\mathbb{R}_+~~\text{and}~~
  \lim_{m\rightarrow 0}\dot{m}f(m)<\infty%~\forall k\in\mathbb{N}
\end{equation}
thus we obtain
\begin{equation}
  \int_0^\infty m^k\pdv{~}{m} \qty(\dot{m}f)\dd m =
-k\int_0^\infty m^{k-1}\dot{m}f\dd m.
\end{equation}
This is the growth and evaporation rate for all moments $k\ge 0$.  For
$k=0$ (i.e. in case of the number concentration), we find
\begin{equation}
  \int_0^\infty \pdv{~}{m}\qty(\dot{m}f)\dd m =\dot{m}f(m)\eval_0^\infty
  =-\dot{m}f(0), 
\end{equation}
which is the evaporation rate of the system. Obviously, it strongly
depends on the distribution $f$ and not on its moments. This
investigation is subject of future research.
For the remaining two terms on the right hand side (sedimentation/sources) there is no general
simplification possible, we will treat sedimentation in
Appendix~\ref{appA_sedimentation}.

\subsection{Nucleation for ensemble of solution droplets}
\label{appA_nucleation}

The amount of newly nucleated ice particles from a distribution of
solution droplets $f(r)$ can be derived as follows. Starting with the
probability for freezing a particle of volume $V$ within a time
interval $\Delta t$, i.e.
\begin{equation}
  P=1-\exp(-JV\Delta t)
\end{equation}
we determine the difference of newly nucleated particles as
\begin{equation}
  \Delta n=\int_0^\infty f(r)P(r)\dd r
  =\int_0^\infty f(r)\qty(1-\exp(-JV\Delta t))\dd r. 
\end{equation}
Using an expansion in the small time interval (assuming $\pdv{J}{r}=0$) we obtain
\begin{eqnarray}
  \Delta n=\int_0^\infty f(r)P(r)\dd r
  & = & \int_0^\infty f(r)
        \qty(1-\qty(1-JV\Delta t+\order{\Delta t^2}))\dd r \\
  & = & \int_0^\infty f(r)JV\dd r~\cdot\Delta t+\order{\Delta t^2}\\
  & = & J\frac{4}{3}\pi\int_0^\infty f(r)r^3\dd r~\cdot\Delta t
        +\order{\Delta t^2}\\
  & = & J\frac{4}{3}\pi\mu_{3}[m]~\cdot\Delta t
        +\order{\Delta t^2}%\\
%  & = & J\hat{V}~\cdot\Delta t
%        +\order{\Delta t^2}
\end{eqnarray}
and after dividing by  $\Delta t$ and obtaining the limit $\Delta
t\rightarrow 0$, we find
\begin{equation}
  \dv{n}{t}=J\frac{4}{3}\pi\mu_{3}[m]
\end{equation}
with the general third moment of the underlying size distribution of
the solution droplets. As in former investigations
\citep{spichtinger_gierens2009a}, we use a
lognormal distribution for the radius of aqueous solution droplets of
the form
\begin{equation}
  f_a(r)=\frac{n_a}{\sqrt{2\pi}\log(\sigma_r)r}
  \exp(-\frac{1}{2}\qty(\frac{\log(\frac{r}{r_m})}{\log(\sigma_r)})^2)
\end{equation}
with the modal radius $r_m=r_{\text{sol}}$ of the distribution
(i.e. the modal radius of solution droplets). Since we do not
explicitly treat the aerosol properties, we make the assumptions that
(a) the type and size of the aerosol does not change, and (b) the
amount of nucleated ice crystals is always several orders smaller than
the aerosol concentration $n_a$, thus $n_a$ remains constant. According to the
investigations by \citet{spichtinger_gierens2009a} we specify
$r_m=r_{\text{sol}}=\SI{75e-9}{\metre}$ and the geometric standard
deviation $\sigma_r=1.5$. Using the analytical description of general
moments we can derive the mean volume of a solution droplet as
\begin{equation}
  V_{\text{sol}}=\frac{4}{3}\pi r_\text{sol}^3\cdot c_{\text{sol}},~~
  c_{\text{sol}}=\exp(\frac{9}{2}\left(\log\sigma_r\right)^2)
\end{equation}
The typical mass of a freshly nucleated ice particle can be set to
$m_{\text{nuc}}=\SI{e-16}{\kilo\gram}$.
Finally, we can assume a typical concentration of solution droplets in
the upper troposphere in order of $N_a\sim\SI{300}{\centi\metre^{-3}}=
\SI{3e8}{\metre^{-3}}$. Using the density $\rho$ as obtained by the ideal gas
law from temperature and pressure, we can determine
$n_a=\frac{N_a}{\rho}$.\\
\textbf{Remark:} For comparison with the reference values as
determined by \citet{kaercher_lohmann2002}, we often use an
unrealistically high aerosol number concentration
$N_a=\SI{10000}{\centi\metre^{-3}}$. Note, that the resulting scaling
of the nucleation term only marginally affects the nucleation events
\citep[see extended discussion in][]{spichtinger_etal2023}.

%\medskip
%
In summary, the nucleation term for the number
concentration can be expressed as
\begin{equation}
  \text{Nuc}_n=\pdv{n}{t}\eval_{\text{nucleation}}
  =  n_a V_{\text{sol}}J.                                     
\end{equation}
For the mass concentration rate $\text{Nuc}_q$, we multiply $\text{Nuc}_n$ by the constant
mass $m_{\text{nuc}}$. The nucleation rate is determined 
\citep[cf.][]{spichtinger_etal2023}
by 
\begin{equation}
    J(s,T)=J_0\exp(p_{\text{1e}}(T)(s-p_2(T))),~~
    J_0=\SI{e16}{\metre^{-3}\second^{-1}}
\end{equation}
and polynomials 
\begin{equation}
    p_1(T)=a_0+a_1T,~ p_{\text{1e}}(T)=\log(10)p_1(T),~ 
    p_2(T)=a_{s2}T^2+a_{s1}T+a_{s0}
\end{equation}
with coefficients
\begin{equation}
    a_0=-50.4085,~
    a_1=\SI{0.9368}{\per\kelvin}
 \end{equation}
   and
\begin{equation}    
    a_{s2}=\SI{-1.36989e-5}{\kelvin^{-2}},~
    a_{s1}=\SI{0.00228}{\kelvin^{-1}},~
    a_{s0}=1.67469.
\end{equation}

\subsection{Parameters and functions}
\label{appA_functions}

The ideal gas constants are given by 
\begin{equation}
   R_a\approx
  \SI{287.058}{\joule\per\kilogram\per\kelvin},~~
  %R_{\ce{H2O}}=
  R_v\approx\SI{461.553}{\joule\per\kilogram\per\kelvin},
\end{equation}
the acceleration by gravity is determined as $g=\SI{9.81}{\metre\per\second^2}$. For the latent heat of sublimation we use the formulation by \citet{murphy_koop2005}
\begin{equation}
    L(T)=L_{\text{mol}}(T)/M_{\text{mol},v},
    L_{\text{mol}}(T)=l_0+l_1T+l_2T^2+l_3\exp(-(T/T_l)^2)
\end{equation}
with coefficients
\begin{equation}
\begin{array}{ccc}
    l_0=\SI{46782.5}{\joule\,\mol^{-1}}, &
    l_1=\SI{35.8925}{\joule\,\mol^{-1}\kelvin^{-1}}, &
    l_2=\SI{-0.07414}{\joule\,\mol^{-1}\kelvin^{-2}}, \\
    l_3=\SI{541.5}{\joule\,\mol^{-1}},&
    T_l=\SI{123.75}{\kelvin},&
    M_{\text{mol},v}=\SI{18.01528e-3}{\kilo\gram\,\mol^{-1}}.
\end{array}
%    l_0=\SI{46782.5}{\joule\mol^{-1}},~
%    l_1=\SI{35.8925}{\joule\mol^{-1}\kelvin^{-1}},~
%    l_2=\SI{-0.07414}{\joule\mol^{-1}\kelvin^{-2}},~
%    l_3=\SI{541.5}{\joule\mol^{-1}},~
%    T_l=\SI{123.75}{\kelvin},~
%    M_{\text{mol},v}=\SI{18.01528e-3}{\kilo\gram\mol^{-1}}.
\end{equation}
%We assume a constant latent heat of sublimation of water $L=\SI{2.83e6}{\joule\,\kilo\gram^{-1}\kelvin^{-1}}$. 
The saturation vapor pressure over hexagonal ice \citep[cf.][]{murphy_koop2005} is expressed as 
\begin{equation}
    p_{\text{si}}(T)=
    p_{\text{unit}}\exp(b_0+b_1/T+b_2\log(T/T_{\text{unit}})+b_3T),~
    p_{\text{unit}}=\SI{}{\pascal},~
    T_{\text{unit}}=\SI{}{\kelvin}
\end{equation}
with
\begin{equation}
    b_0=9.550426,~b_1=-\SI{5723.265}{\kelvin},~
    b_2=3.53068,~b_3=\SI{-0.00728332}{\per\kelvin}.
\end{equation}
The heat conduction of air \citep[cf.][]{dixon2007} is given by 
\begin{equation}
    K_T(T)=\frac{a_KT^{b_K}}{T+T_K\cdot 10^{\frac{c_K}{T}}}
\end{equation}
with coefficients
\begin{equation}
    a_K=\SI{0.002646}{\watt\,\metre^{-1}\kelvin^{1-b_K}},
    ~~b_K=\frac{3}{2}, ~~T_K=\SI{245}{\kelvin},
    ~~c_K=\SI{-12}{\kelvin}. 
\end{equation}

\subsection{Growth for an ensemble of spherical ice particles}
\label{appA_growth}

Using the description of the mass growth rate for a single particle of
mass $m$, i.e.,
\begin{equation}
  \dot{m}=4\pi C D_v f_kG_v(s-1)f_v,
\end{equation}
the averaged growth rate for the first moment, i.e. the mass
concentration $q=\mu_1[m]$, is then given by 
\begin{equation}
  \text{Dep}_q=\int_0^\infty  \dot{m}f(m)\dd m =
  \int_0^\infty 4\pi C D_v f_kG_v(s-1)f_v f(m)\dd m.
\end{equation}
We now apply several simplifications for deriving a simplified version
of the growth rate. We neglect the corrections for the kinetic regime
($f_k=1$) and for ventilation ($f_v=1$). In addition, we assume
spherical particles, which in turn leads to a capacity factor
$C(m)=r=c\,m^\frac{1}{3}$ (see
Equation~\eqref{eq:mass_radius_sphere}). Finally, we fix the type of the
mass distribution as a lognormal distribution (see
Equation~\eqref{eq:lognormal_mass}), thus we can use the analytical
description of the general moments. This leads to the following
representation of the growth term as
\begin{eqnarray}
  \text{Dep}_q&=&\int_0^\infty f(m)4\pi C D_vf_k G_v(s-1)f_v\,\dd m\\
  &\approx&
            4\pi G_vD_v\int_0^\infty f(m) C(m)\dd m\cdot (s-1) \\
  &=& 4\pi G_vD_v\, c\, r_0^{-\frac{1}{9}}
      n^{\frac{2}{3}}q^\frac{1}{3} (s-1)\\
  &=& B_q^*\, c\, r_0^{-\frac{1}{9}}
      n^{\frac{2}{3}}q^\frac{1}{3} (s-1) = B_qn^{\frac{2}{3}}q^\frac{1}{3} (s-1).
\end{eqnarray}

\subsection{Sedimentation for an ensemble of particles}
\label{appA_sedimentation}

For the sedimentation of the averaged quantities, we rewrite the
sedimentation part in terms of a transport equation. We investigate
the vertical sedimentation flux of a quantity $\psi$ (i.e. a general
moment $\mu_k$), i.e. $j_{\psi}=\rho \psi v_{\psi}$ with the
respective vertical velocity $v_{\psi}$, thus the advection
equation can be written as
\begin{equation}
  \pdv{(\rho \psi)}{t} = \pdv{j_{\psi}}{z}.
\end{equation}
For the averaged variables $n=\mu_0,q=\mu_1$, we define the fluxes as 
\begin{eqnarray}
  j_n& \coloneqq & \int_0^\infty \rho f(m)v_t(m)\dd m
                     \stackrel{!}{=}  \rho n v_n \\
  j_q& \coloneqq & \int_0^\infty \rho f(m)mv_t(m)\dd m
                     \stackrel{!}{=}  \rho q v_q 
\end{eqnarray}
expressed by averaged velocities
\begin{eqnarray}
  v_n & = & \frac{1}{n}\int_0^\infty f(m)v_t(m)\dd m\\
  v_q & = & \frac{1}{q}\int_0^\infty f(m)mv_t(m)\dd m.
\end{eqnarray}
Note that from the formulation of the sedimentation as a transport
equation it is obvious that for positive quantities $n,q$ in the time
evolution they will never take a value $=0$.

\clearpage
\section{Derivation of the scalar function for the equilibrium states}
\label{appB}
In this section, we present the calculations for the derivation of the scalar function $f(x)$ as defined in Equation~\eqref{eq:scalar_s} whose roots are in one-to-one correspondence with the equilibrium points of the system. In fact, the roots of $f(x)$ are precisely the values $s_0$ of the supersaturation at an equilibrium point and the values of the number and mass concentrations at the equilibrium point can be uniquely calculated from $s_0$.
We first observe that
\begin{equation}
  n^{\frac{2}{3}}q^{\frac{1}{3}}=n\bar{m}^{\frac{1}{3}},
  n^{\frac{1}{3}}q^{\frac{2}{3}}=n\bar{m}^{\frac{2}{3}}.
\end{equation}

We use the condition for the equilibrium state (i.e. $\dv{\psi}{t}=0$) for deriving expressions for $n$ and $\bar{m}$. We start with the equation for $n$: 
\begin{eqnarray}
  A_n\exp(p_{1e}(s-p_2))
  -FC_nn^{\frac{1}{3}}q^{\frac{2}{3}}=0
  &\Leftrightarrow&
  A_n\exp(p_{1e}(s-p_2)) = FC_nn\bar{m}^{\frac{2}{3}}\\
                                           &\Leftrightarrow& 
  n=\frac{A_n}{FC_n}\exp(p_{1e}(s-p_2))\bar{m}^{-\frac{2}{3}} 
\end{eqnarray}
Then we find an expression for $\bar{m}$ by dividing 
Equation~\eqref{eq:3D_system_q_reduced} by $q\neq 0$
\begin{eqnarray}
  B_qn_i^{\frac{2}{3}}q_i^{-\frac{2}{3}}(S_i-1)-
  FC_qn^{-\frac{2}{3}}q^{\frac{2}{3}} =0 
  &\Leftrightarrow &
  B_q\bar{m}^{-\frac{2}{3}}(s-1) = FC_q \bar{m}^{\frac{2}{3}}\\
    &\Leftrightarrow &
  B_q(s-1) = FC_q \bar{m}^{\frac{4}{3}}\\
  &\Leftrightarrow &
  \bar{m}^{\frac{4}{3}}= \frac{B_q}{FC_q}(s-1)\\
  &\Leftrightarrow & 
  \bar{m}=\left(\frac{B_q}{FC_q}(s-1)\right)^{\frac{3}{4}} 
\end{eqnarray}
This produces the equations for the values of $n,q$ at the equilibrium states:
\begin{equation}
\label{eq:n_equilibrium}
  n=\frac{A_n}{FC_n}\exp(p_{1e}(s-p_2))
  \left(\frac{B_q}{FC_q}(s-1)\right)^{-\frac{1}{2}}
\end{equation}
and 
\begin{equation}
\label{eq:q_equilibrium}
  q=n\bar{m}=\frac{A_n}{FC_n}\exp(p_{1e}(s-p_2))
  \left(\frac{B_q}{FC_q}(s-1)\right)^{\frac{1}{4}}.
\end{equation}
The equilibrium state of the saturation ratio can be obtained by 
\begin{eqnarray}
  A_ss-B_{s2}n_i\bar{m}_i^\frac{1}{3}(s-1) = 0 &\Leftrightarrow&\\
  A_ss-B_{s2}\frac{A_n}{FC_n}\exp(p_{1e}(s-p_2))
  \left(\frac{B_q}{FC_q}(s-1)\right)^{-\frac{1}{4}}(s-1) =0
\end{eqnarray}
This last equation is used for the definition of the scalar function $f(x)$ for $x=s$, i.e., 
\begin{equation}
  f(x)=ax-\dfrac{b\left(x-1\right)\exp(c\left(x-x_0\right))}
  {\sqrt[4]{d\left(x-1\right)}}
\end{equation}
with the coefficients
\begin{equation}
  a=A_s, ~~
  b=B_{s2}\frac{A_n}{FC_n}, ~~
  c=p_{1e},~~
  d=\frac{B_q}{FC_q},~~
  x_0=p_2.
\end{equation}
Rearranging the equation and the coefficients leads to the definition of $f(x)$ as in Equation~\eqref{eq:scalar_s} and if $s_0$ is the supersaturation at any equilibrium point of the ODE system, it satisfies 
$f(s_0)=0$. Vice versa, given an arbitrary root $s_0$ of the function $f(x)$, we set the number concentration $n_0$ and mass concentration $q_0$ to be given by Equations~\eqref{eq:n_equilibrium} and~\eqref{eq:q_equilibrium} for $s = s_0$; then $x= (n_0, q_0, s_0)^T$ is an equilibrium point of the ODE system which can be verified by a short calculation.

\clearpage
\section{Estimates for the contraction}
\label{appC}
In the following, we show that the function $h(x)$ as defined in Equation~\eqref{eq:h_contraction} satisfies 
\begin{equation}
    h([s_\text{min},s_\text{max}])
\subset [s_\text{min},s_\text{max}]
\end{equation}
for $s_\text{min} = 1.37$ and $s_\text{max} = 1.59$.
First, we estimate the ranges of functions $\sigma$ and $p_{1e}$,
respectively. For the relevant ranges of temperature ($190\le T\le
\SI{240}{\kelvin}$) and pressure ($200\le p\le
\SI{300}{\hecto\pascal}$), we find
\begin{equation}
  \sigma_1\le \sigma\le \sigma_2,~~\pi_1\le \frac{1}{p_{1e}(T)}\le \pi_2
\end{equation}
for parameters
\begin{equation}
  \sigma_1 = 1.412801,~\sigma_2 =  1.583606,~~
  \pi_1 = 0.002489,~\pi_2 =  0.003404.
\end{equation}
Next, we would like to estimate the term
\begin{equation}
  \gamma_1\le
  \frac{1}{p_{1e}}\qty(\log(w)+\frac{3}{4}\log(F))
  \le \gamma_2
  \label{eq:estimate_gammas}
\end{equation}
by lower and upper bounds $\gamma_1$, $\gamma_2$
for the relevant ranges
\begin{equation}
0.01=F_\text{min}\le F\le F_\text{max}=1,~~ 
\SI{0.0005}{\metre\per\second}=w_\text{min}\le w\le w_\text{max}=\SI{2}{\metre\per\second}.
\end{equation}
Therefore, we set
\begin{eqnarray}
%  \gamma_\text{max}
  \gamma_2
  & = & 0.002360 ~ \ge~
  \pi_2\qty(\log(w_\text{max})+\frac{3}{4}\log(F_\text{max})) \\
%  \gamma_\text{min}
  \gamma_1
  & = &  -0.037631 ~ \le ~
  \pi_2\qty(\log(w_\text{min})+\frac{3}{4}\log(F_\text{min})),
\end{eqnarray}
which then satisfy Equation~\eqref{eq:estimate_gammas}.
Thus, we find the estimation 
\begin{equation}
  1.375170 = \sigma_1+\gamma_1\le
  \sigma+\frac{1}{p_{1e}}\qty(\log(w)+\frac{3}{4}\log(F))\le
  \sigma_2+\gamma_2 = 1.585966
\end{equation}
and hence get
\begin{equation}
  h(x)=\kappa_0+\frac{1}{p_{1e}}\log(\frac{x}{(x-1)^{\frac{3}{4}}})
\end{equation}
for some (variable) $\kappa_0\in[ 1.375170 , 1.585966]$.
Let us now specify the following interval
\begin{equation}
  [s_\text{min},s_\text{max}]
  ~\text{with}~s_\text{min}=1.37,~  s_\text{max}=1.59.
\end{equation}
%
%In the following, we will show that the function $h$ is a contraction on this interval and that it is the maximal interval for possible values of the saturation ratio $s_0$ at an equilibrium point of the system, i.e. for $s_0$ we always have 
%\begin{equation}
%    s_0\in [s_\text{min},s_\text{max}] = [1.37, 1.59].
%\end{equation}
Calculating all possible values of
$\log(\frac{x}{(x-1)^{\frac{3}{4}}})$ on the interval
$[s_\text{min},s_\text{max}]$ we obtain
\begin{equation}
  0.002139\le\frac{1}{p_{1e}}\log(\frac{x}{(x-1)^{\frac{3}{4}}})\le 0.003670
\end{equation}
such that
\begin{eqnarray}
  h\qty([s_\text{min},s_\text{max}]) &\subset &
  [ 1.375170 , 1.585966] + [0.002139, 0.003670] \\
  &= &[1.377309, 1.589636]\subset
  [s_\text{min},s_\text{max}].
\end{eqnarray}

\clearpage
\section{Full example simulations}
\label{appD}

We show all variables for the example simulations in Section~\ref{sec:equilibrium_states}, i.e. for $T=\SI{225}{\kelvin}$, $p=\SI{200}{\hecto\pascal}$, $F=1$. 
\begin{figure}[h]
  \centering
  \includegraphics[width=0.33\linewidth]
  {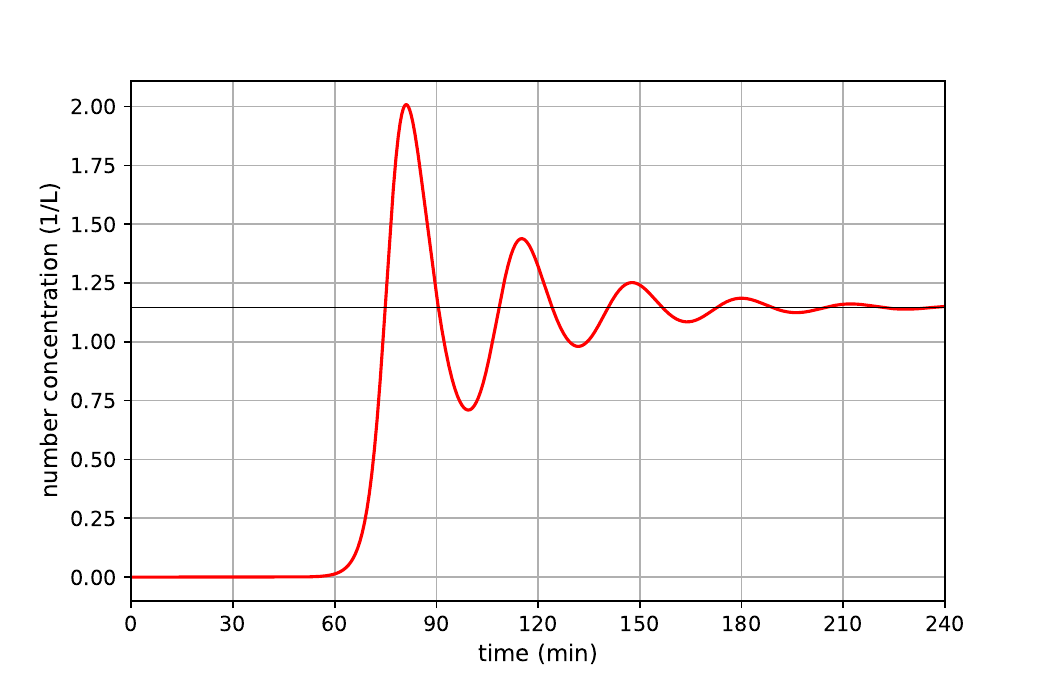}%    
  \includegraphics[width=0.33\linewidth]
  {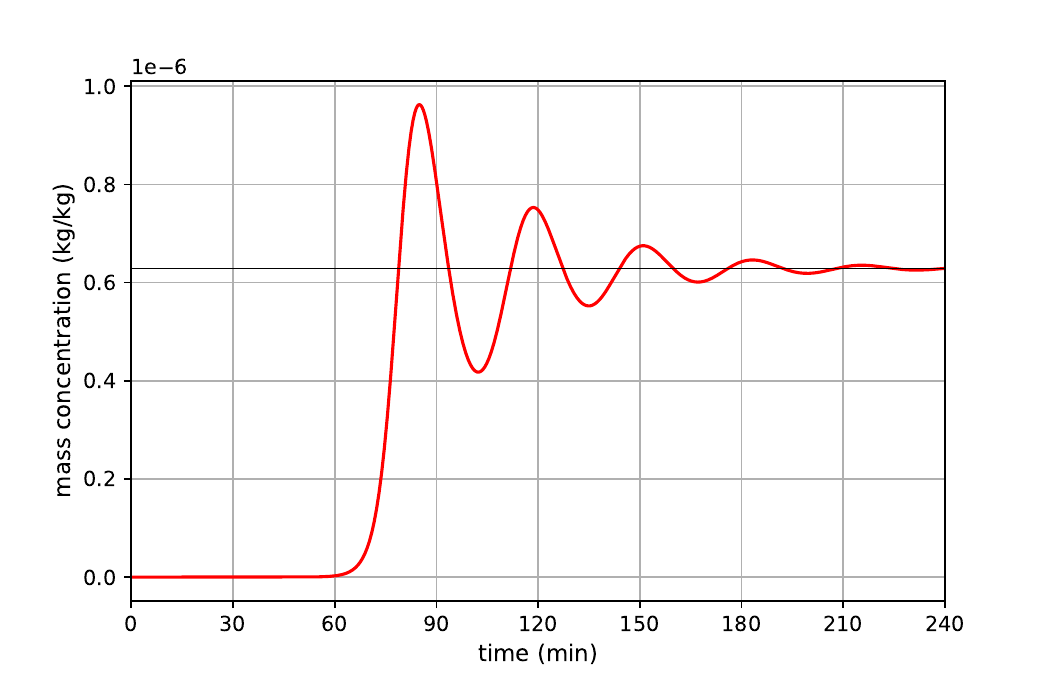}%    
   \includegraphics[width=0.33\linewidth]
  {model2_si_w001_no.pdf}%    
  \caption{Examples of solutions for $w=\SI{0.01}{\metre\per\second}$. 
    Left: number concentration, middle: mass concentration, right: saturation ratio, red line; black lines indicate the equilibrium values.}
 \label{fig:examples_solutions_full001}
\end{figure}
\begin{figure}[h]
  \centering
  \includegraphics[width=0.33\linewidth]
  {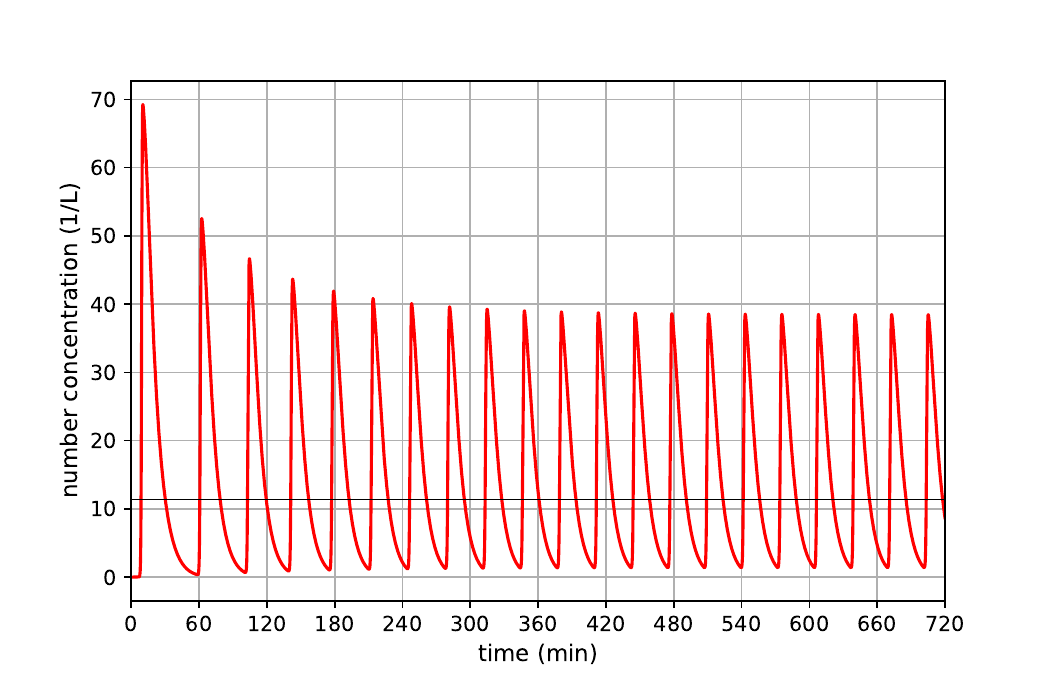}%    
  \includegraphics[width=0.33\linewidth]
  {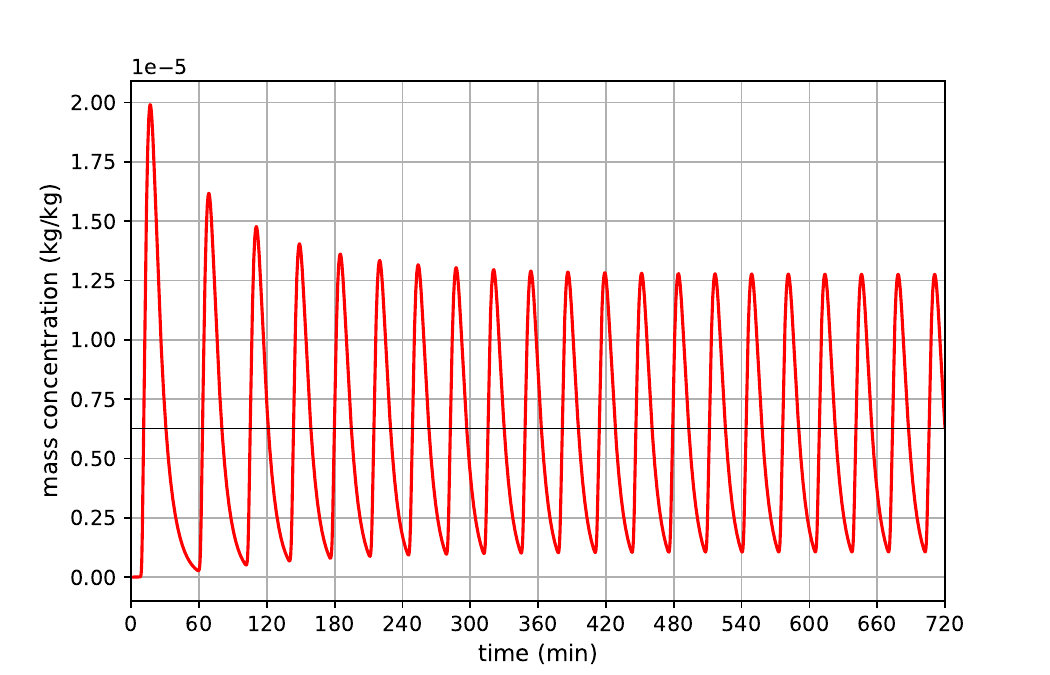}%    
   \includegraphics[width=0.33\linewidth]
  {model2_si_w010_no.pdf}%    
  \caption{Examples of solutions for $w=\SI{0.10}{\metre\per\second}$. 
    Left: number concentration, middle: mass concentration, right: saturation ratio, red line; black lines indicate the equilibrium values.}
 \label{fig:examples_solutions_full010}
\end{figure}
\begin{figure}[h]
  \centering
  \includegraphics[width=0.33\linewidth]
  {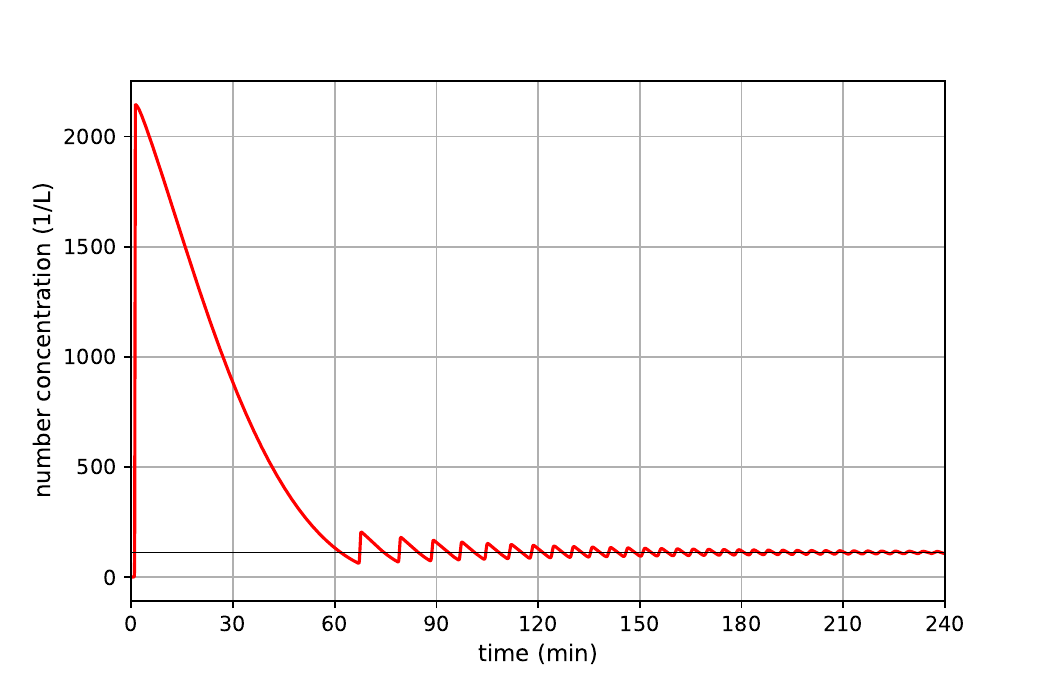}%    
  \includegraphics[width=0.33\linewidth]
  {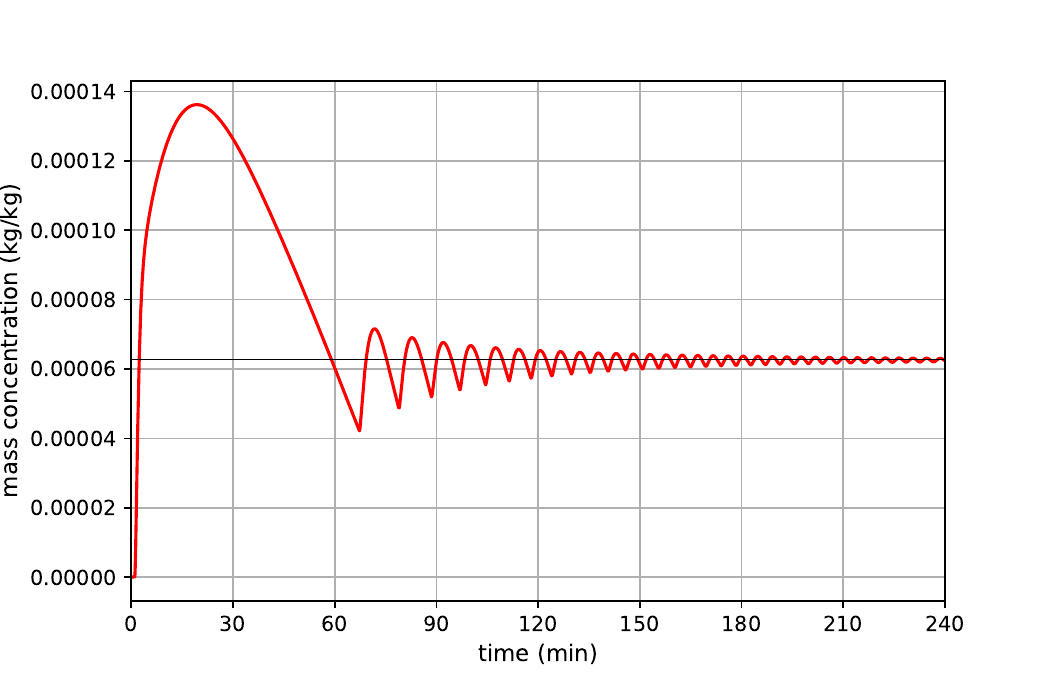}%    
   \includegraphics[width=0.33\linewidth]
  {model2_si_w100_no.pdf}%    
  \caption{Examples of solutions for $w=\SI{1.00}{\metre\per\second}$. 
    Left: number concentration, middle: mass concentration, right: saturation ratio, red line; black lines indicate the equilibrium values.}
 \label{fig:examples_solutions_full100}
\end{figure}
%

% Bibliographie ausgeben
\clearpage
\printbibliography

@Article{joos_etal2014,
AUTHOR = {Joos, H. and Spichtinger, P. and Reutter, P. and Fusina, F.},
TITLE = {Influence of heterogeneous freezing on the microphysical and radiative properties of orographic cirrus clouds},
JOURNAL = {Atmospheric Chemistry and Physics},
VOLUME = {14},
YEAR = {2014},
NUMBER = {13},
PAGES = {6835--6852},
URL = {https://www.atmos-chem-phys.net/14/6835/2014/},
DOI = {10.5194/acp-14-6835-2014}
}

@Article{spichtinger_kraemer2013,
AUTHOR = {Spichtinger, P. and Kr\"amer, M.},
TITLE = {Tropical tropopause ice clouds: a dynamic approach to the mystery of low crystal numbers},
JOURNAL = {Atmospheric Chemistry and Physics},
VOLUME = {13},
YEAR = {2013},
NUMBER = {19},
PAGES = {9801--9818},
URL = {https://www.atmos-chem-phys.net/13/9801/2013/},
DOI = {10.5194/acp-13-9801-2013}
}

@Article{joos_etal2009,
AUTHOR = {Joos, H. and Spichtinger, P. and Lohmann, U.},
TITLE = {Orographic cirrus in a future climate},
JOURNAL = {Atmospheric Chemistry and Physics},
VOLUME = {9},
YEAR = {2009},
NUMBER = {20},
PAGES = {7825--7845},
URL = {https://www.atmos-chem-phys.net/9/7825/2009/},
DOI = {10.5194/acp-9-7825-2009}
}

@Article{spichtinger_gierens2009a,
AUTHOR = {Spichtinger, P. and Gierens, K. M.},
TITLE = {Modelling of cirrus clouds – Part 1a: Model description and validation},
JOURNAL = {Atmospheric Chemistry and Physics},
VOLUME = {9},
YEAR = {2009},
NUMBER = {2},
PAGES = {685--706},
URL = {https://www.atmos-chem-phys.net/9/685/2009/},
DOI = {10.5194/acp-9-685-2009}
}

@Article{baumgartner_spichtinger2018,
author="Baumgartner, Manuel and Spichtinger, Peter",
title="Diffusional growth of cloud particles: existence and uniqueness of solutions",
journal="Theoretical and Computational Fluid Dynamics",
year="2018",
month="Feb",
day="01",
volume="32",
number="1",
pages="47--62",
issn="1432-2250",
doi="10.1007/s00162-017-0437-x",
url="https://doi.org/10.1007/s00162-017-0437-x"
}

@article{baumgartner_spichtinger2017,
	doi = {10.1515/mcwf-2017-0004},
	year = 2017,
	month = {nov},
	publisher = {Walter de Gruyter {GmbH}},
	volume = {3},
	number = {1},
	author = {Manuel Baumgartner and Peter Spichtinger},
	title = {Local Interactions by Diffusion between Mixed-Phase Hydrometeors: Insights from Model Simulations},
	journal = {Mathematics of Climate and Weather Forecasting}
}

@article{spichtinger_cziczo2010,
	doi = {10.1029/2009jd012168},
	year = 2010,
	month = {jul},
	publisher = {Wiley-Blackwell},
	volume = {115},
	number = {D14},
	author = {P. Spichtinger and D. J. Cziczo},
	title = {Impact of heterogeneous ice nuclei on homogeneous freezing events in cirrus clouds},
	journal = {Journal of Geophysical Research}
}

@article{hall_pruppacher1976,
Author = {Hall, W.D. and Pruppacher, H.R.},
Title = {Survival of Ice Particles Falling From Cirrus Clouds in Subsaturated Air},
Journal = {Journal of the  Atmospheric Sciences},
Year = {1976},
Volume = {33},
Number = {10},
Pages = {1995-2006},
DOI = {10.1175/1520-0469(1976)033<1995:TSOIPF>2.0.CO;2},
ISSN = {0022-4928}
}

@Book{dixon2007,
  author = 	 {Dixon, J.},
  title = 	 {The Shock Absorber Handbook},
  publisher = 	 {Wiley},
  year = 	 {2007},
  pages=         {432},
  isbn=          {978-0-470-51020-9}
}

@article{koop_etal2000,
  author = {Koop, T. and Luo, B. and Tsias, A. and Peter, T.},
  journal = {Nature},
  pages = {611--614},
  title = {Water activity as the determinant for homogeneous ice nucleation in aqueous solutions},
  volume = {406},
  year = {2000}
}

@Article{murphy_koop2005,
  author    ={Murphy, D. and Koop, T.},
  title     ={Review of the vapour pressure of ice and supercooled water for atmospheric applications},
  year      =  {2005},
  journal   ={Quarterly Journal of the Royal Meterorological Society},
  volume    ={131},
  pages     ={1539-1565}
}

@Article{chen_etal2000,
  author = 	 {Chen, T. and Rossow, W. and Zhang, Y.},
  title = 	 {Radiative effects of cloud-type variations},
  journal = 	 {J. Climate},
  year = 	 2000,
  volume =	 13,
  pages =	 {264-286}
}

@Article{kay_etal2006,
  author = 	 {Kay, J. E. and  Baker, M. and Hegg, D.},
  title = 	 {Microphysical and dynamical controls on cirrus cloud optical depth distributions},
  journal = 	 {Journal of Geophysical Research},
  year = 	 {2006},
  OPTkey = 	 {},
  volume = 	 {111},
  OPTnumber = 	 {},
  pages = 	 {D24205},
  OPTmonth = 	 {},
  OPTnote = 	 {},
  OPTannote = 	 {},
  doi=           {10.1029/2005JD006916}
}

@article{kaercher_lohmann2002,
author = {Bernd K{\"a}rcher and Ulrike Lohmann},
title = {A parameterization of cirrus cloud formation: Homogeneous freezing of supercooled aerosols},
journal = {Journal of Geophysical Research: Atmospheres},
volume = {107},
number = {D2},
year = {2002},
doi = {10.1029/2001JD000470}
}

@article{spreitzer_etal2017,
author = {Spreitzer, Elisa Johanna and Marschalik, M. Patrik and Spichtinger, Peter},
title = {Subvisible cirrus clouds -- a dynamical system approach},
journal = {Nonlinear Processes in Geophysics},
volume = {24},
year = {2017},
number = {3},
pages = {307--328},
doi = {10.5194/npg-24-307-2017}
}

@Article{fu_liou1993,
  author = 	 {Fu,Q. and Liou,K.N.},
  title = 	 {Parameterization of the Radiative Properties of Cirrus Clouds},
  journal = 	 {J. Atmos. Sci.},
  year = 	 1993,
  volume =	 50,
  number =	 13,
  pages =	 {2008-2025},
  doi =          {10.1175/1520-0469(1993)050<2008:POTRPO>2.0.CO;2}
}

@article{kraemer_etal2016,
Author = {Kr\"amer, Martina and Rolf, Christian and Luebke, Anna and
                  Afchine, Armin and Spelten, Nicole and Costa, Anja
                  and Meyer, Jessica and Zoeger, Martin and Smith,
                  Jessica and Herman, Robert L. and Buchholz, Bernhard
                  and Ebert, Volker and Baumgardner, Darrel and
                  Borrmann, Stephan and Klingebiel, Marcus and
                  Avallone, Linnea},
Title = {A microphysics guide to cirrus clouds - Part 1: Cirrus types},
Journal = {Atmospheric Chemistry and Physics},
Year = {2016},
Volume = {16},
Number = {5},
Pages = {3463-3483},
DOI = {10.5194/acp-16-3463-2016}
}

@article{wernli_etal2016,
Author = {Wernli, Heini and Boettcher, Maxi and Joos, Hanna and Miltenberger,
   Annette K. and Spichtinger, Peter},
Title = {A trajectory-based classification of ERA-Interim ice clouds in the
   region of the North Atlantic storm track},
Journal = {Geophysical Research Letters},
Year = {2016},
Volume = {43},
Number = {12},
Pages = {6657-6664},
Month = {JUN 28},
DOI = {10.1002/2016GL068922}
}

@article{zhang_etal1999,
Author = {Zhang, Y and Macke, A and Albers, F},
Title = {Effect of crystal size spectrum and crystal shape on stratiform cirrus
   radiative forcing},
Journal = {Atmospheric Research},
Year = {1999},
Volume = {52},
Number = {1-2},
Pages = {59-75},
Month = {AUG},
DOI = {10.1016/S0169-8095(99)00026-5}
}

@article{gallo_etal2019,
Author = {Gallo, Paola and Loerting, Thomas and Sciortino, Francesco},
Title = {Supercooled water: A polymorphic liquid with a cornucopia of behaviors},
Journal = {Journal of Chemical Physics},
Year = {2019},
Volume = {151},
Number = {21},
Month = {DEC 7},
DOI = {10.1063/1.5135706}
}

@Article{kraemer_etal2020,
  author = 	 {Krämer, M. and Rolf, C. and Spelten, N. and  Afchine, A. and
                  Fahey, D. and Jensen, E. and Khaykin, S. and Kuhn,
                  T. and Lawson, P. and Lykov, A. and Pan, L. L. and Riese, M. and
                  Stroh, F. and Thornberry, T. and Wolf, V. and Woods, S. and
                  Spichtinger, P. and Quaas, J. and Sourdeval, O.},
  title = 	 {A Microphysics Guide to Cirrus - Part II:
                  Climatologies of Clouds and Humidity from
                  Observations},
  journal = 	 {Atmospheric Chemistry and Physics Disscussions},
  year = 	 {2020},
  doi =          {10.5194/acp-2020-40},
  OPTkey = 	 {},
  OPTvolume = 	 {},
  OPTnumber = 	 {},
  OPTpages = 	 {},
  OPTmonth = 	 {},
  OPTnote = 	 {},
  OPTannote = 	 {}
}

@book{book_pruppacher_klett2010,
  title = {Microphysics of Clouds and Precipitation},
  author = {Pruppacher, Hans R. and Klett, James D.},
  year = {2010},
  ISBN = {0-7923-4211-9},
  series = {Atmospheric and Oceanographic Sciences Library},
  volume = {18},
  publisher = {Kluwer Academic Publishers},
  address = {Dordrecht}
}

@article{kajikawa_heymsfield1989,
Author = {Kajikawa, M and Heymsfield, AJ},
Title = {Aggregation of Ice Crystals in  Cirrus},
Journal = {Journal of the Atmospheric Sciences},
Year = {1989},
Volume = {46},
Number = {20},
Pages = {3108-3121},
Month = {OCT 15},
DOI = {10.1175/1520-0469(1989)046<3108:AOICIC>2.0.CO;2},
ISSN = {0022-4928}
}

@article{koehler1936,
	author = {K\"ohler, H},
	journal = {Transactions of the Faraday Society},
	pages = {1152--1161},
	title = {The nucleus in and the growth of hygroscopic droplets},
	volume = {32},
	year = {1936}
}

@article{khain_etal2015,
Author = {Khain, A. P. and Beheng, K. D. and Heymsfield, A. and Korolev, A. and
   Krichak, S. O. and Levin, Z. and Pinsky, M. and Phillips, V. and
   Prabhakaran, T. and Teller, A. and van den Heever, S. C. and Yano, J.
   -I.},
Title = {Representation of microphysical processes in cloud-resolving models:
   Spectral (bin) microphysics versus bulk parameterization},
Journal = {Reviews of Geophysics},
Year = {2015},
Volume = {53},
Number = {2},
Pages = {247-322},
Month = {JUN},
DOI = {10.1002/2014RG000468}
}

@Article{grulich_etal2021,
title = {Automatic shape detection of ice crystals},
journal = {Journal of Computational Science},
volume = {54},
pages = {101429},
year = {2021},
issn = {1877-7503},
doi = {https://doi.org/10.1016/j.jocs.2021.101429},
url = {https://www.sciencedirect.com/science/article/pii/S1877750321001137},
author = {Lucas Grulich and Ralf Weigel and Andreas Hildebrandt and Michael Wand and Peter Spichtinger}
}

@Article{spichtinger_etal2023,
AUTHOR = {Spichtinger, P. and Marschalik, P. and Baumgartner, M.},
TITLE = {Impact of formulations of the homogeneous nucleation rate on ice nucleation events in cirrus},
JOURNAL = {Atmospheric Chemistry and Physics},
VOLUME = {23},
YEAR = {2023},
NUMBER = {3},
PAGES = {2035--2060},
URL = {https://acp.copernicus.org/articles/23/2035/2023/},
DOI = {10.5194/acp-23-2035-2023}
}

@book{lamb_verlinde2011,
place={Cambridge},
title={Physics and Chemistry of Clouds},
publisher={Cambridge University Press},
author={Lamb, Dennis and Verlinde, Johannes},
year={2011},
doi={10.1017/CBO9780511976377}
}

@article {bailey_hallett2009,
      author = "Matthew P. Bailey and John Hallett",
      title = "A Comprehensive Habit Diagram for Atmospheric Ice
                  Crystals: Confirmation from the Laboratory, AIRS II,
                  and Other Field Studies",
      journal = "Journal of the Atmospheric Sciences",
      year = "2009",
      publisher = "American Meteorological Society",
      address = "Boston MA, USA",
      volume = "66",
      number = "9",
      doi = "10.1175/2009JAS2883.1",
      pages=      "2888 - 2899",
      url = "https://journals.ametsoc.org/view/journals/atsc/66/9/2009jas2883.1.xml" 
}

@book{book_walter_1998,
  title={Ordinary differential equations},
  author={Walter, Wolfgang},
  volume={182},
  pages={XI,384},
  year={1998},
  publisher={Springer Science \& Business Media},
  series={Graduate Texts in Mathematics},
  doi={10.1007/978-1-4612-0601-9}
}

@article{hanke_porz2020,
author = {Hanke, Martin and Porz, Nikolas},
title = {Unique Solvability of a System of Ordinary Differential Equations Modeling a Warm Cloud Parcel},
journal = {SIAM Journal on Applied Mathematics},
volume = {80},
number = {2},
pages = {706-724},
year = {2020},
doi = {10.1137/19M1267751}
}

@Article{         harris2020,
 title         = {Array programming with {NumPy}},
 author        = {Charles R. Harris and K. Jarrod Millman and St{\'{e}}fan J.
                 van der Walt and Ralf Gommers and Pauli Virtanen and David
                 Cournapeau and Eric Wieser and Julian Taylor and Sebastian
                 Berg and Nathaniel J. Smith and Robert Kern and Matti Picus
                 and Stephan Hoyer and Marten H. van Kerkwijk and Matthew
                 Brett and Allan Haldane and Jaime Fern{\'{a}}ndez del
                 R{\'{i}}o and Mark Wiebe and Pearu Peterson and Pierre
                 G{\'{e}}rard-Marchant and Kevin Sheppard and Tyler Reddy and
                 Warren Weckesser and Hameer Abbasi and Christoph Gohlke and
                 Travis E. Oliphant},
 year          = {2020},
 month         = sep,
 journal       = {Nature},
 volume        = {585},
 number        = {7825},
 pages         = {357--362},
 doi           = {10.1038/s41586-020-2649-2},
 publisher     = {Springer Science and Business Media {LLC}},
 url           = {https://doi.org/10.1038/s41586-020-2649-2}
}

@Book{kondepudi_prigogine2014,
  author = 	 {Kondepudi, Dilip and Prigogine, Ilya},
  ALTeditor = 	 {},
  title = 	 {Modern Thermodynamics: From Heat Engines to Dissipative Structures},
  publisher = 	 {John Wiley \& Sons},
  year = 	 {2014},
  pages =        {560},
  isbn=          {9781118371817},
  OPTkey = 	 {},
  OPTvolume = 	 {},
  OPTnumber = 	 {},
  OPTseries = 	 {},
  OPTaddress = 	 {},
  edition = 	 {2nd},
  OPTmonth = 	 {},
  OPTnote = 	 {},
  OPTannote = 	 {}
}

@Article{gallo_etal2016,
author = {Gallo, Paola and Amann-Winkel, Katrin and Angell, Charles Austen and Anisimov, Mikhail Alexeevich and Caupin, Frédéric and Chakravarty, Charusita and Lascaris, Erik and Loerting, Thomas and Panagiotopoulos, Athanassios Zois and Russo, John and Sellberg, Jonas Alexander and Stanley, Harry Eugene and Tanaka, Hajime and Vega, Carlos and Xu, Limei and Pettersson, Lars Gunnar Moody},
title = {Water: A Tale of Two Liquids},
journal = {Chemical Reviews},
volume = {116},
number = {13},
pages = {7463-7500},
year = {2016},
doi = {10.1021/acs.chemrev.5b00750}
}

@book {kuznetsov1998,
    AUTHOR = {Kuznetsov, Yuri A.},
     TITLE = {Elements of applied bifurcation theory},
    SERIES = {Applied Mathematical Sciences},
    VOLUME = {112},
   EDITION = {Second},
 PUBLISHER = {Springer-Verlag, New York},
      YEAR = {1998},
     PAGES = {xx+591},
      ISBN = {0-387-98382-1},
   MRCLASS = {37Gxx (34-04 34C23 37-04 37M20)},
  MRNUMBER = {1711790},
}

@book{pruessner2012,
place={Cambridge},
title={Self-Organised Criticality: Theory, Models and Characterisation},
publisher={Cambridge University Press},
author={Pruessner, Gunnar},
year={2012},
isbn={9780511977671},
doi={10.1017/CBO9780511977671}
}

@Article{kienast_etal2013,
AUTHOR = {Kienast-Sj\"ogren, E. and Spichtinger, P. and Gierens, K.},
TITLE = {Formulation and test of an ice aggregation scheme for two-moment bulk microphysics schemes},
JOURNAL = {Atmospheric Chemistry and Physics},
VOLUME = {13},
YEAR = {2013},
NUMBER = {17},
PAGES = {9021--9037},
URL = {https://acp.copernicus.org/articles/13/9021/2013/},
DOI = {10.5194/acp-13-9021-2013}
}

@Book{golubitsky_schaeffer1985,
  author = 	 {Golubitsky, Martin and Schaeffer, David G.},
  ALTeditor = 	 {},
  title = 	 {Singularities and Groups in Bifurcation Theory},
  publisher = 	 {Springer New York, NY},
  year = 	 {1985},
  OPTkey = 	 {},
  OPTvolume = 	 {},
  OPTnumber = 	 {},
  OPTseries = 	 {Applied Mathematical Sciences},
  OPTaddress = 	 {},
  OPTedition = 	 {},
  OPTmonth = 	 {},
  OPTnote = 	 {},
  OPTannote = 	 {},
  isbn=          {978-0-387-90999-8},
  doi=           {10.1007/978-1-4612-5034-0}
}

@article{libbrecht2005,
doi = {10.1088/0034-4885/68/4/R03},
url = {https://dx.doi.org/10.1088/0034-4885/68/4/R03},
year = {2005},
month = {mar},
publisher = {},
volume = {68},
number = {4},
pages = {855},
author = {Libbrecht, Kenneth G},
title = {The physics of snow crystals},
journal = {Reports on Progress in Physics}
}

@article{rosemeier_spichtinger2020,
url = {https://doi.org/10.1515/mcwf-2020-0104},
title = {Pattern formation in clouds via Turing instabilities},
author = {Juliane Rosemeier and Peter Spichtinger},
pages = {75--96},
volume = {6},
number = {1},
journal = {Mathematics of Climate and Weather Forecasting},
doi = {doi:10.1515/mcwf-2020-0104},
year = {2020}
}

@Article{stubenrauch_etal2017,
AUTHOR = {Stubenrauch, C. J. and Feofilov, A. G. and Protopapadaki, S. E. and Armante, R.},
TITLE = {Cloud climatologies from the infrared sounders AIRS and IASI: strengths and applications},
JOURNAL = {Atmospheric Chemistry and Physics},
VOLUME = {17},
YEAR = {2017},
NUMBER = {22},
PAGES = {13625--13644},
URL = {https://acp.copernicus.org/articles/17/13625/2017/},
DOI = {10.5194/acp-17-13625-2017}
}

@article{khvorostyanov1995,
title = {Mesoscale processes of cloud formation, cloud-radiation interaction, and their modelling with explicit cloud microphysics},
journal = {Atmospheric Research},
volume = {39},
number = {1},
pages = {1-67},
year = {1995},
issn = {0169-8095},
doi = {https://doi.org/10.1016/0169-8095(95)00012-G},
url = {https://www.sciencedirect.com/science/article/pii/016980959500012G},
author = {V.I. Khvorostyanov}
}

@article{wacker1992,
  volume = {65},
  title = {Structual stability in cloud physics using parameterized microphysics},
  author = {Ulrike Wacker},
  year = {1992},
  pages = {231--242},
  journal = {Beitr{\"a}ge zur Physik der Atmosph{\"a}re}
}

@article{wacker1995,
  author = {Wacker, Ulrike},
  title = {Competition of Precipitation Particles in a Model with Parameterized Cloud Particles},
  journal = { J. Atmos. Sci.},
  volume = {52},
  number = {14},
  year = {1995},
  pages = {2577--2589}
}

@Article{wacker2006,
  AUTHOR = {Wacker, U.},
  TITLE = {Nonlinear effects in a conceptual multilayer cloud model},
  JOURNAL = {Nonlinear Processes  in Geophysics},
  VOLUME = {13},
  YEAR = {2006},
  NUMBER = {1},
  PAGES = {99--107},
  optURL = {http://www.nonlin-processes-geophys.net/13/99/2006/},
  DOI = {10.5194/npg-13-99-2006}
}

@article{hauf1993,
    author = "Thomas Hauf",
    title = "Microphysical kinetics in phase space: The warm rain process ",
    journal = "Atmospheric Research ",
    volume = "29",
    number = "1–2",
    pages = "55 - 84",
    year = "1993",
    note = "",
    issn = "0169-8095",
    doi = "10.1016/0169-8095(93)90037-O",
    optURL = "http://www.sciencedirect.com/science/article/pii/016980959390037O",
    
}

@Article{koren_feingold2011,
  author = 	 {Koren, Ilan and Feingold, Graham},
  title = 	 {Aerosol--cloud--precipitation system as a predator-prey problem},
  journal = 	 {Proceedings of the National Academy of Sciences},
  year = 	 {2011},
  OPTkey = 	 {},
  volume = 	 {108},
  number = 	 {30},
  pages = 	 {12227--12232},
  OPTmonth = 	 {},
  OPTnote = 	 {},
  OPTannote = 	 {}
}

@Article{feingold_koren2013,
  author = 	 {Feingold, G and Koren, I},
  title = 	 {A model of coupled oscillators applied to the aerosol--cloud--precipitation system},
  journal = 	 {Nonlinear Processes in Geophysics},
  year = 	 {2013},
  OPTkey = 	 {},
  volume = 	 {20},
  number = 	 {6},
  pages = 	 {1011--1021},
  OPTmonth = 	 {},
  OPTnote = 	 {},
  OPTannote = 	 {},
publisher={Copernicus GmbH}
}

@book{guckenheimer_holmes2002,
  year = {2002},
  address = {New York},
  author = {Guckenheimer, John and Holmes, Philip},
  edition = {7th},
  pages = {XVI, 462},
  isbn = {978-0-387-90819-9},
  number = 42,
  publisher = {Springer},
  series = {Applied mathematical sciences},
  title = {Nonlinear oscillations, dynamical systems, and bifurcations of vector fields},
  doi = {10.1007/978-1-4612-1140-2}
}

% % Was ist noch zu tun?
% \todototoc
% \listoftodos

%TC:endignore

%%%%%%%%%%%%%%%%%%%%%%%%%%%%%%%%%%%%%%%%%%%%%%%%%%%%%%%%%%%%%%%%%%%%%%%%%%%%%%%%
%% =============================================================================
%% =============================================================================
\end{document}